\documentclass[preprint,8pt]{elsarticle}

\usepackage{float}
\usepackage{color}
\usepackage{graphicx}
\usepackage{subfigure}
\usepackage{subcaption}
\usepackage[margin=2.5cm]{geometry}
\usepackage{empheq}
\usepackage{stmaryrd}
\usepackage{algorithm}
\usepackage{algpseudocode}
\usepackage{tabularx}
\usepackage{placeins}
\usepackage{tikz}
\usetikzlibrary{arrows.meta}

\usepackage[colorlinks=true,
            linkcolor=black,
            citecolor=black,
            urlcolor=black]{hyperref}
\usepackage{amssymb}
\usepackage{amsmath}
\usepackage{mathtools}
\usepackage{booktabs}

\usepackage{amsthm}

\journal{Material Review}

\begin{document}
\let\WriteBookmarks\relax
\def\floatpagepagefraction{1}
\def\textpagefraction{.001}

\begin{frontmatter}



\title{
Triboelectric nanogenerators across length scales: linking materials, multiscale modelling and applications
}

\author[GCEC]{MD Tanzib Ehsan Sanglap\corref{cor1}} 
\author[DSD]{Ans Al Rashid\corref{cor1}}
\author[DFE]{Abdullah Sayam\corref{cor1}}
\author[LIFD]{Injamamul Arief}
\author[GCEC]{Andrei G. Shvarts}
\author[GCEC]{Lukasz Kaczmarczyk}
\author[DSD]{Muammer Koc}
\author[SW]{Mokarram Hossain\corref{cor2}} 
\cortext[cor1]{Authors have equal contributions}
\cortext[cor2]{Corresponding author: \texttt{mokarram.hossain@swansea.ac.uk}}

\affiliation[GCEC]{organization={Glasgow Computational Engineering Centre (GCEC), James Watt School of Engineering, University of Glasgow},
            city={Glasgow},
            postcode={G12 8QQ},
            country={UK}}
            
\affiliation[DSD]        
{organization={Division of Sustainable Development, College of Science and Engineering, Hamad Bin Khalifa University},
            city={Doha},
            postcode={34110},
            country={Qatar}}
\affiliation[DFE]        
{organization={Department of Fabric Engineering, Bangladesh University of Textiles},
            city={Dhaka},             
            postcode={1208},
            country={Bangladesh}}
\affiliation[LIFD]        
{organization={Leibniz-Institut für Polymerforschung Dresden e.V. (IPF)},
            city={Dresden },             
            postcode={01069},
            country={Germany}}
\affiliation[SW]        
{organization={Department of Mechanical Engineering, Swansea University},
            city={Swansea},             
            postcode={SA1 8EN},
            country={UK}}

\begin{abstract}
Triboelectric nanogenerators (TENGs) offer a versatile route for converting
low-frequency mechanical energy into electrical output and for enabling
self-powered sensing across a broad range of device dimensions. Their
performance is governed by strongly interconnected factors, including
triboelectric material selection, interfacial charge generation, surface
morphology, mechanical contact, device architecture, environmental
conditions, and external circuit behaviour. However, these aspects are often reviewed independently, limiting the development of an integrated design perspective. In contrast, the current review brings together recent progress in TENG materials, computational modelling, and applications across length scales. First, triboelectric materials are examined in terms of fluorinated polymers, silicone and elastomeric polymers, thermoplastics, flexible polymer films, hybrid multifunctional composites, inorganic materials, and natural or biodegradable materials.  Second, computational approaches are reviewed from device-scale electrostatic and circuit models to contact mechanics, explicit rough-surface simulations, molecular dynamics, and first-principles methods. Particular attention is given to the limitations associated with prescribed surface charge density, idealised geometries, nominal contact-area assumptions, and coupling between mechanical, electrostatic, and circuit responses. Third, TENG applications are classified according to macroscale, milliscale, and microscale device dimensions, highlighting how dimensional constraints influence material choice, architecture, mechanical integration, and functionality. The review identifies the development of experimentally validated multiscale models that couple real contact evolution, contact-dependent charge generation and decay, electrostatic induction, and circuit dynamics as a central requirement for predictive TENG design. Integrating material engineering, computational modelling, experiments and scale-dependent application requirements will be essential for translating TENG technologies from individual proof-of-concept devices to
reliable and optimised self-powered systems.
\end{abstract}

\begin{graphicalabstract}
    \centering
    \vspace{1cm}
    \includegraphics[width=\textwidth]{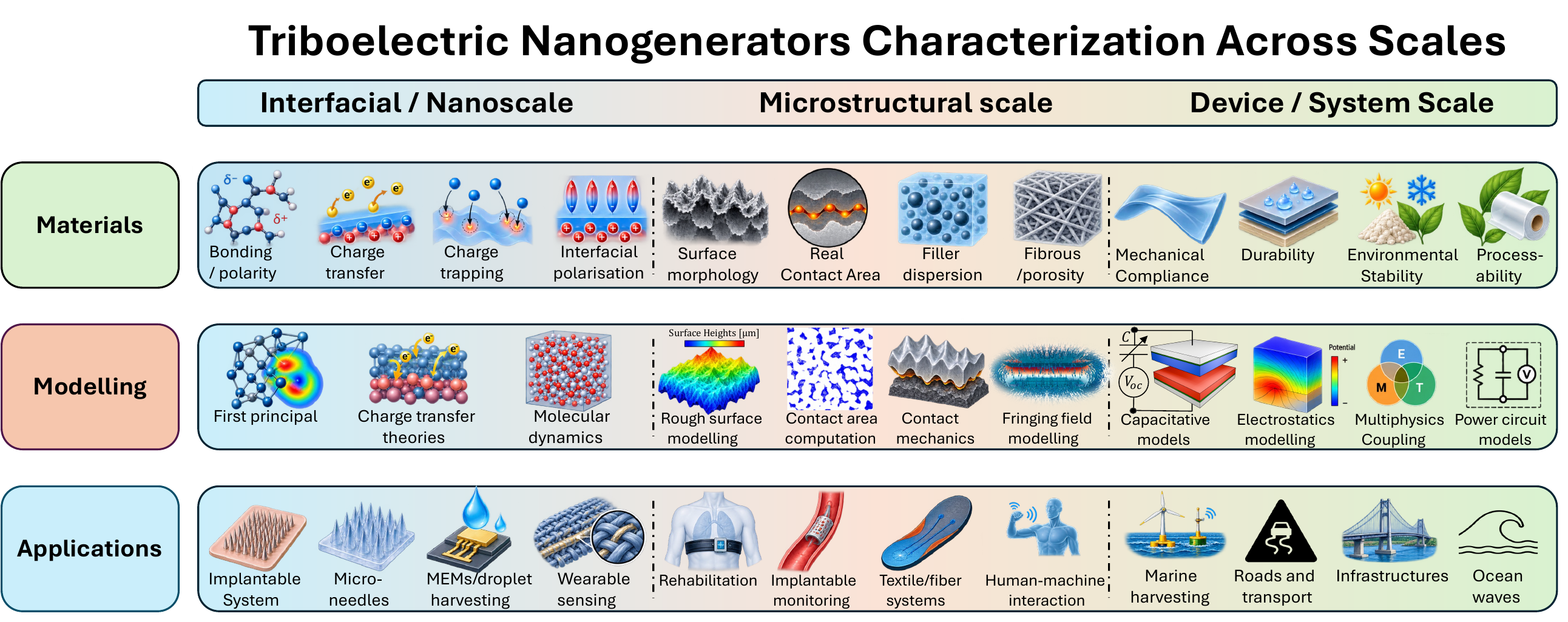}\

\end{graphicalabstract}


\begin{highlights}

\item Triboelectric materials are examined across interfacial and device scales.

\item TENG models are assessed by their inputs, predictions and validation 

\item Real contact area links roughness to load-dependent electrical output.

\item Identifies routes towards device dimensions shape material selection and application requirements.

\end{highlights}


\end{frontmatter}



\section{Introduction}
Triboelectric nanogenerators (TENGs) enable the harvesting of widely distributed mechanical energy to electrical output via contact electrification and electrostatic induction. They have gained unprecedented attention in recent times thanks to their wide material compatibility, simple architectures, low-frequency operation, mechanical adaptability, and high potential for self-powered sensing and energy harvesting in environments where typical power sources are difficult to use \cite{zhang2020material,udofa2025recent}. With the development of the field, it has been recognised that a single design parameter cannot determine TENG performance; rather, the interplay among material chemistry, interfacial charge physics, device structure, and system-level integration is the key. Accordingly, the modern TENG-related literature can be categorised into three domains: the development of high-performance triboelectric materials, the improvement of analytical and numerical simulation tools, and the further extension of the application field from microscale sensors to large-area and macroscale energy systems \cite{chen2020universal,xu2022dynamics}.

The output performance of TENGs is strongly influenced by the triboelectric characteristics of contact surfaces, dielectric properties, charge-trapping ability, surface roughness, flexibility, durability, and environmental stability. Therefore, the material portfolio for TENGs is rapidly expanding and includes fluorinated polymers, elastomers, engineering polymers, natural and biodegradable materials, inorganic functional constituents, and composites to enhance charge generation \cite{zhu2024biopolymer,li2024review,davoudi2024review}. The optimisation of materials is not limited to open-circuit voltage or short-circuit current; it also extends to real-world applications such as flexible electronics, biocompatible systems, harsh environments, textiles, sustainable devices, etc. \cite{davoudi2024review,dien2025energy}. Therefore, the material development for TENGs reflects a broader transition from proof-of-concept to more application-oriented and performance-engineered systems. 

In addition, computational and theoretical modelling have become essential for understanding, predicting, and optimising TENG behaviour. These models have shown that the charge transfer, potential distribution, and output current are geometry-dependent, voltage-dependent, and charge-density-dependent \cite{chen2020universal,xu2022dynamics,zhang2020theories,shao2020theoretical}. Modelling approaches are especially crucial for TENG operations because they encompass multiscales ranging from nanoscale charge transfer and interfacial states to microscale surface roughness and local field distributions, mesoscale device mechanics, and macroscale packaging and deployment. Existing modelling approaches have provided important insight into TENG behaviour. Analytical and equivalent-circuit models have been widely used to describe idealised contact-separation, sliding, single-electrode, and freestanding-layer TENGs~\citep{XUVoc, dharmasena2017triboelectric,dharmasena2018nature}. Finite element simulations have further enabled the analysis of electric potential, electric field, capacitance, and charge distribution in more complex geometries~\citep{sanglap2026multiphysics,shvarts2021finite,venugopal2022effective, mulvihill2025test}. More recently, contact mechanics, rough-surface modelling, molecular dynamics, density functional theory, and circuit simulations have been introduced to investigate specific aspects of TENG operation. However, most existing models still rely on prescribed surface charge density, simplified contact assumptions, idealised geometries, or sequential coupling between mechanical and electrical fields. However, the literature addresses TENG materials, modelling methods, and applications separately, without considering them as interdependent components of a complete design framework \cite{udofa2025recent,zhang2020material}. 

Over the past decade, several review articles have summarised the current state of the art in the field of TENGs. Table \ref{tab:teng_review_articles} compares representative reviews published during the past decade, their contributions, and the current gaps.
\begin{table}[!b]
\centering
\caption{Review articles on triboelectric nanogenerators and their limitations.}
\label{tab:teng_review_articles}
\resizebox{\textwidth}{!}{%
\begin{tabular}{
    c
    p{0.34\textwidth}
    p{0.40\textwidth}
    c
}
\toprule
\textbf{Year} &
\textbf{Scopes / Contributions} &
\textbf{Limitations / Gaps} &
\textbf{Ref.}
\\
\midrule

2020 &
Material selection for TENGs, flexibility, stability, sustainability, and application-related material requirements &
Limited attention to coupling material properties with modelling and device-level requirements &
\cite{zhang2020material}
\\

2020 &
Theoretical foundations of TENG operation and analytical models for established working modes &
Limited discussion of material-specific parameters, realistic multiphysics simulations, and translation to applications &
\cite{zhang2020theories}
\\

2020 &
Mathematical modelling of TENGs and theoretical treatment of different operating modes &
Limited coverage of material microstructure, computational implementation, and application-driven multiscale modelling &
\cite{shao2020theoretical}
\\

2022 &
Dynamic behaviour of TENGs and dynamic output optimisation &
Limited comparison of material families and interfacial phenomena &
\cite{xu2022dynamics}
\\

2023 &
Fundamental mechanisms, materials, device structures, power management, sensing, energy harvesting, integrated systems, and commercial opportunities &
Comprehensive but not organised around explicit interactions among materials, computational modelling, and applications across scales &
\cite{choi2023recent}
\\

2023 &
Environmentally friendly natural materials and green TENG applications &
Restricted to natural materials, and modelling is outside the scope &
\cite{liu2023environmentally}
\\

2024 &
Material design for high-performance TENGs and performance improvement strategies &
Detailed materials analysis, but analytical and modelling approaches and applications are not classified &
\cite{li2024review}
\\

2024 &
Biopolymer-based TENGs and their applications in healthcare and environmental monitoring &
Restricted to biopolymers, with limited computational modelling analysis &
\cite{zhu2024biopolymer}
\\

2024 &
Energy harvesting, self-powered sensing, wearable electronics, IoT-related applications, environmental considerations, and practical challenges &
Applications are emphasised, while material--model relationships and computational design approaches receive comparatively limited attention &
\cite{davoudi2024review}
\\

2025 &
Working mechanisms, material developments, structural configurations, application trends, current challenges, and future opportunities &
Broad review without a dedicated multiscale or integrated analysis of modelling methods and material behaviour &
\cite{rajasekar2025recent}
\\

2026 &
Working principles, operating modes, materials, structural designs, environmental effects, performance factors, and applications &
Broad coverage but limited critical integration of computational modelling with material-selection decisions and scale-dependent applications &
\cite{udofa2025recent}
\\

\bottomrule
\end{tabular}%
}
\end{table}
For instance, Zhang and Olin \cite{zhang2020material} comprehensively reviewed materials used in TENGs, highlighting their relationships to power density, stability, flexibility, and sustainability across various applications. Further, the theoretical background of TENGs was summarised by Zhang et al. \cite{zhang2020theories}, with an emphasis on the analytical understanding of device physics and the development of a systematic modelling framework of output performance. Shao et al. \cite{shao2020theoretical} reviewed theoretical modelling of TENGs, with a focus on formal analytical models and predictive methods. An et al. \cite{an2021advances} explored recent developments and future directions for TENGs in self-powered systems, including their operating modes, device architectures, and applications on autonomous platforms. Xu et al. \cite{xu2022dynamics} discussed the dynamics of TENGs, including the behaviour of springs, magnetic fields, structural movement, and optimisation of output under dynamic operating conditions. Han et al. \cite{han2021paper} reviewed energy harvesting and interactive applications of paper-based TENGs, and Liu et al. \cite{liu2023environmentally} reviewed naturally occurring materials used as TENGs. This review discusses modelling approaches for TENGs across scales, from contact-electrification mechanisms to device-level material selection and design. It highlights the strengths and limitations of existing approaches and identifies promising routes towards predictive multiscale TENG applications. In a recent review, Li et al. \cite{li2024review} emphasized the design of materials for high-performance TENGs, particularly on how to enhance the output by optimizing charge generation and charge loss. Zhu et al. \cite{zhu2024biopolymer} discussed the use of biopolymers in TENGs, highlighting their sources, sustainability, and health and environmental applications. Davoudi et al. \cite{davoudi2024review} reviewed recent applications and challenges of TENGs, while Dien et al. \cite{dien2025energy} focused on energy storage, power management, and system-level applications of TENGs. Rajasekar and Sivakumar \cite{rajasekar2025recent} discussed recent trends and prospects in advanced materials and emerging applications. Recently, Udofa et al. \cite{udofa2025recent} reviewed progress in TENGs by discussing their theoretical background, operating modes, material selection, and environmental influences. 

The reviews mentioned above cover various critical aspects of TENGs, but the literature remains scattered across materials, theory, dynamics, sustainability, power management, and applications. So, there is a lack of literature that examines the choice of materials, how they are used, and the importance of that choice across scales (macro, micro, and nano).  Therefore, the current study aims to systematically review TENGs, including materials, computational modelling, and their applications across length scales. First, it presents the materials landscape of TENGs, particularly covering organic/polymeric, inorganic, natural/biodegradable, and composite materials. Secondly, it covers the computational modelling landscape, including theoretical formulations, finite element simulations, multiphysics methods, and their roles in predicting TENG behaviour. Thirdly, it organises applications across the length scales ranging from microscale to mesoscale.

\section{Recent Materials in TENGs}

Materials used in triboelectric nanogenerators can be broadly categorised into organic/polymeric, inorganic, natural, biodegradable, and composite materials, each offering distinct advantages in charge generation, mechanical compliance, stability, and device applicability. Organic and polymeric materials remain the most widely used class for TENG fabrication because they can be easily processed into films, fibres, porous membranes, and stretchable architectures, while offering broad tunability of triboelectric polarity and mechanical behaviour. Their widespread use in TENGs also reflects their compatibility with wearable electronics, flexible devices, and scalable fabrication routes \cite{liu2026unveiling}. The triboelectric response of these materials is governed by physicochemical phenomena occurring across different length scales. At the molecular and nanoscale levels, chemical bonding and surface functional groups influence charge transfer and retention. At the microstructural level, surface morphology, porosity, filler distribution, and interfacial polarization affect local contact characteristics and dielectric behavior. At the bulk-material scale, mechanical properties, viscoelastic response, structural integrity, and environmental stability influence material behavior under repeated mechanical loads. The relative importance of these phenomena varies across material categories, underscoring the need to understand multiscale structure–property relationships in triboelectric materials \cite{zhang2020material,li2024review}. Primarily, various materials for TENGs are discussed in detail in the following sections. On top of that, Table S1 provides an overview of different classes of triboelectric materials and their applications from recent studies. 
\subsection{Fluorinated Polymers}
\begin{figure}[!b]
    \centering
    \includegraphics[width=0.7\linewidth]{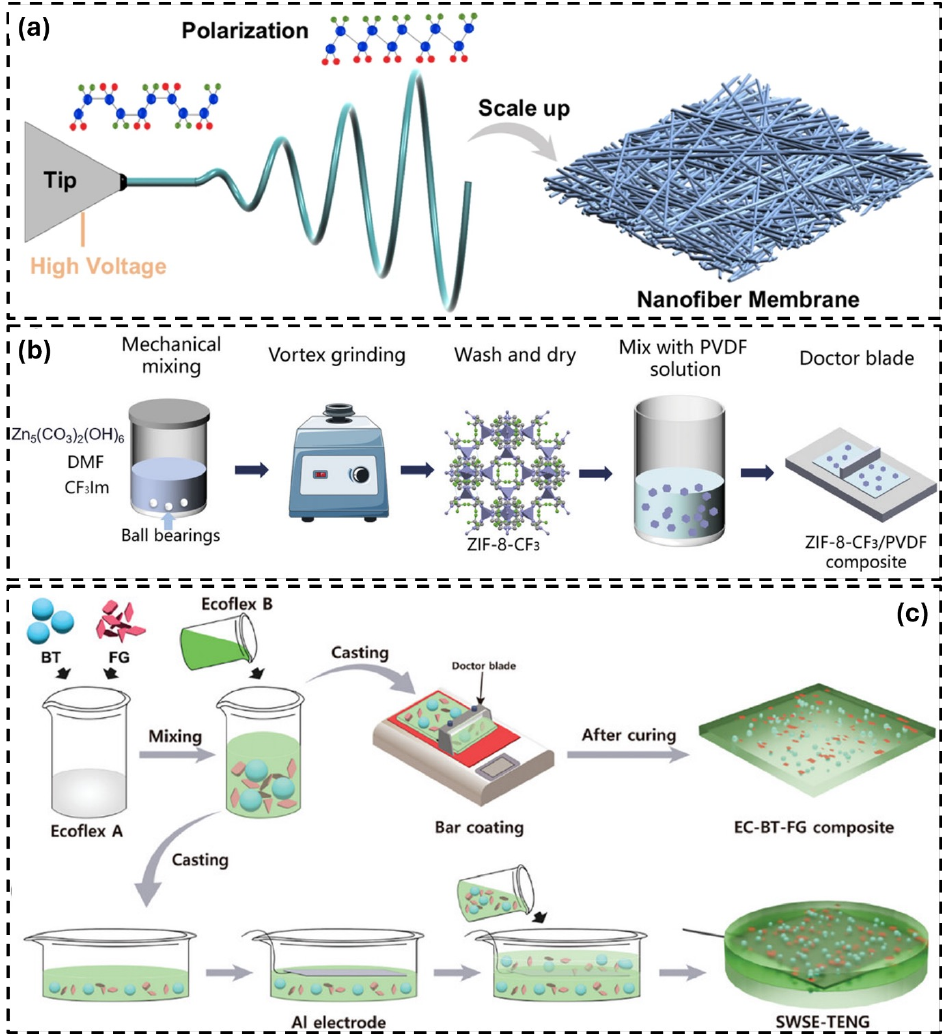}
    \caption{Fluorinated Polymers-Based TENGs (a) Schematic diagram of Self-Polarization process in the Electrospinning of PVDF-Based Polymers (Adapted from Ref. \cite{liu2026unveiling}). (b) Schematic Diagram of the Preparation Procedures for $ZIF-8-CF_3$ and $ZIF-8-CF_3$/PVDF Composite Film (Adapted from Ref. \cite{ye2025noncontact}). (c) An illustration of the steps involved in the fabrication of SWSE-TENG (Adapted from Ref. \cite{yoon2025creating}).}
    \label{fig:Self-Polarization_Process}
\end{figure}
Fluorinated polymers are the most effective tribo-negative materials because fluorinated groups increase electron affinity and favour charge accumulation during repeated contact electrification. Recent studies have attempted to chemically modify PVDF- or PTFE-based tribolayers to enhance their triboelectric potential. For instance, Liu et al. \cite{liu2026unveiling} investigated the synergistic contribution of C-F and C-Cl bonds within PVDF copolymers in altering triboelectric polarity. Therein, PVDF-based copolymers were spin-coated using high-voltage polarisation to produce nanofibre membranes, as illustrated in Figure~\ref{fig:Self-Polarization_Process}(a). PVDF copolymers achieved the highest triboelectric charge density of 310 $\mu C/cm^2$. Fluorine-functionalised coordination polymer eutectic materials have also been introduced as alternative high-output triboelectric materials, resulting in a short-circuit current of 47.18 $\mu$A at 5~Hz, a charge density of 91.53 $\mu C/m^2$ and a power density of 1584.04 mW/m² \cite{zhao2025fluorinated}. Recent studies on fluorinated polymers have also included the multifunctionality and durability of TENG devices. For instance, Ye et al. \cite{ye2025noncontact} synthesised $ZIF-8-CF_3$ nanoparticles to develop a fluorinated MOF/PVDF-based noncontact rotational TENG (as illustrated in Figure~\ref{fig:Self-Polarization_Process}(b)) that delivered ~280 V and ~19.0 $\mu$A, demonstrating that fluorinated framework-assisted systems can combine strong electrical output with reduced mechanical abrasion during operation. Likewise, Mahapatra et al. \cite{mahapatra2026fluorinated} improved the triboelectric performance of PDMS through fluorinated silane functionalisation and fabricated single-electrode devices for biomechanical energy harvesting and sensing.

PDMS films spin-coated on ITO/PET substrates were surface-functionalised with 1H, 1H, 2H, 2H-perfluorooctyl-trichloroethoxysilane (FOTES), resulting in increases in open-circuit voltage from 91.5 to 390 V and short-circuit current from 10 to 65 $\mu A$. Yoon et al. \cite{yoon2025creating} utilised barium titanate and graphite fluorinated polymer in the Ecoflex matrix to develop a single-electrode TENG (SWSE-TENG). The synthesis process of SWSE-TENG is reported in Figure 1(c), which delivered an open-circuit voltage of 1000 V and a short-circuit current of 25 $\mu A$. Fluorinated polymers have been explored for applications in developing TENGs for smart and washable flooring \cite{yoon2025creating}, composite polyimide films \cite{hu2026development}, and wearable e-textiles/bioelectronics \cite{zhao2026fluorinated}. Fluorinated polymers remain key tribo-negative materials; however, their performance depends not only on fluorination but also on interfacial chemistry, wear resistance, and device scalability \cite{liu2026unveiling,ye2025noncontact,hu2026development}. At the molecular level, fluorine-containing groups influence electron acquisition and charge retention, as demonstrated by C–Cl and C–F bonds in PVDF copolymers \cite{liu2026unveiling}. Nanofibrous architectures and surface functionalization assist in the modification of triboelectric interface characteristics at the microstructural level \cite{liu2026unveiling,mahapatra2026fluorinated}. Wear resistance and mechanical integrity at larger material scales become important for preserving material functionality under repeated contact \cite{ye2025noncontact,hu2026development}.

\subsection{Silicone and Elastomeric Polymers}
\begin{figure}[!b]
    \centering
    \includegraphics[width=0.6\linewidth]{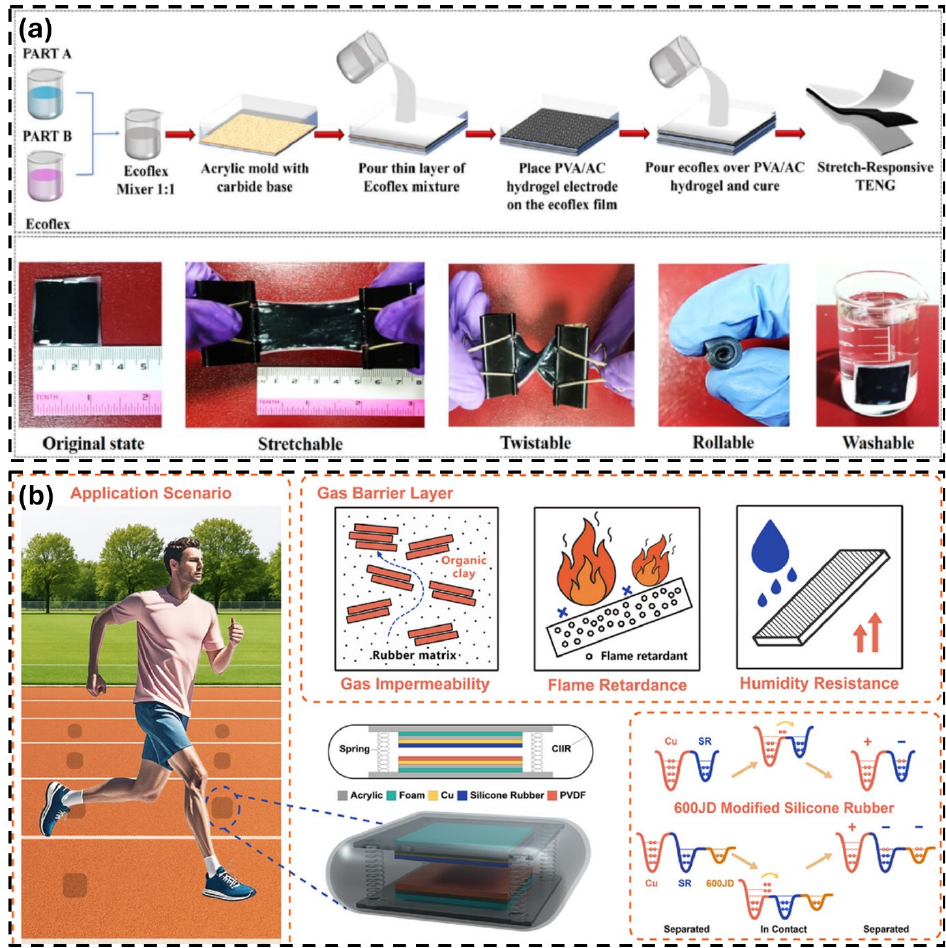}
    \caption{Silicone and Elastomers Based TENGs (a) Fabrication process of stretch-responsive TENG using a layer-by-layer approach and different mechanical distortions (Adopted from Ref. \cite{viswanathan2025stretch}). (b) Structural Design, Performance, and Applications of HTV Silicone TENG (Adopted from Ref. \cite{wei2026ambient})}
    \label{fig:Fabrication_Process}
\end{figure}
Silicone and elastomeric polymers are ideal materials for wearable and deformable TENGs due to their low modulus, high stretchability, and conformal contact behaviour, while their major limitation is the limited triboelectric performance in neat elastomers. Therefore, recent literature has focused on developing silicone/elastomeric composites. For example, Karimmirza and Manavizadeh \cite{karimmirza2025high} utilised Ag and Cu nanoparticles as reinforcement to silicone, in which an Ag-based TENG delivered 710 V, 35 $\mu A$, and 10.08 mW, with a normalised power density of 356 $\mu W/cm^2$ and about 93.5$\%$ output retention after $30,000$ cycles. This study also concluded that nanoparticle dispersion within the silicone matrix drives the triboelectric performance and durability of TENGs. In another study, stretch-responsive Ecoflex-based devices with porous PVA/activated-carbon hydrogel electrodes were developed, which demonstrated that elastomeric tribolayers can be paired with soft conductive media to preserve compliance and sensing capability under deformation, as illustrated in Figure~\ref{fig:Fabrication_Process}(a) \cite{viswanathan2025stretch}. Environmental stability has also been a major area of exploration in elastomeric TENGs. Wei et al. \cite{wei2026ambient} developed conductive-filler-modified silicone-rubber-based TENGs with gas-barrier encapsulation, specifically for ambient-stable footstep energy conversion, as demonstrated in Figure~\ref{fig:Fabrication_Process}(b).

High-performance functional MAX-phase nanofillers \cite{kumar2026boosting}, self-healing ionic elastomers \cite{wang2026flexible}, and organic-silicone hybrid adhesive elastomers \cite{bhattacharyya2025synthesis} have been used to enhance charge generation, self-repair capability, and interface adhesion in soft triboelectric devices. In addition, the literature reports several studies on the performance assessment of elastomer-based composites. A dual-filler PDMS-based TENG by Le et al. \cite{lee2025boosting} achieved 468 V, 43.04 µA, and 7.83 $W/m^2$, while Zhang et al. \cite{zhang2025flexible} utilised BaTiO$_3$ and carbon spheres in Ecoflex to produce a flexible TENG, which produced 532 V and 72 $\mu A$ as electrical output. More recently, MXenes have also been explored for their utilisation in the synthesis of elastomeric composites. For instance, Ecoflex/MXene composite films with orientated fillers resulted in improved triboelectric performance, with open-circuit voltage increased by $25.2\%$, short-circuit current by $34\%$, and charge density by $31\%$ compared to unorientated filler composites \cite{wang2026electric}, while $Ti_2C$ MXene-embedded PDMS revealed an open-circuit voltage, short-circuit current, and power density of 105 V, 6.78 $\mu A$, and 15.01 $\mu W /cm^2$, respectively \cite{chittibabu2026biomorphic}. 

In silicone and elastomeric polymers, incorporating functional nanofillers modifies the electrical characteristics of silicone-based tribolayers \cite{kumar2026boosting,lee2025boosting}. At the microstructural level, nanoparticle dispersion and filler orientation influence the electrical response of the composites, as demonstrated in metal-nanoparticle/silicone and MXene/Ecoflex systems \cite{karimmirza2025high,wang2026electric}. At larger scales, stretchability and the ability to accommodate mechanical deformation are key characteristics of soft elastomers, while self-healing ionic elastomers offer an additional approach to improving material recoverability \cite{viswanathan2025stretch,mandal2024self,wang2026flexible}. Silicone and elastomeric materials are an appropriate choice for applications that require mechanical compliance, conformal contact, and durable wearable integration, even if other material classes may achieve higher peak outputs under rigid conditions \cite{lee2025boosting,zhang2025flexible}.

\subsection{Thermoplastics}
\begin{figure}[!b]
    \centering
    \includegraphics[width=1\linewidth]{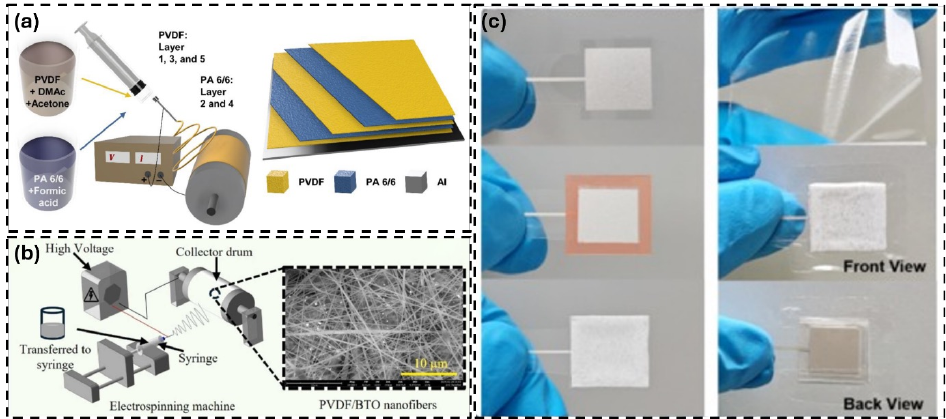}
    \caption{Thermoplastics Based TENGs (a) Electrospinning Process to Produce Multi-Electrospun Layers (Adopted from Ref. \cite{kim2025triboelectric}). (b) Electrospinning Process for Developing PVDF/BTO Nanofibers for AF-TENG (Adopted from Ref. \cite{bairagi2026electro}). (c) Fabrication Process of MEM-TENG (Adopted from Ref. \cite{wang2025flexible}).}
    \label{fig:Electrospinning_Process}
\end{figure}

Thermoplastics are also important in TENGs owing to their low cost, commercial availability, and compatibility with several processing techniques, including electrospinning, coating, and lamination. These materials play a vital role in the structural design and hybridisation of TENG devices. For instance, PVDF and PA 6/6 were hybridised via electrospinning to enhance contact through engineered architectures, as illustrated in Figure~\ref{fig:Electrospinning_Process}(a) \cite{kim2025triboelectric}. Similar approaches have been used for different materials, such as PVDF-BTO and TPU (Figure~\ref{fig:Electrospinning_Process}(b)) \cite{bairagi2026electro}, TiO$_2$/Nylon 66 composite nanofibers \cite{moon2026enhanced}, and recycled polypropylene from food packaging \cite{esmeria2025triboelectric}, to produce fibrous membranes from engineering polymers for TENG fabrication. In addition, efforts have been made to develop bioinspired and thermally stable TENG systems using hollow-array PVDF/MWNT membranes for musculoskeletal disorder monitoring (Figure~\ref{fig:Electrospinning_Process}(c)) \cite{wang2025flexible}, a cordierite-induced polyimide material for high-temperature vibration sensing \cite{han2025thermal}, and a polyimide/copper mesh no-gap device for force-sensing and integrated energy harvesting systems \cite{tran2025high}. For thermoplastic materials, electrospinning forms fibrous architectures, while incorporating functional constituents and engineered hollow structures modifies morphology and interfacial characteristics \cite{kim2025triboelectric,bairagi2026electro,moon2026enhanced}. At larger material scales, mechanical integrity and thermal stability become important considerations, as illustrated by polyimide-based materials modified with inorganic constituents or structured conductive layers \cite{han2025thermal,tran2025high}. Although their performance does not match that of the elastomeric composites, these materials offer scalable, structurally versatile, and practical manufacturing routes \cite{moon2026enhanced,han2025thermal,tran2025high}.

\subsection{Flexible Polymer Films}
\begin{figure}[!b]
    \centering
    \includegraphics[width=0.8\linewidth]{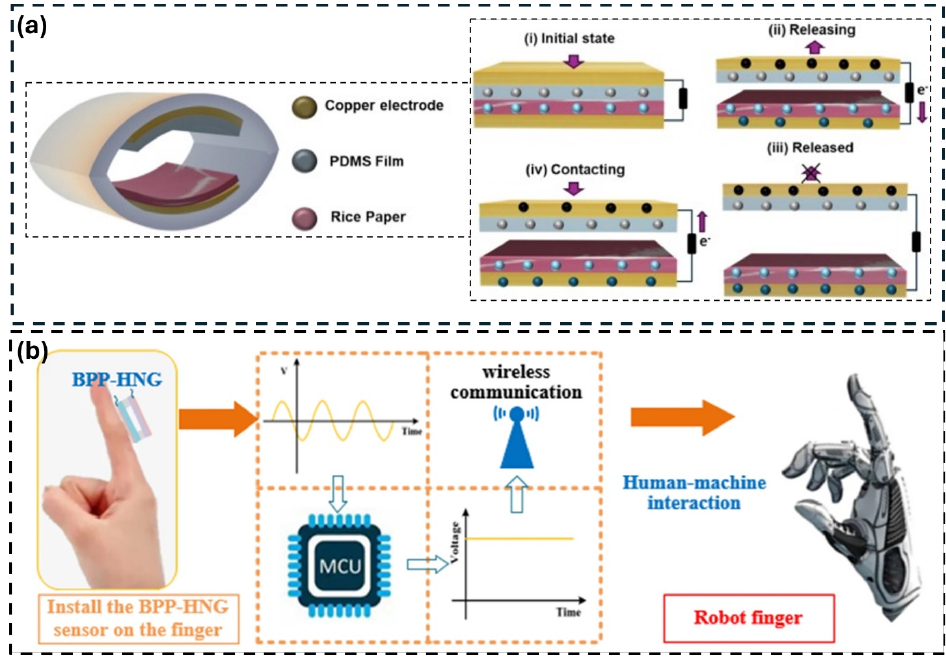}
    \caption{Flexible Polymer Films Based TENGs (a) Schematic Diagram and Working Mechanism of RP-PDMS TENG (Adopted from Ref. \cite{panda2026triboelectric}). (b) Schematic Diagram of BPP-HNG for Controlling Robotic Motion (Adopted from Ref. \cite{dong2025flexible})}
    \label{fig:RP_BPP_TENG}
\end{figure}

Flexible polymer films used as triboelectric materials are particularly important for lightweight, self-powered wearables/electronics. Recent research has extensively explored charge transfer potential and environmental durability of polymer-based flexible films \cite{yang2026mechanisms}. Yang et al. \cite{yang2026mechanisms} investigated the polarity-driven and material-dependent charge transfer for PVDF with different polymeric interfaces to illustrate the impact of polymer polarity and interface chemistry on triboelectric performance. In addition, degradable and hybrid flexible films are also gaining attention as triboelectric materials. For instance, a fully degradable spent coffee grounds and PCL-based TENG delivered a volumetric power density of 12.28 $mW/m^3$ \cite{duan2025fully}, while a flexible BTO-PVDF/PDMS piezo-triboelectric hybrid film demonstrated application in a real-time gesture monitoring system (Figure~\ref{fig:RP_BPP_TENG}(b)) \cite{dong2025flexible}. Likewise, all-cellulose thin-film \cite{chen2026all} and rice paper/PDMS \cite{panda2026triboelectric} TENGs demonstrate the growing use of flexible films for environmental compatibility and stability, including operation under natural wind, body motion, and extreme humidity. The schematic diagram and working principle of the rice paper/PDMS device by Panda et al. \cite{panda2026triboelectric} are illustrated in Figure~\ref{fig:RP_BPP_TENG}(a), which reported an output voltage of about 305 V, a current of 2.6 $\mu A$, and a charge of 29 nC under humid conditions, highlighting the importance of encapsulation and moisture management in practical film-based TENGs.  For flexible polymer films, polarity and chemical composition influence charge transfer, as demonstrated for PVDF interfaces with different polymers \cite{yang2026mechanisms}. At the microstructural level, multilayer configurations and the incorporation of functional constituents can modify the electromechanical response of hybrid polymer films \cite{dong2025flexible}. At larger material scales, flexibility and mechanical strength influence material integrity, while protective encapsulation can improve the stability of film-based assemblies under humid conditions \cite{chen2026all,panda2026triboelectric}.

\subsection{Hybrid Multifunctional Composites}
Hybrid multifunctional materials can be engineered strategically for their performance rather than improving individual constituents. Recently, TENGs combining different materials, including polymers and inorganic materials, have demonstrated improved dielectric constant, charge trapping, interfacial polarisation, conductivity, durability, and multifunctionality \cite{ye2025noncontact,karimmirza2025high,lee2025boosting}. Dual-filler PDMS \cite{lee2025boosting} and Ecoflex/BaTiO$_3$/Carbon sphere \cite{zhang2025flexible} TENGs have been reported to exhibit amplified polarisation, resulting in improved triboelectric performance. 

In addition, the utilisation of conductive fillers such as Ag nanoparticles, CNTs, MXenes, and carbon-based materials has proved to improve charge transfer and interfacial interactions in MOF/PVDF \cite{das2026microstructurally}, MXene/PVDF \cite{dev2026advanced}, MXene/Chitosan/CNT \cite{liu2025self}, silk-fibrin/MOF \cite{chakraborty2026natural}, and cellulose/graphene \cite{wang2025enhancing} TENGs. However, an excess concentration of fillers can lead to agglomeration, increased stiffness, charge leakage, and a conductive network that reduces triboelectric charge retention. Interactions between functional constituents and their organisation across different structural scales influence the properties of hybrid composites. At the nanoscale, the incorporation of dielectric and conductive fillers modifies the electrical characteristics of polymer-based composites \cite{lee2025boosting,zhang2025flexible}. At the microstructural level, filler orientation, distribution, and the formation of layered or hybrid architectures influence charge-storage characteristics and the overall electrical response \cite{das2026microstructurally,dev2026advanced,liu2025self}. At larger material scales, filler incorporation can introduce competing effects involving electrical performance and mechanical characteristics, particularly when increasing filler concentration promotes agglomeration or changes the conductivity of the composite \cite{lee2025boosting,wang2026electric}. Therefore, it is essential to balance the conductive network of fillers with triboelectric performance to maximise the output~\cite{lee2025boosting,wang2026electric}.

\begin{figure}[!b]
    \centering
    \includegraphics[width=0.9\linewidth]{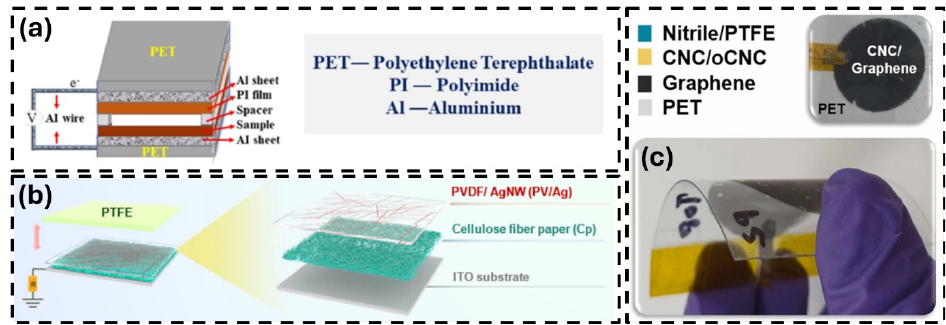}
    \caption{Hybrid Multifunctional Composites-Based TENGs (a) Schematic diagram of TENG Device (Adopted from Ref. \cite{dev2026advanced}). (b) Configuration of the TENG comprising a Cp-PV/Ag Layer on ITO glass paired with a PTFE Sheet (Adopted from Ref. \cite{yamklang2026agricultural}). (c) CNC/Graphene TENG (Adopted from Ref. \cite{alghamdi2025triboelectric})}
    \label{fig:Hybrid_Multifunctional_Composites}
\end{figure}
\subsection{Inorganic Materials}
Inorganic materials are effective when used as active coatings, membranes, structured interfaces, or function-enhancing phases, rather than as bulk replacements for polymers \cite{hao2025triboelectric,potu2023enhancing}. Polyoxometalate nanomaterials have been explored for gait-monitoring TENGs \cite{hao2025triboelectric}, while MOF-modified ZnO nanosheets have been developed to improve output through surface-state engineering \cite{potu2023enhancing,fan2012flexible}. Recent literature reviews also highlighted that perovskites and other emerging inorganic materials are becoming increasingly important for enhancing dielectric response and multifunctionality in TENGs \cite{liu2026research,jin2025perovskite}. Edge-effect-based liquid-metal solid-liquid TENGs \cite{ren2026edge} and microstructurally engineered MXene/MOF membranes \cite{das2026microstructurally} further highlight the applications of inorganic compounds toward unconventional solid-liquid interfaces and membrane-driven architectures for self-powered devices. In addition, Deb et al. \cite{deb2026zno} demonstrated the use of ZnO nanorod-modified cotton fabrics as a tribo-positive material to improve triboelectric performance without compromising wearability. At the nanoscale, surface modification and incorporation of functional inorganic phases can alter interfacial electrical characteristics, as shown by MOF-modified ZnO structures \cite{potu2023enhancing}. Nanosheets, nanorods, and engineered inorganic-containing membranes offer different surface architectures and interfacial arrangements \cite{das2026microstructurally,hao2025triboelectric,deb2026zno}. At larger material scales, the dielectric characteristics of inorganic constituents and their compatibility with polymeric matrices become important considerations for the effective behaviour of hybrid materials \cite{dev2026advanced,liu2026research,jin2025perovskite}. Inorganic materials are particularly important when dielectric amplification, charge trapping, conductivity modulation, or interfacial specialisation is desired \cite{liu2026research,dev2026advanced,das2026microstructurally}.

\subsection{Natural and Biodegradable Materials}
With growing interest in biodegradable and natural materials, there has been increased interest in developing sustainable TENGs. In addition, there have been increasing efforts to use natural waste as tribolayers in TENGs \cite{alghamdi2025triboelectric,liu2025self}. Cellulose and chitosan have been demonstrated as prominent materials in recent studies where a cellulose nanocrystal/graphene TENG was developed for biomechanical energy harvesting from piano playing \cite{alghamdi2025triboelectric}, and MXene/chitosan/CNT films have also demonstrated their potential for wearable health monitoring platforms \cite{liu2025self}. Gu et al. \cite{gu2025biodegradable} developed a biodegradable silk fibroin-based TENG regulated by interfacial and ionic polarisation, which achieved 63 V, 2.4 $\mu A$, and 828.8 $mW/m^2$. In addition, degradable materials and plant-protein-based devices have been explored for green electronics and sustainable agriculture \cite{xue2025triboelectric,jiang2022plant}. Recent reviews have also highlighted the rapid expansion of natural biowaste- and polybutylene adipate terephthalate (PBAT)-based TENGs \cite{shivanna2026pbat,troncoso2023lignocellulosic}. Within biodegradable materials, chitosan/PVA fibrous membranes \cite{ankit2025engineering}, chitosan/PVA nanofibres \cite{ankit2025scalable}, and edible microbial cellulose-based TENGs \cite{ankit2026edible} have demonstrated their potential for energy harvesting and self-powered sensing. Cellulose-based and pectin-rich TENGs derived from watermelon skin \cite{kim2025sustainable}, agricultural waste-based cellulose paper modified with PVDF/Ag nanowires \cite{yamklang2026agricultural}, silk-fibroin/MOF composites \cite{chakraborty2026natural}, and cellulose enhanced by dual interfacial polarisation \cite{wang2025enhancing} have also been presented. At the molecular and interfacial levels, chemical functionalization of cellulose influences surface charge characteristics and charge trapping, while interfacial and ionic polarization have been demonstrated in silk-fibroin-based materials \cite{wang2025enhancing,gu2025biodegradable}. At the microstructural level, cellulose-based composites and electrospun chitosan/PVA membranes showed how fibrous morphology and material distribution influence triboelectric performance \cite{alghamdi2025triboelectric,ankit2025engineering,ankit2025scalable}. At larger scales, forming mechanically robust cellulose films and modifying cellulose surfaces to improve hydrophobicity highlight the importance of material integrity and surface protection \cite{chen2026all,yamklang2026agricultural}. However, the output performance strongly depends on structural optimisation, charge trapping, and functional hybridisation, in addition to the sustainability aspect of these materials.

\section{Towards Predictive Modelling of TENG}
\begin{figure}[htbp]
\centering
\resizebox{\textwidth}{!}{%
\begin{tikzpicture}[x=0.65cm,y=0.98cm]

\tikzset{
  root/.style={
    draw,
    rounded corners,
    align=center,
    font=\normalsize\bfseries,
    text width=3.4cm,
    minimum height=1.5cm
  },
  box/.style={
    draw,
    rounded corners,
    align=center,
    font=\footnotesize,
    text width=2.1cm,
    minimum height=1.6cm
  },
  arrow/.style={
    -{Latex[length=1.8mm]},
    thick
  }
}

\node[root] (root) at (0,0) {Existing modelling approaches};

\node[box] (continuum) at (-12,-2.4) {Continuum /\\device-scale models};
\node[box] (interface)  at (-4,-2.4)  {Coupled interface and\\roughness models};
\node[box] (atomistic)  at (4,-2.4)   {Atomistic /\\molecular models};
\node[box] (circuit)    at (12,-2.4)  {Circuit models};

\node[box] (electro)  at (-14,-5.0) {Electrostatic\\FEM / BEM};
\node[box] (contact) at (-10,-5.0) {Contact and\\deformation};

\node[box] (micro) at (-6,-5.0) {Idealised\\microtextures};
\node[box] (rough) at (-2,-5.0) {Measured or random\\rough surfaces};

\node[box] (md)  at (2,-5.0) {Molecular dynamics\\MD / ReaxFF};
\node[box] (dft) at (6,-5.0) {First-principles\\DFT};

\node[box] (circuit_eqv)   at (10,-5.0) {Equivalent ODE\\circuit  models};
\node[box] (circuit_power) at (14,-5.0) {Power management\\circuit};

\draw[arrow] (root.south) -- (continuum.north);
\draw[arrow] (root.south) -- (interface.north);
\draw[arrow] (root.south) -- (atomistic.north);
\draw[arrow] (root.south) -- (circuit.north);

\draw[arrow] (continuum.south) -- (electro.north);
\draw[arrow] (continuum.south) -- (contact.north);

\draw[arrow] (interface.south) -- (micro.north);
\draw[arrow] (interface.south) -- (rough.north);

\draw[arrow] (atomistic.south) -- (md.north);
\draw[arrow] (atomistic.south) -- (dft.north);

\draw[arrow] (circuit.south) -- (circuit_eqv.north);
\draw[arrow] (circuit.south) -- (circuit_power.north);

\end{tikzpicture}
}
\caption{Broader classification of existing numerical modelling approaches for TENGs.}
\label{fig:teng_modelling_branches}
\end{figure}
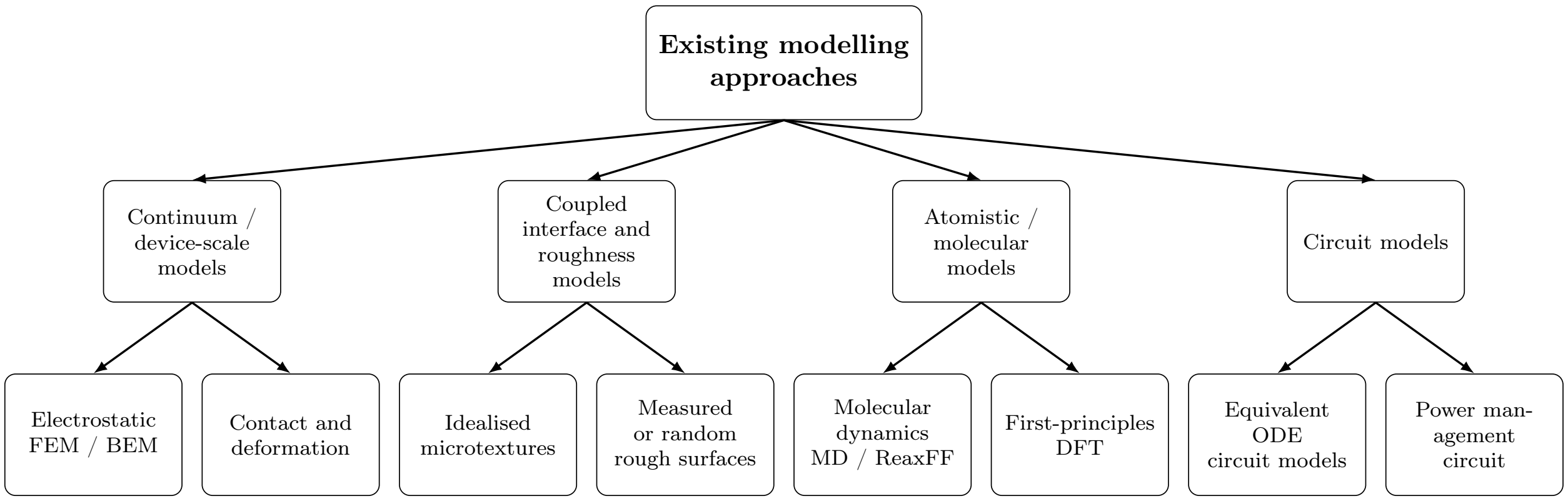

TENG output largely depends on various interacting physical processes. To accurately capture distinct and multiphysical behaviour, numerical modelling has become an influential tool for understanding device behaviour and guiding design optimisation. Existing TENG modelling approaches can be broadly summarised into continuum or device-scale models, coupled interface and roughness models, atomistic or molecular models and circuit-based models, as shown in Figure~\ref{fig:teng_modelling_branches}.These can be further broadly categorised into macro-, micro-, and nanoscale descriptions according to the characteristic dimensions of the features explicitly resolved. At the macroscale, models describe overall device deformation, electrostatic induction and circuit response to predict voltage, current and power. At the microscale, models resolve surface features and asperity deformation to determine how roughness and loading influence real contact area and local interfacial gaps. At the nanoscale, atomistic and molecular models examine interfacial structure and interactions, while suitable electronic-structure methods investigate electron-transfer mechanisms based on first principles~\cite{fatti2026first}. This distinction is useful because each scale requires different abstractions of physical detail, and explicitly resolving atomic interactions throughout an entire device scale is computationally impractical. Therefore, these individual physics sometimes overlap and are coupled as multiphysics models, which can span between micro- and nanoscale physics across all three scales. For example, micro/nano-scale roughness couples with macro-scale electrostatics, which can observe elemental-scale electric field bending. These interfacial quantities can then inform device-scale output voltage and current.

Further, analytical and equivalent-circuit models describe the electrical response of contact-separation and sliding-mode TENGs using quantities such as a time-varying capacitance, an open-circuit voltage source and an external load resistance~\citep{niu2013theoretical, XUVoc, dharmasena2018unified, shao2020theoretical}. These approaches provide physical insight at relatively low computational cost, although their applicability depends on assumptions concerning geometry, contact, motion and charge retention. The treatment of finite-size and fringing effects also varies between formulations.

From a numerical modelling perspective, a major unresolved challenge is developing a fully predictive TENG simulation framework in which the surface charge density is not prescribed but instead emerges from the evolving contact state, material properties, surface chemistry, humidity, roughness, and charge decay. Most existing models, therefore, combine predictive electrostatic or circuit calculations with phenomenological assumptions for contact electrification. This means that while current numerical models can often predict the potential distribution, capacitance, contact area, voltage, current, or power under prescribed conditions, the direct coupling among mechanical contact, charge generation, electrostatic induction, and circuit response for actual TENG surfaces remains only partially resolved. The following sections assess these approaches through three criteria: the quantities they predict, the inputs they require, and the conditions under which their predictions have been validated.

Early modelling approaches mainly focused on electrostatic behaviour and macroscale equivalent-circuit descriptions. These models are useful for estimating open-circuit voltage, transferred charge, current, and power output under idealised operating conditions. For example, analytical and circuit-based models have been widely used to describe contact-separation and sliding-mode TENGs, where the device is represented using a time-varying capacitance, an open-circuit voltage source, and an external load resistance \citep{niu2013theoretical, XUVoc, dharmasena2018unified, shao2020theoretical}. Although these approaches provide physical insight and low computational cost, they often rely on simplified geometries, e.g., nominal flat surfaces, load-independent separation distances, ideal contact assumptions, and neglect fringing effects, electrode conductivity, and charge leakage. Nonlinear behaviour, such as plastic deformation and hyperelastic material behaviour is further simplified or omitted.
\begin{figure}[!b]
    \centering
    \includegraphics[width=1\linewidth]{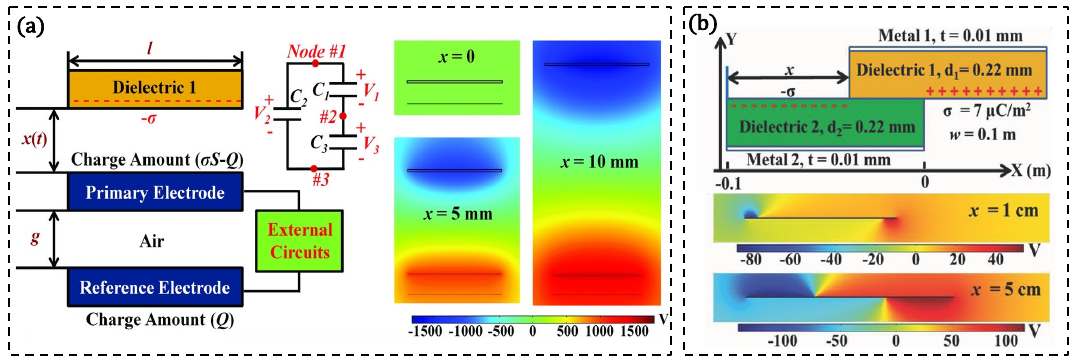}
    \caption{Non-contact electrostatics FEM simulation as a standard for validation of different analytical solutions, showing potential distribution of different TENG dielectric configurations: (a) single and double-electrode vertical contact-separation method at different air gaps (reproduced from \cite{niu2014theoretical}) (b) lateral sliding TENG layers at different sliding distances (reformatted from \citep{niu2013theory})}
    \label{fig:analytical verification}
\end{figure}
To overcome these limitations, finite element method (FEM) simulations have been introduced to solve the electrostatic field distribution in TENGs with more realistic geometries. Commercial multiphysics platforms such as COMSOL have been commonly used to analyse the potential distribution, electric field, capacitance variation, and charge transfer process \citep{shao2020theoretical, chen2020universal}. Furthermore, FEM simulations have frequently been used as benchmarks for
verifying analytical TENG models. Some significant contributory outcomes for different TENG modes are shown in Figure~\ref{fig:analytical verification}. However, several FEM-based electrostatic studies model triboelectric charging using prescribed surface charge densities and compute the resulting electrical response for specified geometries or motions~\citep{niu2013theory,kim2017morphology,chen2020universal}. This approach describes electrostatic induction but does not, by itself, predict the load-dependent real contact area arising from asperity deformation. Therefore, an additional mechanical contact model is required to capture this dependence even for macroscale investigations.

More advanced multiphysics models couple electrostatics with mechanical contact, surface roughness, material deformation, and circuit dynamics. This is particularly important because the electrical output of a TENG is strongly affected by the actual contact area between rough triboelectric surfaces at the microscale, rather than solely by the nominal contact area. When two rough triboelectric layers are pressed together, contact occurs locally at asperities, leaving gaps elsewhere within the nominal contact area. Tribocharges are generated only within these contact regions and therefore depend largely on the real contact area. Using the flexoelectric model~\cite{mizzi2022flexoelectricity}, these charges can also attain near the contact regions  Therefore, contact mechanical models and rough-surface descriptions are increasingly used to connect mechanical loading conditions with charge generation and electrostatic response. For example, Xu et al~\cite{XUVoc} used Persson’s contact theory \citep{persson2002elastic} to predict load-dependent electrical behaviour, while Sanglap et al~\citep{sanglap2026multiphysics} coupled FEM-based contact analysis of explicitly represented rough surfaces with electrostatic simulations and circuit dynamics.

Despite these developments, fully coupled and predictive simulation frameworks for TENGs remain limited for nanoscale investigation. Many existing models consider only a single physical process, such as electrostatics or circuit response, while others rely on simplified coupling assumptions among contact, charge generation, and electrical output. A robust modelling framework should therefore account for the interaction between mechanical contact, surface roughness, electrostatic induction, external circuit, and molecular behaviour. Such a framework is necessary for predicting TENG performance under realistic operating conditions and for supporting the design of efficient devices with complex geometries and triboelectric materials. Therefore, the following sections review three complementary aspects of TENG modelling in the literature: device-scale electrostatic modelling, microscale modelling of contact and deformation at triboelectric interfaces, and nanoscale molecular dynamics and atomistic simulations of contact electrification. The review discusses their capabilities and limitations in linking interfacial processes to device-level performance, while recognising the overlap between these scales.

\subsection{Device scale Electrostatics modelling}
A significant part of the numerical simulation literature focuses on predicting the spatial distribution of electric potential, electric field, induced charge, and capacitance in complex TENG geometries. Compared with closed-form analytical descriptions, finite element method (FEM)-based electrostatic simulations are more flexible because they can account for fringing fields, multilayer dielectric structures, curved surfaces, finite-size electrodes, non-uniform electric fields, and complex device configurations. In most electrostatic TENG simulations, the problem is formulated under a quasi-electrostatic assumption because the operating frequency of TENGs is usually much lower than the frequency range at which electromagnetic wave propagation effects become significant. The electric field is therefore expressed in terms of the scalar electric potential, $\mathbf{E} = - \nabla \phi$  where $\phi$ is the electric potential. For a dielectric medium with permittivity $\varepsilon$, the electric displacement field is given by the constitutive relation $\mathbf{D} = \varepsilon \mathbf{E}$. Using Gauss's law, the governing equation for the electrostatic potential becomes
\begin{equation}
\nabla \cdot  \mathbf{D} = \rho_f ,
\label{eq}
\end{equation}
where $\rho_f$ is the free charge density. In many TENG simulations, free charge is not distributed throughout the dielectric domain. Instead, the triboelectric charge is commonly prescribed as a surface charge density at the contact interface between two triboelectric materials. 

At a charged interface $\Gamma_T$ separating two material
regions, the normal electric displacement satisfies
\begin{equation}
    \hat{\mathbf n}_{\Gamma} \cdot
    \llbracket \mathbf{D} \rrbracket = \sigma_T,
    \qquad \text{on } \Gamma_T,
    \label{eq:surface_charge_condition}
\end{equation}
where $\llbracket \mathbf{D} \rrbracket
=\mathbf{D}_2-\mathbf{D}_1$,
$\hat{\mathbf n}_{\Gamma}$ points from region 1 to region 2,
and $\sigma_T$ is the signed free surface charge density. The total free charge on an electrode can subsequently
be obtained from the integral form of Gauss's law.
Let $\Omega_e$ be a control volume enclosing one electrode
and no other free charges. Then
\begin{equation}
    \oint_{\partial\Omega_e}
    \mathbf{D}\cdot\hat{\mathbf n}_e\,\mathrm{d}S
    = Q_e(t),
    \label{eq:electrode_charge}
\end{equation}
where $\hat{\mathbf n}_e$ is the outward unit normal to
the closed control surface $\partial\Omega_e$. For an initially neutral electrode that exchanges charge
only through its circuit connection, $Q_e(t)$ represents
the net charge transferred into the electrode. Taking
current as positive into the electrode gives
$I_e(t)=\mathrm{d}Q_e/\mathrm{d}t$.

Furthermore, the electric field can also generate mechanical traction on charged or polarised interfaces. This electromechanical coupling is commonly expressed using the Maxwell stress tensor, which relates the electric field to the electromagnetic stress acting on a material boundary \cite{landau1984electrodynamics,jackson1999classical,suo2008nonlinear}. The electrostatic part of the Maxwell stress tensor $\mathbf{T}^{M}$ is expressed in terms of $\mathbf{E}$ and the second-order identity tensor $\mathbf{I}$ such that,
\begin{equation}
    \mathbf{T}^{M}
    =
    \varepsilon
    \left(
    \mathbf{E}\otimes\mathbf{E}
    - \frac{1}{2}|\mathbf{E}|^2\mathbf{I}
    \right).
    \label{eq:maxwell_stress}
\end{equation}

Similar Maxwell-stress-based formulations have been used in dielectric elastomers, electroadhesion, electret systems, and coupled electromechanical finite-element models \cite{pelrine2000high,gao2011modeling,tuncer2005modeling,jia2016numerical}, where the electric field acts mainly in the normal direction; the normal component of the electrostatics traction from equation \ref{eq:maxwell_stress} reduces to an electrostatic pressure proportional to the electric-field energy density such that \citep{dorfmann2017nonlinear, wissler2007electromechanical}:
\begin{equation}
    p_e = \frac{1}{2}\varepsilon |\mathbf{E}|^2 .
\end{equation}

Various studies have used electrostatic FEM simulations to visualise electric potential and electric-field distributions due to equal and opposite charges in the contacting tribolayer and to examine how device geometry influences the induced electrical response~\citep{yan2022design}. Because open-source simulation frameworks for TENG modelling remain limited, many numerical studies in the literature rely on COMSOL Multiphysics, particularly for electrostatic FEM and coupled multiphysics simulations. For example, Wang et al.~\cite{wang2013sliding} performed simulations to verify the in-plane charge-separation mechanism in sliding-mode TENGs, considering the relative sliding between nylon and polytetrafluoroethylene (PTFE) surfaces, and observed the transferred charge quantity. Bai et al. \cite{bai2013cylindrical} observed the possibility of harvesting mechanical energy by TENGs from rotational motion through simulation. Similar finite-element simulations have also been used to study potential distributions at different electrode positions in non-contact rotating TENGs, considering different air buffer thicknesses between layers~\citep{zhang2020non}. Such models are particularly useful when the geometry is too complex for one-dimensional analytical assumptions or when fringing fields and finite-size electrode effects become important. 
\begin{figure}[!b]
    \centering
    \includegraphics[width=\linewidth]{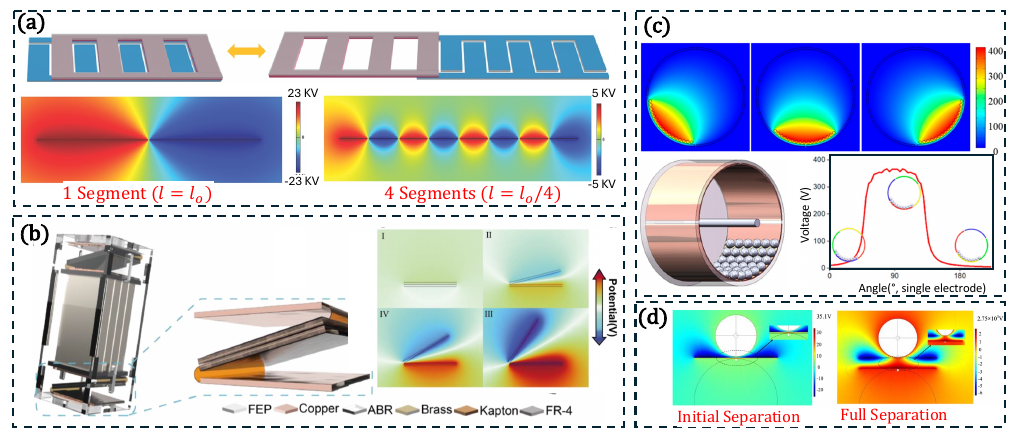}
    \caption{Representative electrostatic FEM simulations used to interpret electric-potential evolution in different TENG architectures. (a) Grating-structured sliding TENG showing how segmentation modifies the spatial potential distribution and reduces the potential drop across each repeated unit (adapted from Xie et al.~\citep{xie2014grating}). (b) V-shaped contact-separation TENG, where prescribed geometric states are used to visualise the redistribution of electric potential during pressing and releasing~\citep{shang2023high}. (c) Rolling or ball-motion TENG, showing the variation of electric potential and output voltage with the angular position of the charged moving body~\citep{qin2023ball}. (d) TENG simulation illustrating the change in potential distribution between two triboelectric layers as the separation distance evolves~\cite{zhang2024simulation}}.
    \label{fig:Electro_2}
\end{figure}

More recent electrostatic simulation studies have extended simple static potential calculations in two main directions. The first direction is dynamic electrostatic FEM, where the geometry is updated during device operation. Chen et al.~\cite{chen2020universal} proposed a dynamic FEM framework in which moving-mesh and air-buffer-layer strategies were introduced to avoid topology-change problems during contact and separation. A similar dynamic tendency can also be observed in recent rotational TENG studies, where moving-mesh, electrostatic and external power management circuit simulations are combined to describe time-dependent electrical behaviour~\citep{heo2024ultrahigh}.  
These studies illustrate advances in the treatment of motion and time-dependent electrical behaviour within FEM-based TENG models. Alongside these developments, alternative numerical formulations have also been investigated. Gravesen et al.~\cite{gravesen2023quantitative} discuss a boundary-element formulation for quantitative calculation and optimisation of spherical TENGs, including a comparison between BEM and FEM for a finite circular capacitor benchmark. Therefore, BEM currently appears as a specialised approach for idealised electrostatic TENG configurations rather than a widely adopted device-level simulation method. 

The applicability of BEM also depends on the assumptions underlying the formulation. A standard Laplace-based electrostatic formulation treats each region as homogeneous, linear and isotropic. Multi-domain formulations accommodate regions with different constant permittivities, whereas continuously varying properties, nonlinear material behaviour or volume charge require additional treatment. Hybrid FEM–BEM provides one option: FEM represents regions requiring volumetric modelling, while BEM represents the surrounding homogeneous exterior~\citep{kim2025coupling, comsol_bem_fem_electrostatics_learning_center}. However, it may require increased computational resources due to the requirement of solving problems with a large number of DOFs.

The second direction is application-specific for electrostatic FEM simulation. Figure~\ref{fig:Electro_2} demonstrates how FEM electrostatic simulations provide local field-level insight into charge separation, fringing effects, electrode configuration, and geometry-dependent voltage generation under prescribed operating states. In these studies, COMSOL or similar finite element software is commonly used to compute electric potential distributions, transferred charge, capacitance, output voltage and field evolution at selected operating states. Examples include simulating soft-contact TENGs, biomedical TENG sensors~\citep{venugopal2022effective}, ball motion between different  electrodes in hollow-drum TENGs~\citep{qin2023ball}, multilayer high-storable-power TENGs~\citep{shang2023high}, contact-separation TENGs for magnetic-energy harvesting~\citep{zhang2024simulation}, MXene/PVDF-based wearable TENGs~\citep{shuvo2024finite}, and biodegradable flexible TENGs~\citep{liu2024biodegradable}.Arief et al.~\cite{arief2022elastomeric,arief2026metasurface} investigated microwell-patterned fluoroelastomer TENGs with laser-induced graphene electrodes  using experimentally calibrated charge density. Their COMSOL simulations represented the surface texture using periodic two-dimensional square wells and evaluated electrostatic potential distributions at prescribed separation distances.  These models are useful for interpreting device mechanisms and optimising geometry or material parameters and example of incorporating engineered surface features into electrostatic modelling. Furthermore, modelling TENG electrodes also requires additional numerical treatments of the solution.  Sanglap et al.~\cite{sanglap2026multiphysics} modelled a highly conductive floating electrode response by imposing an equipotential electrode condition and determining the unknown electrode potential from electrostatic charge-balance relations. This provides an efficient way to represent electrode behaviour without explicitly solving charge transport inside the metallic electrode. However, most of the  device-level electrostatic simulations in the literature remain quasi-static or sequential-state electrostatic simulations, where the surface charge density, material permittivity and mechanical motion are prescribed rather than solved from a fully coupled formulation.  Šutka et al.~\cite{vsutka2026electric} experimentally demonstrated that an applied electric field can alter both the magnitude and polarity of persistent charge generated during polymer contact electrification. Their complementary FEM simulations examined strain-gradient-induced polarisation, supporting a possible flexoelectric contribution. These findings motivate charge-evolution laws that connect contact mechanics and interfacial electric fields to charge transfer and retention.

\subsection{Suitable numerical contact and deformation modelling of TENG microscale interfaces}
Prescribing device motion does not establish how much of a rough triboelectric surface actually makes contact. Local gaps, contact pressures and contact patches depend on both surface topography and material deformation. Contact mechanics therefore provides the relation between applied loading and the interfacial state required by the electrical model. This section summarises the applicable constitutive relations, material models, and numerical approaches that can be applied for more accurate TENG modelling. Mechanical contact and deformation modelling is essential for contact-mode TENGs because the electrical response depends strongly on the real contact area, local pressure distribution, deformation of surface textures, and possible adhesion between the triboelectric layers. For a generic deformable triboelectric layer occupying the domain $\Omega$, and assuming quasi-static equilibrium  the governing equation for a mechanical problem can be written as
\begin{equation}
    \nabla \cdot \boldsymbol{\sigma}(\mathbf{u}) + \mathbf{b}= \mathbf{0}
    \quad \text{in } \Omega ,
    \label{eq:mechanical_equilibrium}
\end{equation}
where $\mathbf{u}$ denotes the displacement field  $\boldsymbol{\sigma}(\mathbf{u})$ is the Cauchy stress tensor obtained from the chosen mechanical constitutive law. Contact between the triboelectric surface and the opposing surface is described using the normal gap function $g_n(\mathbf{u})$ and the normal contact pressure $p_n$.

For frictionless and non-adhesive contact, the contact constraints are expressed through the Karush–Kuhn–Tucker (KKT) conditions with the gap function $g_n$ and negative compressive pressure $p_n$ such that~\citep{bonet2008nonlinear, popov2019handbook, shvarts2021finite}:
\begin{equation}
    g_n(\mathbf{u}) \geq 0,
    \qquad
    p_n \leq 0,
    \qquad
    p_n g_n(\mathbf{u}) = 0
    \quad \text{on } \Gamma_c ,
    \label{eq:kkt_contact_conditions}
\end{equation}
where $\Gamma_c$ denotes the potential contact boundary. These conditions enforce non-penetration between the contacting surfaces, allow only compressive contact pressure, and ensure that contact pressure develops only when the normal gap is closed. 

For small deformation, the strain tensor is commonly defined as \begin{equation}
\boldsymbol{\varepsilon} = \frac{1}{2} \left( \nabla \mathbf{u} + \nabla \mathbf{u}^{T} \right). 
\end{equation} 
For an isotropic linear elastic material, the stress-strain relation is given by Hooke's law: 
\begin{equation} 
\boldsymbol{\sigma} = \lambda \, \mathrm{tr}(\boldsymbol{\varepsilon})\mathbf{I} + 2\mu \boldsymbol{\varepsilon}, \label{eq:linear_elasticity} \end{equation} 
where $\lambda$ and $\mu$ are the Lamé parameters. 
This model is suitable when the deformation of the triboelectric layer remains small. However, many TENGs use soft polymers such as PDMS, Ecoflex, or silicone rubber, which may undergo large deformation during contact. In this case, finite-strain constitutive models are more appropriate. One such model is based on the Hencky, or logarithmic, strain, where the Hencky strain model~\citep{hencky1928uber} is defined 
\begin{equation} 
\boldsymbol{\varepsilon}^{H} = \frac{1}{2}\ln \mathbf{C}, \qquad \mathbf{C}=\mathbf{F}^{T}\mathbf{F}, 
\label{eq:hencky_strain} \end{equation}
where $\mathbf{C}$ is the right Cauchy-Green deformation tensor.  This formulation is useful because it allows a Hookean-type relation to be written in logarithmic strain space while still accounting for finite deformation. 
Further, for hyperelastic materials, the stress is derived from a strain-energy density function $\Psi(\mathbf{F})$, for example, 
\begin{equation}
\mathbf{S} = 2\frac{\partial \Psi}{\partial \mathbf{C}}, \label{eq:hyperelastic_stress} 
\end{equation} where $\mathbf{S}$ is the second Piola–Kirchhoff stress tensor. For a hyperelastic material, the strain-energy density function can be expressed in terms of the first three invariants of the right Cauchy-Green deformation tensor $\mathbf{C}$, such that
\begin{equation}
    \Psi(\mathbf{C}) = \Psi (I_1, I_2, I_3),
\end{equation}
where $I_1, I_2$ and $I_3$ are three principal invariants of $\mathbf{C}$. Various hyperelastic material models are available, and the choice of model depends on the material type, deformation range, and available experimental data. For PDMS, Ecoflex, silicone rubber, and other elastomeric tribolayers, Neo-Hookean or Mooney-Rivlin models may be sufficient under moderate strain\citep{mooney1940,rivlin1951}, whereas Ogden, Yeoh, Gent, or Arruda-Boyce models become more appropriate when large stretch, strain-stiffening, or limiting-chain extensibility is important \citep{ogden1984nonlinear,ogden1972large,yeoh1990characterization,arruda1993three,gent1996new}. Although numerical investigation with full hyperelastic or Hencky-strain contact formulations remains limited in TENG simulation, they are widely relevant for soft and stretchable layers, where large deformation of elastomeric layers directly affects the contact state and electrical output~\citep{hossain2013more,zhou2020ultrastretchable,karthikeyan2025optimising,song2018textile, dong2022stretchable,cao2018lantern, hossain2015eight}.

Furthermore, frictional contact in TENG interfaces is also important in sliding-mode, freestanding-layer, and rotating devices \cite{niu2013theory,wang2013sliding,he2020boosting, kumar2023mechanics}. Sliding introduces tangential traction, slip history, frictional dissipation, wear, surface damage, and changes in real contact area, all of which can affect charge generation \cite{yang2024anisotropic}. Therefore, frictional TENG models should couple normal pressure, tangential traction, surface evolution, and charge-generation history. In a Coulomb-type model, the tangential traction $\mathbf{t}_t$ is bounded by the normal contact pressure through a friction law. For a Coulomb-type friction model, the admissible tangential traction is bounded by
\begin{equation}
    \|\mathbf{t}_t\|
    \leq
    \mu_f |p_n|
    \quad \text{on } \Gamma_c ,
    \label{eq:coulomb_friction_bound}
\end{equation}

where $\mu_f$ is the coefficient of friction.  Let $\mathbf{v}_t$ denote the tangential velocity of the surface receiving $\mathbf{t}_t$ relative to the opposing surface. The sticking and sliding conditions are then~\citep{wriggers2006computational,popov2010contact,kim2014introduction,yastrebov2016sliding}.  The sticking and sliding conditions are~\citep{wriggers2006computational, popov2010contact,kim2014introduction, yastrebov2016sliding}
\begin{align}
    \|\mathbf{t}_t\| &< \mu_f |p_n|,
    && \text{sticking} (\dot{\mathbf{g}}_t = \mathbf{0}), \\
    \|\mathbf{t}_t\| &= \mu_f |p_n|    \frac{\dot{\mathbf{g}}_t}
    {\|\dot{\mathbf{g}}_t\|},
    && \text{sliding} ((\dot{\mathbf{g}}_t > \mathbf{0}).
\end{align}

In the above, $\dot{\mathbf{g}}_t$ is the relative tangential slip rate. The sticking condition includes the limiting case $\lVert \mathbf{t}_t \rVert = \mu_f |p_n|$ with zero relative tangential velocity. During sliding, the traction reaches the friction limit and opposes the relative tangential motion. However, for sliding-mode TENGs, a classical Coulomb law alone is unlikely to be predictive because the same sliding process where a high frictional coefficient is involved also modifies surface morphology and wear and affects the charge-generation history.

A classical Coulomb law can serve as the tangential contact law within a coupled TENG model. However, it does not itself describe wear, changes in surface morphology or charge-generation history. Where these processes influence sliding-mode TENG performance, additional evolution laws and coupling to a charge-generation model are needed. Studies of adhesive viscoelastic sliding contact have also shown that contact area and sliding resistance depend on sliding velocity through the interaction between adhesion and viscoelastic dissipation~\citep{mandriota2024modelling,mastropasqua2025adhesion}. These studies identify mechanical mechanisms that may require consideration when modelling deformable triboelectric interfaces.

The numerical treatment of contact constraints is a separate modelling choice. Laursen and Simo~\cite{laursen1993continuum} developed a continuum-based finite element formulation for large-deformation contact incorporating Coulomb friction, while Cavalieri and Cardona~\cite{cavalieri2013augmented} combined an augmented-Lagrangian approach with a mortar formulation for large-sliding contact. These approaches address contact discretisation and constraint enforcement; the constitutive laws governing friction, surface evolution and charge generation must be specified separately.  Such promising frameworks offer numerical simulation capabilities for sliding-mode TENG models in which mechanical contact and surface evolution are treated in a coupled manner. 

In adhesive contact, the contact boundary condition is generally modified because the interface can sustain both compressive and tensile tractions. Instead of the non-adhesive KKT condition, the normal contact pressure is described by a traction-separation relationship,
\begin{equation}
    p_n = p_{\mathrm{adh}}(g_n)
    \quad \text{on } \Gamma_c ,
\end{equation}
where $p_{\mathrm{adh}}$ is the adhesive interaction pressure. Adhesive contact models have been used to investigate how van der Waals forces, electrostatic attraction, texture geometry, and contact pressure influence the effective contact area and electrical performance of vertical contact-separation TENGs~\citep{jin2016contact,yang2018comprehensive,yang2019fundamental,yang2021controlled,wang2021investigation}. In some cases, the Lennard-Jones potential-like relation is widely used to represent the force~\citep{yu2004adhesive}.  

However, for soft polymeric TENG materials, such as PDMS and silicone rubber, the mechanical response can also be viscoelastic rather than purely elastic. In a linear viscoelastic material, the stress depends on the strain history and may be written as 
\begin{equation} 
\boldsymbol{\sigma}(t) = \int_{0}^{t} \mathbb{C}(t-\tau) : \frac{\partial \boldsymbol{\varepsilon}(\tau)}{\partial \tau} \, d\tau , \label{eq:linear_viscoelasticity} 
\end{equation} 
where $\mathbb{C}(t)$ is the time-dependent relaxation tensor. Equivalently, the strain may be expressed using a creep compliance tensor $\mathbb{J}(t)$ as \begin{equation} 
\boldsymbol{\varepsilon}(t) = \int_{0}^{t} \mathbb{J}(t-\tau) : \frac{\partial \boldsymbol{\sigma}(\tau)}{\partial \tau} \, d\tau . 
\label{eq:creep_compliance} 
\end{equation} 
These relations show that a viscoelastic formulation can be used to investigate cyclic contact-separation motion when the contact pressure evolves with loading time, frequency, and previous contact cycles. Beyond classical linear viscoelastic models such as the Maxwell, Kelvin-Voigt, and standard linear solid models, several nonlinear formulations can be considered for soft polymeric TENGs. These include the Pipkin-Rogers nonlinear integral model~\citep{pipkin1968nonlinear}, Schapery's  model~\citep{schapery1969characterization}, and different finite-strain semisoft viscoelastic formulations that can be investigated (see ~\citep{simo1987finite,reese1998theory, bergstrom1998constitutive, hossain20203d, zhou2025modified, wineman2009nonlinear}).   Such behavior is important for PDMS, silicone, vitrimer, and shape-memory-polymer-based TENGs, where creep, stress relaxation, hysteresis, or time-dependent recovery can modify the contact area~\citep{kumar2023multiscale} and, consequently, the voltage, transferred charge, current, and output power~\citep{wang2021investigation,li2019viscoelasticink,xiong2019selfrestoring,cai2022selfhealable,niu2022thermadapt,ye2023vitrimer}. More broadly,~\cite{kumar2023multiscale} experimentally showed that real contact area and multiscale surface roughness strongly influence TENG output under different forces and frequencies. Overall, numerical contact and deformation modelling provides the mechanical link between applied force, surface texture, real contact area, and electrical response in TENGs. Linear elastic models are suitable for small deformation and simplified rough-contact descriptions, hyperelastic and Hencky-strain formulations are more appropriate for large deformation of soft elastomers, adhesive contact models are needed for micro-/nano-textured interfaces, and viscoelastic models are required when time-dependent polymer deformation affects cyclic TENG operation.

\subsubsection{Contact simulation of micro-nano scale textured geometries}
\begin{figure}[!b]
    \centering
    \includegraphics[width=1\linewidth]{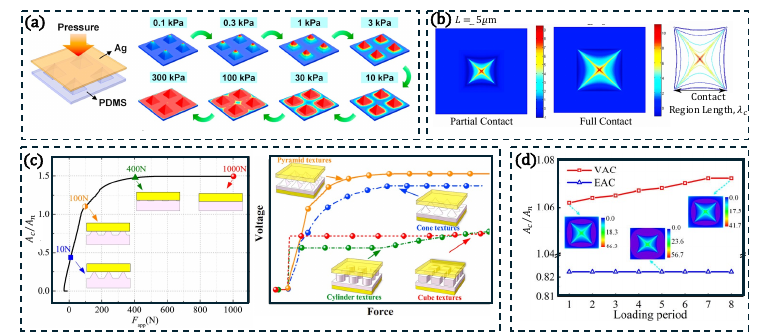}
    \caption{Contact and deformation models for micro-/nano-textured TENG interfaces. 
    (a) Pyramid-array deformation under increasing pressure~\citep{seol2015impact}. (b) Transition from partial to full adhesive contact and extraction of the contact-region length $\lambda_c$~\citep{jin2016contact}. (c) Influence of van der Waals adhesion, electrostatic attraction, texture geometry, and applied force on contact area and voltage~\citep{yang2018comprehensive}.(d) Viscoelastic--adhesive contact (VAC) predicts a larger contact area than elastic--adhesive contact (EAC) during cyclic loading~\citep{wang2021investigation}.}

    \label{fig:micro_nano}
\end{figure}
Most numerical and theoretical investigations of contact-area evolution in TENGs have focused on idealised micro-/nano-textured surfaces with simple periodic geometries, as shown in Figure \ref{fig:micro_nano}. Typical examples include pyramid arrays, micro-grooves, line patterns, cube patterns, and other regular surface textures~\citep{seol2015impact,jin2016contact,yang2018comprehensive,yang2019fundamental,yang2021controlled,wang2021investigation}. These simulations highlight the importance of real contact area in TENG performance but also reveal a common limitation: most models use idealised periodic textures rather than measured multiscale rough surfaces.
For the linear elastic case,~\cite{seol2015impact} used FE deformation analysis to study how contact pressure deforms pyramid-array structures and affects the output voltage of a TENG. Their model showed that the open-circuit voltage increases with contact pressure and is sublinearly proportional to $p_0^{2/3}$:
\begin{equation}
V_{\mathrm{OC}}
=
2.85
\left(
\frac{\sigma z}{\varepsilon_r \varepsilon_0}
\right)
\left(
\frac{z p_0}{EL}
\right)^{2/3},
\end{equation}
where $\sigma$ is the uniform surface charge density, $z$ is the polymer thickness, $\varepsilon_r$ is the relative permittivity,$\varepsilon_0$ is the vacuum permittivity,  $p_0$ is the applied contact pressure, $E$ is Young's modulus, and $L$ is the pyramid unit length.
Jin et al.~\cite{jin2016contact} numerically investigated the adhesive contact behaviour between a rigid metal plate and a pyramid-shaped PDMS microstructured surface. This investigation also showed that increasing contact pressure causes a transition from partial contact to full contact. Using the simulated pressure-dependent contact length $\lambda_c(p_0)$, the open-circuit voltage was related to the real contact area as
\begin{equation}
V_{OC}= \left(\frac{\sigma z}{\varepsilon_r \varepsilon_0}\right)\left(\frac{\lambda_c p_0^{2}}{L^{2}}\right),
\label{eq:jin_voc_contact_area}
\end{equation}
where $\lambda_c$ is the contact-region length of a unit pyramid block obtained from the adhesive contact simulation, and $L$ is the unit pitch of the pyramid array.
Based on the prior investigation by Yang et al.~\cite{yang2019fundamental}, \cite{yang2018comprehensive} included adhesivity into contact analysis to represent van der Waals and electrostatic forces for low charges and used the fast Fourier transform (FFT) technique to analyse the effects of shapes and sizes to find the optimal sizes for high voltage:
\begin{equation}
V_{\mathrm{OC}}
=\left(\frac{\sigma  z}{ \varepsilon_0}\right) \left(\frac{A_c}{A_n}\right).
\label{eq:V_oc}
\end{equation}
Wang et al.~\citep{wang2021investigation} introduced PDMS viscoelasticity through the time-dependent creep compliance $J(t)$ and relaxation modulus $G(t)$ obtained from a fractional derivative Zener model. After fitting the nanoindentation data for  of PDMS, these functions were written as
\begin{align}
J(t)
&=
2.249
-
1.01 E_{0.2126}
\left(
-0.0123 t^{0.2126}
\right),
\\
G(t)
&=
0.4447
+
0.3611 E_{0.2126}
\left(
-0.0223 t^{0.2126}
\right).
\end{align}
These functions were then used in the viscoelastic-adhesive contact model to account for the history-dependent surface deformation of the textured polymer surface. Conceptually, the deformation can be written as
\begin{equation}
u(t) \propto J(t) * p_{\mathrm{ah}}(t),
\end{equation}
where $p_{\mathrm{ah}}(t)$ is the adhesive contact pressure. The numerical results showed that viscoelasticity increases the predicted contact area compared with an elastic-adhesive model, leading to computing the $V_{OC}$ from equation \eqref{eq:V_oc}. The study also showed that pyramid textures with smaller pitch and larger width can improve electrical performance under larger applied loads.

At the same time, simulations of these micro- and nanoanalytical geometries are attractive because their shape can be described mathematically using a small number of parameters, such as pitch, width, height, and side angle. As a result, the contact area can be computed more conveniently either from analytical geometric relations or from reduced numerical contact models. For example, pyramid-array surfaces have been widely used because the transition from partial contact to full contact can be related to the deformation of a single periodic unit cell. In such models, the real contact area is commonly estimated from the contact-region length or the deformed geometry of the texture, and then linked to the open-circuit voltage or transferred charge. This approach provides useful insight into how applied pressure, texture size, adhesion, and viscoelasticity influence the TENG output. However, their quantitative accuracy is nevertheless subject to two distinct limitations. First, simplified contact-area estimates rely on assumptions about the deformation and shape of the contact patches. These assumptions can introduce errors even when the nominal patterned geometry is represented accurately, and these errors can propagate into the predicted electrical response. Second, the idealised geometry may itself inadequately represent real triboelectric surfaces, which can contain random roughness, asperities at multiple scales, manufacturing defects and non-uniform contact patches. Geometric parameterisation therefore simplifies model construction but does not, by itself, ensure accurate prediction of the real contact area.

\subsubsection{Contact area and rough surfaces analysis}
\begin{figure}[t]
    \centering
    \includegraphics[width=1\linewidth]{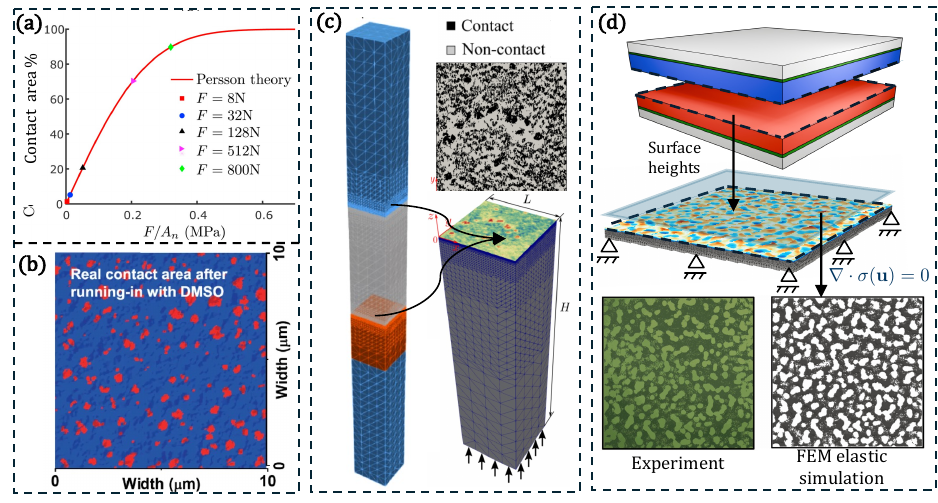}
    \caption{Modelling of real contact area at rough TENG interfaces: (a) Persson et al. \cite{persson2002elastic} contact area formula computed using composite RMS surface gradient $S_{dq}$, in a unified
    analytical model of Xu et al.~\cite{XUVoc}, (b) elastoplastic FEM contact map on randomly generated TENG surface lubricated with dimethyl sulfoxide (DMSO), with red patches indicating real contact area from Zhao et al.~\cite{zhao2026solvent}; (c) representative-element FEM based on the measured
    PET-PDMS roughness and the resulting contact distribution,
    by Shvarts et al.~\cite{shvarts2021finite} and (d) full-specimen representation of measured roughness in the FE mesh and comparison of predicted contact morphology with interference reflection microscopy (IRM) experiment for PET-PVS layers from Sanglap et al.~\cite{sanglap2026multiphysics}.}
    \label{fig:surface_rough}
\end{figure}

Practical TENG surfaces are rarely ideal periodic structures. They usually contain random multiscale roughness, irregular asperity distributions, manufacturing defects, and non-uniform contact patches. Therefore, rough-surface modelling is needed to connect realistic interface morphology with contact area and electrical output. A first route to include roughness in TENG modelling is through statistical rough-contact theories. To observe load-dependent TENG behaviour, Xu et al.~\cite{XUVoc} calculated the real contact area incorporating Persson's contact area formula~\citep{persson2002elastic} which is shown in Figure~\ref{fig:surface_rough}(a). In this analytical model, the load-dependent real contact area is estimated from the surface roughness and then used to modify the electrostatic response. This approach improves upon flat-surface models by introducing the effect of applied force and surface roughness, but it still represents rough contact through an averaged statistical description rather than resolving individual asperity contact patches. Wen et al.~\cite{wen2021load} also followed the same approach in MATLAB simulation and used Persson's contact theory to estimate the load-dependent real contact area. However, asperity-based models are only valid for very light loads (infinitesimal contact). They fail at high pressures because they cannot simulate how contact patches merge into a continuous contact zone~\citep{carley2026multiscale, yastrebov2015infinitesimal, nayak1971random}.

Recently, Zhao et al.~\cite{zhao2026solvent} used elastoplastic FEM to investigate changes in real contact area with penetration depth following solvent-assisted running-in, as shown in Figure~\ref{fig:surface_rough}(b). Their solid-contact analysis uses synthetic rough surfaces with prescribed power spectral densities but does not explicitly couple solid deformation to lubricant flow or the pressurisation of trapped liquid pockets. For lubricated TENG interfaces, these additional mechanisms should influence the real contact area because interfacial fluid pressure can support part of the applied load. Existing formulations for trapped-fluid contact, thin-film flow and partially lubricated rough interfaces~\citep{shvarts2018trapped,shvarts2021computational,alavi2026numerical} provide a mechanical basis for incorporating fluid-supported loading into TENG models. However, predicting electrical output additionally requires coupling to triboelectric charge generation and electrostatics.

A more direct numerical route is to simulate contact using explicit rough-surface representations. Shvarts et al.~\citep{shvarts2021finite} introduced a methodological FE model for contact-separation TENG that accounts for nanoscale surface roughness by considering a representative surface element, which is shown in Figure~\ref{fig:surface_rough}(c). Considering the measured surface roughness from atomic force microscopy, it shows qualitative agreement with Persson's~\citep{persson2002elastic} contact area formula. More recently, Sanglap et al.~\citep{sanglap2026multiphysics} proposed a multiphysics FE model for rough-surface representation and computed the active Gauss-point contact-area integration shown in figure~\ref{fig:surface_rough}(d). Through homogenisation, this framework coupled rough-contact mechanics with an electrostatic model accounting for fringing-field effects. The predicted load-dependent real contact area at the PET--PVS interface agreed closely with interference reflection microscopy measurements. The contact-area predictions also showed a transition between the trends described by the models of Bush et al.~\cite{bush1975elastic} and Persson~\cite{persson2002elastic}.

Overall, rough-surface contact simulation remains less developed in the TENG literature than electrostatic FEM or idealised microtexture contact analysis. Statistical models are computationally efficient and useful for capturing average roughness effects, but they cannot resolve the spatial distribution of irregular contact patches. Explicit rough-surface FEM models can capture heterogeneous asperity contact and load-dependent real contact area, but they are more computationally expensive and require real or generated surface topographies.

\subsection{Molecular dynamics and atomistic simulations for nanoscale contact electrification}
A resolved contact map identifies where surfaces interact, but it does not by itself determine the amount, polarity or persistence of transferred charge. These quantities also depend on interfacial structure, chemistry and the available charge-transfer mechanisms. Molecular and electronic-structure simulations can help investigate these processes and identify information needed to formulate charging laws for larger-scale models. Molecular dynamics (MD) and atomistic simulations provide a route to investigate contact electrification at the nanoscale, where interfacial charge transfer originates. Unlike device-level continuum FEM models, where the material response is described by constitutive laws such as elasticity or viscoelasticity, atomistic simulations describe material behaviour through interatomic interactions\citep{ zhang2024thermodynamic}.  Applications include reactive simulations of charge transfer and bond scission at polymer-polymer and polymer-metal interfaces~\cite{ratanaporn2024reaxff,verners2022smooth, verners2023mechanisms},
all-atom simulations of ion-mediated contact charging between insulating polymers~\cite{zhang2024thermodynamic} and atomistic investigations of solid--liquid interfaces and droplet-based TENGs ~\cite{xu2019slips,wu2023liquidsolid,meng2024droplet}.

For classical all-atom MD, each atom is represented explicitly, and the interatomic interactions are commonly described using a molecular-mechanics
force field~\cite{cornell1995forcefield,jorgensen1996opls}. Commonly for classical all-atom MD, the force obtained from the interatomic potential energy $U$ for each atom can be explicitly represented by the the total potential energy such that
\begin{equation}
    U
    =
    U_{\mathrm{bond}}
    +
    U_{\mathrm{angle}}
    +
    U_{\mathrm{dihedral}}
    +
    U_{\mathrm{vdw}}
    +
    U_{\mathrm{Coulomb}} .
\end{equation}

Here, $U_{\mathrm{bond}}$, $U_{\mathrm{angle}}$, and $U_{\mathrm{dihedral}}$ describe bonded molecular interactions and stretching, angle bending, and torsional rotation, respectively, while $U_{\mathrm{vdw}}$ and $U_{\mathrm{Coulomb}}$ describe non-bonded van der Waals and electrostatic interactions. This type of model is useful for studying polymer surfaces, water adsorption, ion distribution, and humidity-assisted contact electrification. 

A direct MD-based approach has been reported using reactive molecular dynamics by Ratanaporn et al.\citep{ratanaporn2024reaxff}. Here, the ReaxFF (reactive molecular dynamics) force field and MD, together with a charge equilibration method were used to investigate contact electrification in polyisoprene, a natural rubber polymer, in contact with copper and PTFE. Their simulations examined the evolution of charge transfer at the atomic scale and related the observed charge-transfer direction to the triboelectric behaviour of natural-rubber-based TENGs. Wu et al.~\cite{wu2018insights} studied metal-polymer contact electrification for an Al-PTFE tribopair and analysed how interfacial electronic structure and charge redistribution explain the direction of electron transfer. Wu et al.~\cite{wu2019amorphous} extended this direction to amorphous polymer-metal interfaces, which are more representative of practical polymer TENG materials. Li et al.~\cite{li2020humidity} further considered the effect of humidity on electron transfer between metals and amorphous polymers, showing that environmental water can modify the interfacial charge-transfer process. Tan et al.~\cite{tan2021corrugated} used first-principles calculations to study contact electrification between corrugated two-dimensional materials, showing how surface morphology at the atomic scale affects electron transfer.

More recently, all-atom MD simulations have been used to investigate ion-mediated contact electrification between insulating polymers. Zhang et al.~\cite{zhang2024thermodynamic} suggest that ion-transfer mechanisms can contribute to contact charging between insulating polymers, especially when surface water or humidity is present. MD and atomistic simulations complement continuum TENG models by providing microscopic insight into the origin of triboelectric charge. However, these simulations are still limited by small length scales, short time scales, idealised interfaces, and high computational cost. As a result, they are usually not used to predict complete TENG voltage, current, or power output directly. Instead, they provide mechanistic information that can inform continuum models, for example, by improving assumptions about surface charge density, charge-transfer direction, humidity dependence, or material selection. The reviewed approaches form a hierarchy of descriptions with different predictive capabilities. Their usefulness depends on how clearly model inputs, calculated quantities and coupling assumptions are distinguished. The appropriate level of detail should follow the intended prediction and the mechanisms controlling it. Consequently, assessing a model requires an application context: predicting contact area, explaining a sensing signal and estimating energy delivered to a load impose different modelling and validation requirements.

\section{TENG Applications Across Scales}
TENGs harvest electrical energy from mechanical contact, separation, and sliding, enabling applications ranging from structural health monitoring of bridges to drug delivery inside the human body~\cite{huang2023real,wang2023multifunctional}. It is noteworthy that each length scale imposes its own set of spatial design criteria. For instance, at the macroscale, a TENG must withstand large mechanical displacements, tolerate outdoor environmental exposure, and deliver sufficient instantaneous power to drive wireless sensors and actuators without external energy storage \cite{pu2023triboelectric}. As device dimensions reduce to the millimeter scale, devices must be mechanically compliant, conformable to diverse and irregular surfaces, and self-powered to support long-term operation across wearable, biomedical, structural, and other applications \cite{pandey2024recent}. Finally, at the microscale, the device must additionally be biocompatible, and where implantation is intended, biodegradable and non-toxic \cite{ramaraj2024biocompatible}, so that it can be indelibly attached to the human body without triggering immune rejection and jeopardizing biological tissue mechanics~\cite{zainalabidin2025current}.

Despite the rapidly growing body of literature on TENGs, our review of the existing literature indicates that no previous study has systematically classified these applications using device dimensions as the primary framework. Here, for the first time,  we attempt to present the applications of TENGs by taking length scale through which macroscale, millimetre-scale, and microscale implementations are discussed. We present applications of TENGs by examining device length scales, i.e., macroscale (x>10 mm), milliscale ($1\,\mathrm{mm} < x \leq 10\,\mathrm{mm}$), and microscale ($1\,\mu\mathrm{m} < x \leq 1\,\mathrm{mm}$), followed by comparison along them~(Figure~\ref{fig:length_scale_class}). The summary of reviewed applications across the scales is also presented in supplementary material in Table S2. In adopting these classification criteria, the characteristic active dimension and the overall system dimension were clearly distinguished. The characteristic active dimension refers to the relevant geometric dimension of the functional triboelectric unit responsible for contact electrification and electrostatic induction, such as the lateral size of an active contact region or the diameter of a functional fibre. In contrast, the overall system dimension describes the physical extent of the complete device or integrated assembly, including supporting structures, substrates, and packaging. Accordingly, the macroscale, milliscale, and microscale categories are assigned primarily according to the dimensions of the functional triboelectric unit. It was followed that if one or more characteristic dimensions, including lateral length, height, and diameter (fibre-shaped TENG only) of a TENG device, fall within the stated range, then that device is assigned to the corresponding length-scale category. This distinction reduces ambiguity associated with anisotropic geometries, thin films, fibrous structures, and multiscale architectures, thereby providing a more consistent basis for comparing TENG applications across length scales. The resulting framework unveils how miniaturisation unlocks new functionality and application prospects at each successive step down the length scale.

\begin{figure}[H]
    \centering
    \includegraphics[width=0.85\linewidth]{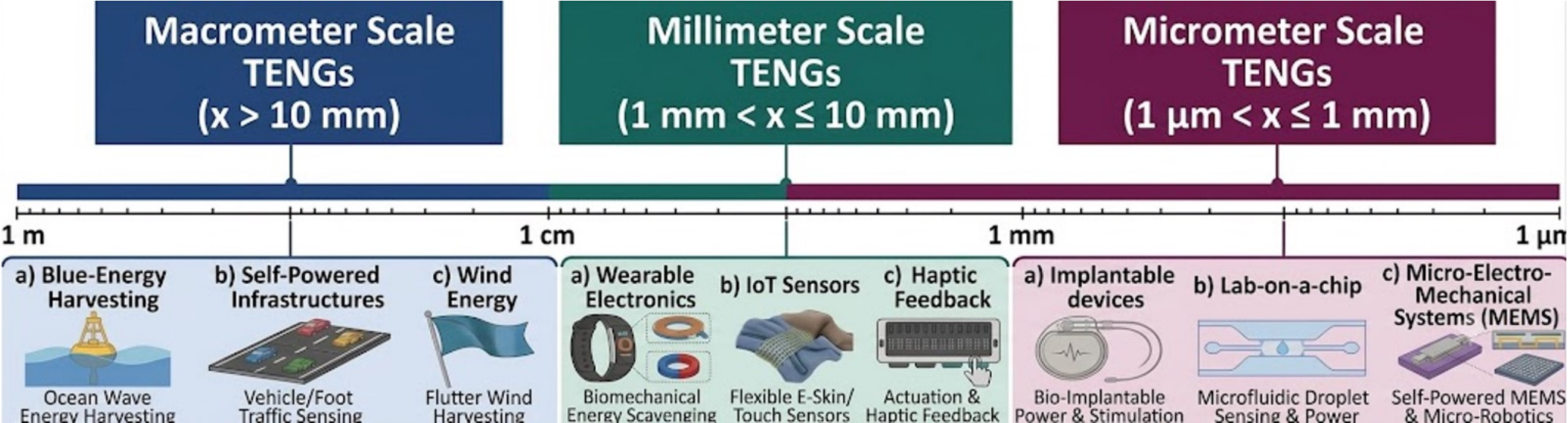}
    \caption{The Length Scale-Wise Classification of TENGs for Different Applications.}
    \label{fig:length_scale_class}
\end{figure}
\subsection{Macroscale Application}
TENGs are predominantly fabricated at the macroscale, with dimensions ranging from over 10 mm to several centimetres, decimetres, and even meters. At the functional-element scale, individual triboelectric components are typically centimetre-sized. At the same time, complete TENG devices or integrated systems can extend to decimetre- or meter-scale dimensions. Unlike micro- and nano-structured TENGs that rely on sub-micron surface texturing to enhance contact electrification, macroscale TENGs achieve comparable or superior output by engineering device-level geometry, including pyramidal arrays, corrugations, origami folds, gear-driven rotors, and magnetically gated multilayer stacks \cite{zhang20233d,so2023flexible,cho2022highly,bhatia2022triboelectric,bhatia2021wearable}. These techniques yield a larger effective contact area and regulate contact–separation kinetics under slow, irregular mechanical inputs, typical of human motion. Besides wearable platforms, macroscale TENGs have been integrated into civil and marine infrastructure, where device dimensions are governed by the host structure. This includes buoy-integrated TENGs used in ocean environments enclosed in ellipsoidal or spherical shells for data collection and meteorological purposes \cite{dip2023triboelectric,feng20220}, bridge-mounted TENGs for providing self-powered structural health monitoring \cite{huang2023real} and a sponge-supported TENG-based wireless network system for monitoring the safety of the railway system \cite{zhang2021sponge}. Synthesis and detection of chemical compounds and degradation of chemical pollutants are other vital aspects of macroscale-based TENGs \cite{tushar2023triboelectric,uddin2025triboelectric}.

\begin{figure}[t]
    \centering
    \includegraphics[width=\linewidth]{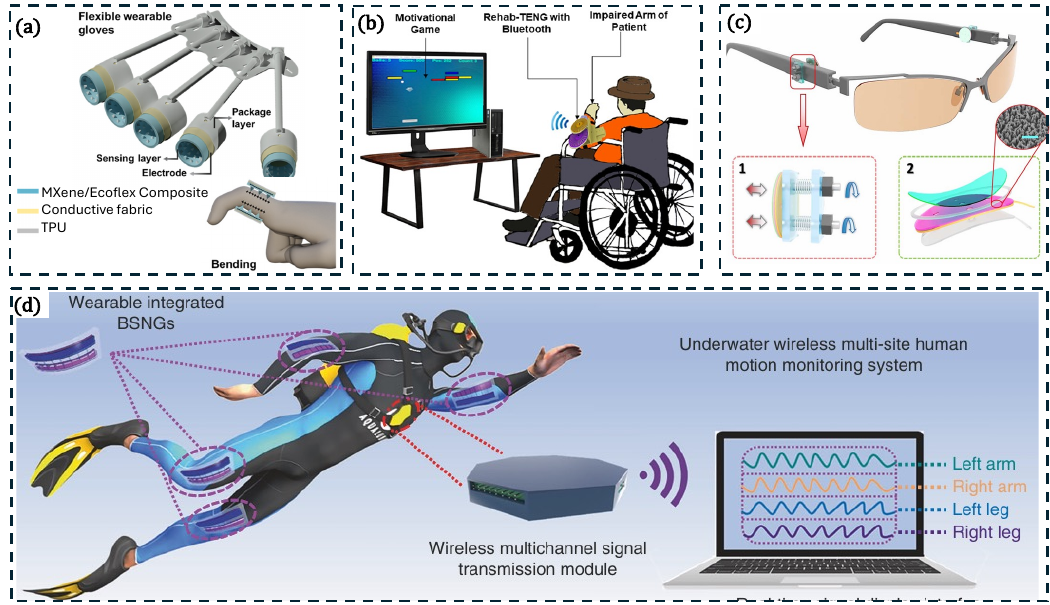}
    \caption{Macroscale TENGs. (a) 3D-printed wearable glove incorporated with STTS )Adopted from Zhang et al. \cite{zhang20233d}), (b) Schematic illustrating the application of the Rehab-TENG attached to a patient's impaired arm during a motivational game to promote functional recovery (Adopted from Bhatia et al.\citep{bhatia2021wearable}(c) Schematic of a pair of conventional eyeglasses integrated with the msTENG (Adopted from Pu et al. \citep{pu2017p}).  (d) Underwater wireless human movement observation system by BSNG (Adopted from Zou et al. ~\citep{zou2019bionic})}
    \label{fig:Macroscale_TENG}
\end{figure}

Geometry-engineered macroscale-based TENG for wearables was reported by Zhang et al. \cite{zhang20233d}. They developed a self-powered toroidal triboelectric sensor (STTS) (Figure~\ref{fig:Macroscale_TENG}(a)) in which an Ecoflex/MXene nanocomposite was cast onto an additive-manufactured mould bearing. The resulting pyramid-textured film was paired with a conductive-fabric electrode and rolled into a toroidal package that could be slipped onto a finger. The toroidal sensors were subsequently assembled with a 3D-printed flexible thermoplastic polyurethane glove into a wearable smart glove capable of producing a peak-to-peak voltage of 19.91 V. The authors demonstrated its use in racing-game control, balance-table manipulation and robotic-hand actuation.

A markedly different macroscale actuation strategy was reported by Kim et al. \cite{kim2025p}, who designed a magnetically actuated TENG (MA-TENG) intended for wireless, medium-independent energy transfer to implantable devices. The device consists of two magnetic-PDMS composite discs, loaded with NdFeB micro-magnetic particles for rapid magnetic responsiveness and BaTiO$_3$ microparticles to raise triboelectric output. An oscillating magnetic field, produced by an external electromagnet, drives the two composite discs into an alternating contact (with a top and a bottom electrode), yielding a dual-layer-rectified output of more than 11 V and a maximum power density of 169.4 $\mu W/cm^2$. The authors further verified the long-term reliability of the device through accelerated thermal ageing and over 6.5 million actuation cycles.

Hierarchical surface-relief engineering was demonstrated by Cho et al. \cite{cho2022highly}, who fabricated a hierarchical-wrinkled-architecture TENG (HWA-TENG) by combining a chemically distinct, sputter-deposited plasma-polymer-fluorocarbon (PPFC) thin film. The resulting device exhibited an approximately 20-fold higher electrical output than the corresponding TENG without the hierarchical wrinkled architecture, while retaining optical transparency and superhydrophobicity. Because the functional layer was only 100 nm thick yet matched the electrical output of 100-µm-thick PTFE, the authors applied it as a transparent triboelectric raindrop-energy harvester mounted on a window-like substrate and as an ultrathin "triboelectric skin" affixed to the fingertip that converted contact with everyday objects into usable electrical energy.

Origami-based folding has also been employed to integrate TENGs into assistive devices that combine mechanical support with self-powered sensing. Bhatia et al. \cite{bhatia2022triboelectric} built a gravity-support exoskeleton for shoulder rehabilitation (Origami-TENG) around a water-bomb-based origami scaffold whose torsion- and tension-spring-assisted folds counteract the weight of an impaired arm, while two foldable, nanopatterned aluminium/PTFE TENGs are embedded within the structure. A related gear-coupled rotational macroscale device called the Rehab-TENG (Figure~\ref{fig:Macroscale_TENG}(b)) was developed by the same group for elbow and shoulder rehabilitation \cite{bhatia2021wearable}. Therein, a free-standing rotor–stator TENG pair is driven through 5:1 or 25:1 gear trains by the patient’s own joint motion, amplifying low-frequency arm movements into a higher-frequency rotational input and thereby increasing both the harvested charge and the resistive exercise load presented to the impaired limb.


Beyond gross-motor harvesting, macroscale TENGs have been adapted to capture subtle physiological motions for diagnostic purposes. Pu et al. \cite{pu2017p} mounted a circular, multilayer mechano-sensational TENG (Figure~\ref{fig:Macroscale_TENG}(c))  (msTENG)—comprising a fluorinated ethylene propylene/indium-tin-oxide film opposed by a natural latex membrane on the inner arm of a pair of eyeglasses to transduce eyelid-blink-induced skin micromotion into an electrical signal, enabling a smart-glasses-based appliance controller and a blink-driven virtual keyboard for users with severe motor impairment. In a parallel effort focused on respiratory diagnostics, He et al. \cite{he2021monitoring} constructed a respiration-monitoring TENG (RM-TENG) from stacked electrospun polyacrylonitrile and polyvinylidene fluoride nanofibrous mats, sandwiched between perforated aluminium electrodes, packaged as a removable, self-powered mask filter. The optimised RM-TENG reproduced the ventilator-set respiratory rate with a near-perfect linear correlation and the inspiratory-to-expiratory time ratio with roughly 94\% agreement, while simultaneously achieving 99 wt\% particulate-filtration efficiency for particles between 0.3 and 5 µm, surpassing several commercial KN95/N95/FFP2 masks in overall filter quality factor.

Besides, mimicking the ion-channel architecture of electric-eel electrocytes, Zou et al. \cite{zou2019bionic} devised a bionic stretchable nanogenerator (BSNG) at the macroscale (10 cm × 6 cm × 8 mm) in which a PDMS–silicone bilayer hosts an array of mechanically gated bionic channels. Under tensile actuation, an electrically conductive liquid (deionised water) is drawn from a reservoir into the open channels, where triboelectrification at the liquid–silicone interface (combined with electrostatic induction) generates an alternating current as the channels open and close in a cyclic manner. Four such BSNG units, worn on the elbows and knees and connected through rectifier bridges, were used to construct a wireless underwater multi-site motion-monitoring system (Figure~\ref{fig:Macroscale_TENG}(d)) that could distinguish different swimming strokes and act as an underwater distress-signalling system. 

\subsection{Milliscale applications}
In recent years, progress in flexible electronics and microfabrication technologies has enabled the development of a new generation of millimetre-scale triboelectric nanogenerators. In contrast to conventional TENGs, which are typically fabricated at the centimetre scale and are often limited by their large footprints, recent advances in microfabrication technologies have enabled the development of millimetre-scale TENGs that can be seamlessly integrated with microelectromechanical systems (MEMS), implantable biomedical devices, wearable electronics,  and sensing platforms  \cite{alzgool2022micro,mousavi2023event}.

Recently, salient strides have been made toward the realisation of MEMS-compatible TENG architecture. One of the earliest demonstrations of MEMS-integrated triboelectric systems was reported by Alzgool et al. \cite{alzgool2022micro}, who devised a micro-triboelectric generator for self-powered shock detection. The device comprised of a suspended aluminium microplate positioned above a polyimide triboelectric layer, fabricated entirely using MEMS processes. The triboelectric voltage induced by the MEMS device was subsequently used to actuate an electrostatic MEMS switch without involving any external power source, demonstrating its ability as a self-powered shock detector for integration into the inner lining of protective helmets. Later, Mousavi et al. \cite{mousavi2023event} rationalised the concept from shock-event detection to intelligent vibration monitoring. Rather than merely demonstrating the fabrication and operation of a MEMS-TENG, they developed an event-driven vibration sensor capable of distinguishing different vibration conditions. The triboelectric output served as a decision-making signal, activating the switching mechanism only when predefined frequency or acceleration thresholds were exceeded. Although miniaturisation offers substantial benefits for system integration, reducing the dimensions of triboelectric devices often leads to decreased displacement amplitude and electrical output.

\begin{figure}[!b]
    \centering
    \includegraphics[width=1\linewidth]{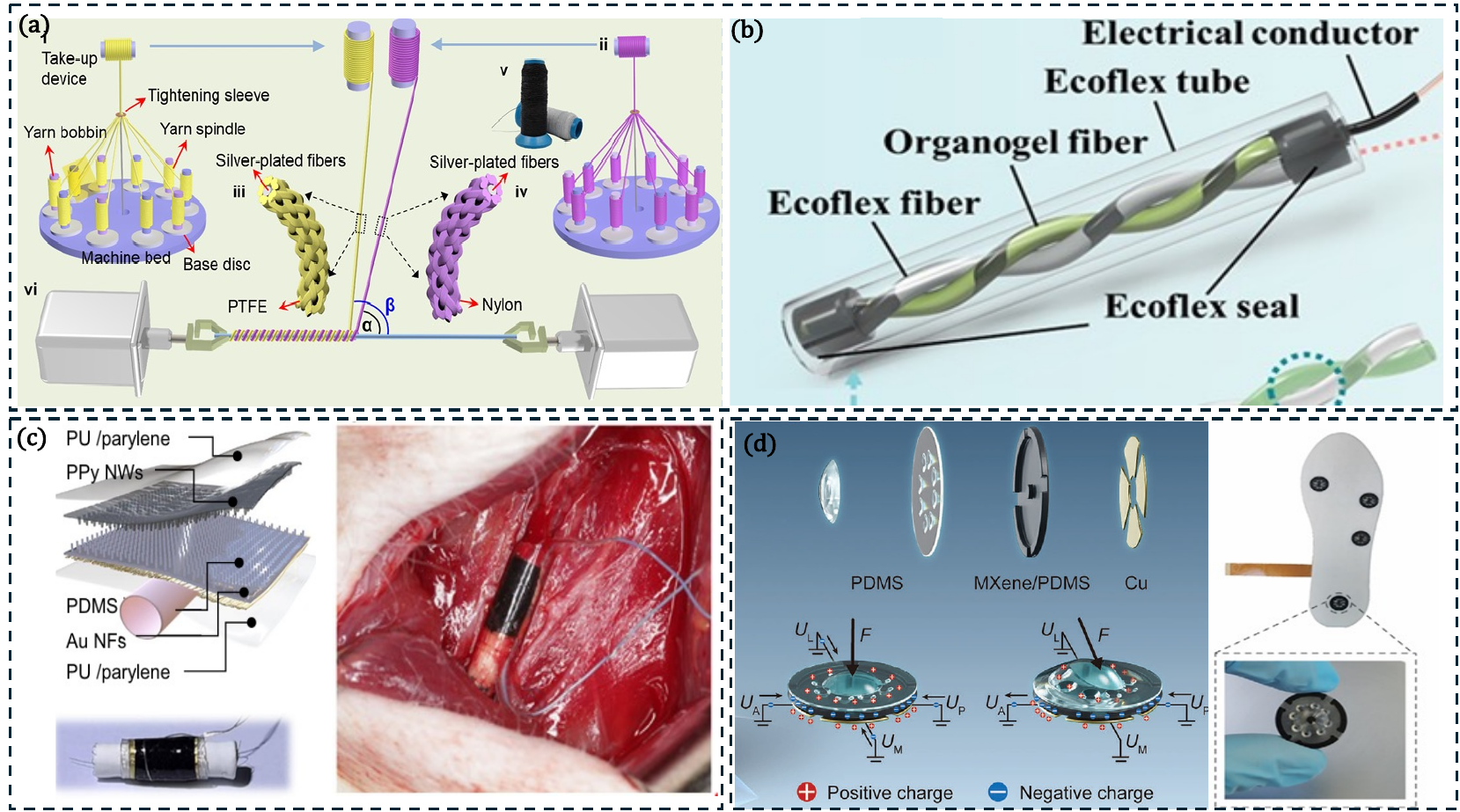}
    \caption{(a) Fabrication process of the helical fibre strain sensor for human respiratory monitoring  (adopted from Ning et al~\cite{ning2022helical}); (b) The diagram of the electronic vascular conduit for postoperative monitoring and suturing of the electronic vascular conduit to the rabbit carotid artery (Adopted from Liu et al~\citep{liu2025electronic}); (c) TENG-based self-powered sensing insole for simultaneous mapping of shear and normal plantar stresses (Adopted from Hu et al\citep{hu2024triboelectric})}.
    \label{fig:mm_scale_TENGs}
\end{figure}

To address these limitations, Mousavi et al. \cite{mousavi2024high} later developed a high-signal-to-noise-ratio MEMS triboelectric motion sensor. They introduced a warped diaphragm formed by a nonuniform residual film stress instead of using orthodox spring-supported movable structures. The dome-shaped diaphragm increased the effective contact–separation displacement and enabled significantly greater charge generation than traditional planar configurations. The high signal-to-noise ratio enabled reliable motion detection without extra signal-conditioning circuits, simplifying the sensor system. Besides experimental developments, the design optimisation of MEMS-scale triboelectric energy harvesters has garnered significant research attention. For instance, Hamid et al.~\cite{ref20192019} proposed a MEMS-scale triboelectric energy harvester ($5000 \mu m \times 3000 \mu m \times 8.5 \mu m$) to power autonomous nanosensors and microsystems. They investigated the impact of variegated design parameters, including proof-mass dimensions, suspension stiffness, electrode spacing, and triboelectric layer geometry, on electrical output performance. The results showed that optimising the structural parameters and resonant frequency significantly enhanced charge transfer. To experimentally demonstrate the feasibility of such miniaturised triboelectric systems, they subsequently developed a novel MEMS high-aspect-ratio energy harvester fabricated using the UV-LIGA technique \cite{hamid2020novel}.

Beyond MEMS-based sensing, millimetre-scale TENGs have gained increasing attention for wearable and textile-based electronic systems because of their lightweight design, conformability, and self-powered operation. Ning et al. \cite{ning2022helical} fabricated a helical fibre triboelectric strain-based sensor for respiratory monitoring (Figure~\ref{fig:mm_scale_TENGs}(a)) by combining nylon fibres and PTFE in a helical configuration. The helical fibre strain sensor (HFSS) was impeccably built so that, with minimal stretching, contact electrification occurred between PTFE and nylon. The HFSS achieved high strain sensitivity and excellent mechanical compliance, which enables the real-time detection of subtle respiratory motions. Building on the concept of fibre-based triboelectric systems, Xu et al. \cite{xu2023multi} proposed a multi-mode, durable fibre triboelectric nanogenerator inspired by the Hookean structure of vascular stents. Unlike conventional fibre TENGs that primarily function as sensors, the biomimetic spring-like architecture enabled simultaneous biomechanical energy harvesting and motion sensing. This fibre-shaped TENG was further applied as an intelligent walking monitoring system in the shoe insole for fitness tracking and injury warning. Further extending the application of fibre TENGs towards implantable bioelectronics, Sheng et al. \cite{sheng2022ultrastretchable} reported an ultra-stretchable organo-gel/silicone fibre-helical triboelectric nanogenerator (OFS-TENG) as shown in Figure~\ref{fig:mm_scale_TENGs}(b) for ligament strain monitoring. The incorporation of a double-helical architecture imparted excellent stretchability, mechanical robustness, and long-term stability under large deformations. Consequently, the fibre TENG was directly sutured onto rabbit patellar ligaments to enable continuous monitoring of ligament deformation and muscle activity during movement.

It is evident that significant advances in biomedical monitoring and implantable bioelectronics have also been achieved with the development of millimetre-scale TENGs. Liu et al. \cite{liu2025electronic} engineered an electronic vascular conduit (Figure~\ref{fig:mm_scale_TENGs}(c)) by integrating a flexible triboelectric sensor on a tissue-engineered vascular graft. The device continuously translated blood flow-induced vascular deformation into electrical signals, thus allowing real-time detection of thrombosis and vascular stenosis in animal models. Moreover, triboelectric devices have begun to evolve beyond sensing platforms toward active therapeutic systems. Kim et al. \cite{kim2026wireless} introduced a wireless, bioresorbable triboelectric nerve-block system for postoperative pain management. The implant comprised biodegradable polymers with juxtaposing triboelectric polarities that generated electrical stimulation under ultrasound excitation to regulate peripheral nerve activity and alleviate pain. 

Additionally, Hu et al. \cite{hu2024triboelectric} proposed a triboelectric insole (Figure~\ref{fig:mm_scale_TENGs}(d))  capable of simultaneously detecting normal and shear plantar stresses. Multiple triboelectric sensing units were rationally distributed within the insole to generate spatial plantar stress maps, allowing for assessment of athletic performance, gait abnormalities, and rehabilitation progress. Coupled with artificial-intelligence-assisted signal analysis, the system demonstrated high effectiveness for sports science and disease diagnosis. Chang et al. \cite{chang2023triboelectric} developed a TENG-based tactile sensor array for pressure distribution monitoring within prosthetic limbs. The device involved a flexible self-powered sensor array embedded within the prosthetic socket–residuum interface, which enabled real-time measurement of pressure variations experienced by amputees during movement. Furthermore, extending pressure mapping towards multimodal sensing, Arief et al.~\cite{arief2026metasurface} developed a three-by-three triboelectric e-skin array containing 2 mm × 2 mm nanocone-patterned PDMS sensing pixels. The platform combined contact mapping, proximity detection and acoustic response, illustrating how millimetre-scale functional units can be integrated into multifunctional sensing arrays.

\subsection{Microscale Applications}

\begin{figure}[!b]
    \centering
    \includegraphics[width=\linewidth]{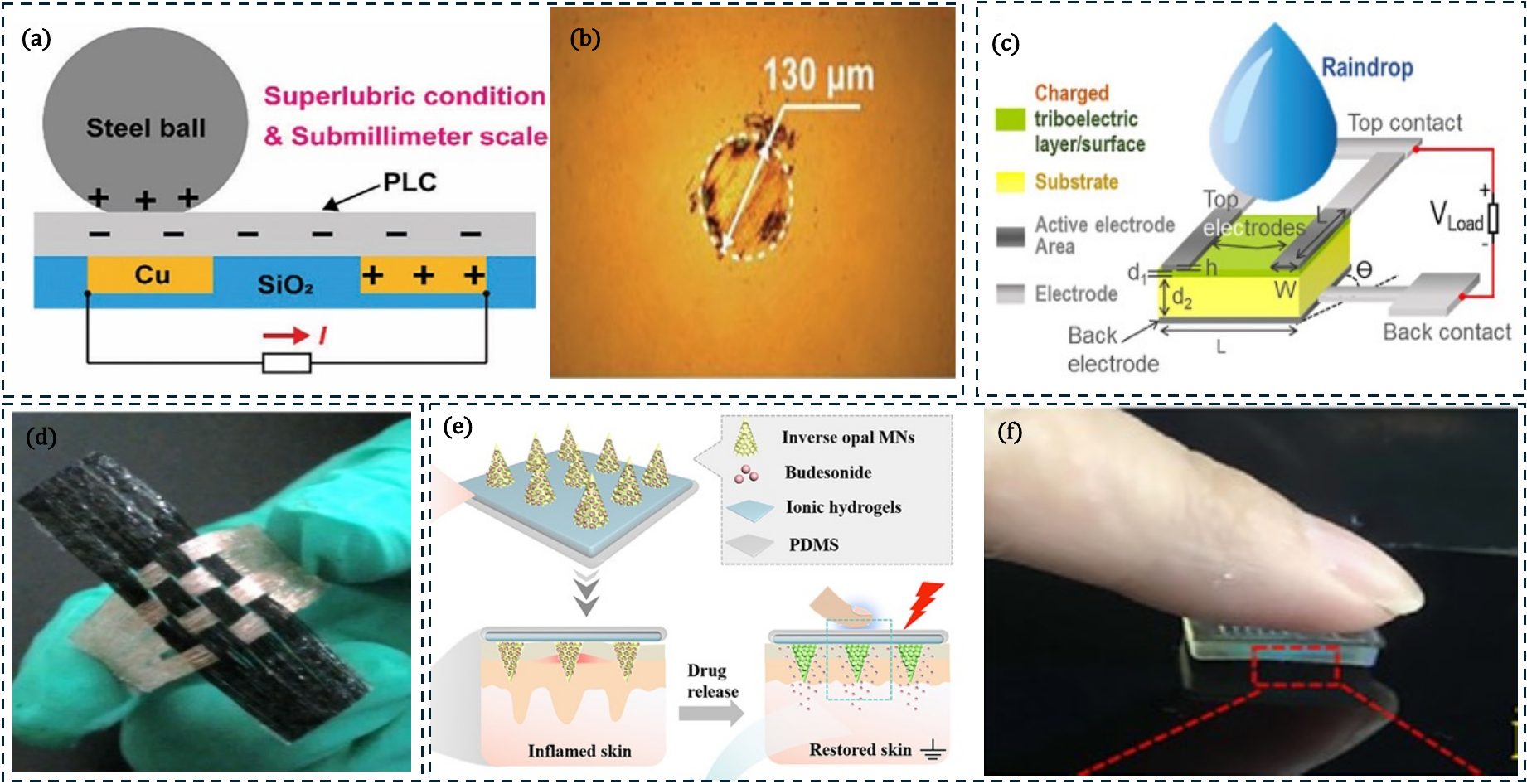}
    \caption{Microscale TENGs:(a) working mechanism of the submillimetre-scale superlubric TENG, and (b) the optical microscope image of the steel ball (Adopted from Bu et al. \citep{bu2024submillimeter}); (c) Working mechanism of single raindrop actuating Trecxels (Adopted from Ref.~\citep{ghaffarinejad2025triboelectric}) (d) Optical image of the 2DW-WTNG (Adopted from Liu et al. \citep{liu2019core};  ) (e) Schematic of multifunctional structural-colour triboelectric microneedle patches (SMP-TENG) for psoriasis therapy, and (f) Optical images of finger contact with the SMP-TENG (Adopted from Wang et al. \citep{wang2023multifunctional}}.
    \label{fig:micro_scale_TENGs_1}
\end{figure}

Microscale TENG devices, with characteristic lateral dimensions ranging from 1 $\mu m$ to 1 mm, have emerged as compact, integrable triboelectrification-based generators that bridge the gap between bulky macroscale energy harvesters and lab-on-chip or implantable microsystems. These miniaturised TENG architectures are designed for easy integration with MEMS, microfluidic platforms, and wearable microneedle arrays, where the available device footprint, and not merely the harvested power, becomes the governing design constraint \cite{chen2020micro, bu2024submillimeter, ghaffarinejad2025triboelectric,ning2021flexible,wang2023multifunctional, liu2019core}. Chen et al. \cite{chen2020micro} reported the first MEMS-fabricated microscale TENG, termed a micro triboelectric ultrasonic device ($\mu$TUD) (Figure~\ref{fig:micro_scale_TENGs_1}(a)).
The device features a fully clamped silicon membrane positioned above a vacuum cavity, which forms a triboelectric pair with an underlying silicon oxide dielectric layer. By patterning a circular diaphragm of only 50 µm in diameter using standard wafer-bonding and reactive-ion-etching processes, the authors pushed the resonant, and hence the operating, frequency of the device into the megahertz range (1.17 MHz). This microscale diaphragm was subsequently used to demonstrate acoustic energy transfer. A second strategy for achieving microscale TENG operation while simultaneously addressing the wear-induced output decay that plagues most contact-mode TENGs is to embed submillimetre interdigital electrodes directly into a flat dielectric surface, as demonstrated in the submillimetre-scale superlubric TENG (SS-TENG)~\cite{bu2024submillimeter}(Figure~\ref{fig:micro_scale_TENGs_1}(a). In the experiment, interdigital Cu electrodes are recessed into a $SiO_2$ layer to form a flush, low-roughness surface (<6 nm), on which a highly hydrogenated diamond-like carbon (DLC) triboelectric film is deposited. The practical sliding contact established with a steel counter-body is confined to a diameter of only 130 µm~(Figure~\ref{fig:micro_scale_TENGs_1}(b). Notably, this submillimetre-scale device overcame the trade-off between electrical output and wear that has been prevalent in previously reported superlubric tribovoltaic generators. The SS-TENG achieved a stable peak power density of 3 $mW/m^2$.

Beyond solid-solid contact architectures, the inception of solid–liquid interactions for droplet energy harvesting by Ghaffarinejad et al. \cite{ghaffarinejad2025triboelectric} introduced the triboelectric pixel, or “Trecxel,” a capacitive unit cell with a square active area of L $\times$ L (L = 900 µm ) that converts the entire kinetic energy of a falling droplet into a microsecond-scale electrical pulse through an instantaneous change in capacitance (Figure~\ref{fig:micro_scale_TENGs_1}(c). Because the Trecxel's submillimetric footprint is comparable to the contact diameter of a single droplet, multiple Trecxels packed into a dense n × m array can be actuated simultaneously by a single impact event, thereby increasing the harvested instantaneous power by up to 67\%. This also demonstrated, for the first time, a microscale building-block strategy for large-area “rain panel” integration.

Liu et al. \cite{liu2019core} twisted enamelled copper and steel micro-wires together with everyday nylon and polyester sewing threads to obtain core-shell composite conductive fibres with diameters of 380 µm and 385 µm, respectively, which were subsequently woven into a two-dimensional structure (2DW-WTNG) (Figure~\ref{fig:micro_scale_TENGs_1}(d)) to harvest body-motion energy through a sliding-mode mechanism. Along similar lines, Ning et al. \cite{ning2021flexible} orderly deposited silver-nanowire/carbon-nanotube conducting layers and a PDMS encapsulation onto a stretchable spandex core to obtain a coaxial fibre-shaped TENG with an overall diameter of 0.63 mm. The device was capable of sustaining over 140\% tensile strain while functioning as a self-powered tactile sensor. In both cases, it is the microscale fibre diameter, essentially not the centimetre-to-metre lengths dictated by the textile or garment application, that defines the device’s characteristic dimension and the parameters.

Finally, the microscale form factor of TENGs has been directly coupled to biomedical microneedle (MN) platforms for self-powered, electrically stimulated transdermal drug delivery. Wang et al. \cite{wang2023multifunctional} fabricated structural-colour triboelectric MN patches (Figure~\ref{fig:micro_scale_TENGs_1}(e) from a budesonide-loaded ionic hydrogel possessing an inverse-opal scaffold, in which each MN tip occupied 830 µm in height with a 360 µm basal diameter. These patches simultaneously puncture the skin, release the drug under the friction-induced electrostatic stimulation generated by contact with the epidermis, and self-report the progress of drug release through a real-time, photonic-band-gap-driven colour shift. Such MN-integrated triboelectric devices in Figure~\ref{fig:micro_scale_TENGs_1}(f) illustrate how confining the contact-mode TENG architecture to the dimensions of an individual microneedle allows triboelectric stimulation, drug delivery and optical self-reporting to be co-located within the same micrometre-scale structure. Although the individual microneedles possess microscale dimensions, the complete microneedle patch may occupy a substantially larger area. Accordingly, the microscale classification refers to the individual functional microneedle elements rather than the overall patch dimensions.

\section{Discussion and prospects}
The reported increases in electrical output becomes a practical design advantage when its benefit persists under the loading, environment and operating lifetime required by the application. Material selection, numerical modelling and device integration must therefore be assessed together. This discussion explains how material choices create modelling requirements, which measurements can resolve critical uncertainties, and how performance and validation targets change across device scales.
\subsection{Selection of materials} TENG performance is strongly influenced by the choice of triboelectric materials, in which several key factors are important, such as: (1) Polarity remains fundamental, i.e., the higher the polarity differences, the better the performance of the TENG. Fluorinated groups strengthen tribo-negative behaviour, while amino-, hydroxyl-, and polysaccharide-rich surfaces serve as tribo-positive counterparts \cite{liu2026unveiling,alghamdi2025triboelectric,wang2025enhancing}. (2) Inorganic fillers such as BaTiO$_3$, TiO$_2$, MXene, etc., result in dielectric improvement and can be used to improve polarisation and electrostatic induction \cite{lee2025boosting,zhang2025flexible}. (3) Interfacial polarisation and charge trapping play a vital role in composite films and layered architectures, such as in fluorinated, cellulose-based, and MXene composites \cite{ye2025noncontact,ren2026edge,gu2025biodegradable,wang2025enhancing}. (4) Surface morphology is modified to enhance effective contact area using porous films, hollow arrays, nanofibrous membranes, and honeycomb structures and reflects the morphology-enhanced performance \cite{wang2026electric,chittibabu2026biomorphic,wang2025flexible}. (5) Environmental protection using encapsulation or humidity/temperature-resistant packaging is also critical to sustain the TENG performance under harsh operating conditions \cite{wei2026ambient,panda2026triboelectric}. Therefore, a thorough consideration of all aspects, i.e., choice of materials (with high polarity differences, mechanical/environmental durability), device architecture/design, and surface morphology, must be carefully selected and taken into consideration to optimise the TENG performance.

\subsection{Promising routes to multiscale-multiphysics modelling}

Although TENG modelling has progressed from electrostatic field calculations to contact, circuit, roughness, and atomistic simulations, several numerical constraints still limit predictive device-scale modelling. A central issue is that the triboelectric surface charge density is usually prescribed as an input parameter rather than obtained from the contact electrification process itself~\citep{liu2022standardized,kumar2023multiscale,hu2024review}. In practice, charge generation depends on the material pair, real contact area, surface chemistry, humidity, charge decay, repeated cycling, and possible ion or electron transfer mechanisms. Moreover, visualisation and standardised quantification of surface charge remain challenging~\citep{liu2022standardized,li2024visualization}. Therefore, future models need charge-generation laws that are either physics-based, data-assisted, or calibrated using direct surface-charge measurements.

Atomistic simulations can reveal electron transfer, ion transfer, humidity effects, and interfacial charge redistribution at the nanoscale, but they cannot directly predict complete TENG voltage, current, or power because of their limited length and time scales~\citep{ratanaporn2024reaxff}. Therefore, the problem with the air gap between molecular-scale charge transfer and device-scale performance remains challenging.   In this context, continuum FEM models can simulate full devices but usually require assumed or fitted charge-density parameters. Furthermore, a very recently opened branch of concepts  flexoelectricity shows that the strain-gradient effect the nanoscale triboelectricity~\citep{mizzi2022flexoelectricity,silavnieks2025flexoelectricity,qiao2021mixed}. To incorporate this, an advanced numerical method, e.g., the mixed finite element method, can be a suitable choice for TENG modelling~\citep{mcbride2020modelling,deng2023mixed,liu2026mixed}. 

Prioritising the representation of real contact area, many earlier models rely on nominal contact area, idealised microtextures, or statistical rough-contact theories, which cannot fully resolve the irregular contact patches formed on practical triboelectric surfaces. Real TENG interfaces often contain random multiscale roughness, asperities, manufacturing defects, and non-uniform contact regions~\citep{kumar2023multiscale,XUVoc,jin2016contact,persson2002elastic}. This issue has recently been addressed by Sanglap et al. \citep{sanglap2026multiphysics}, who introduced explicit rough-surface representation into a multiphysics finite element framework. Their approach provides a more direct link between rough-surface contact mechanics and electrostatic output. However, further developments to include cyclic roughness evolution, wear, humidity-dependent charging, viscoelastic deformation, and long-term circuit response under realistic operating conditions are neccessary~\citep{wang2021investigation,yang2018comprehensive}.

The fully coupled numerical method between mechanical contact and electrostatics also remains numerically limited. Electrostatic FEM simulations often prescribe the device motion or separation distance, while contact models may compute the contact area separately before passing it into the electrical model. One difficulty is that finite-size electrostatic TENG simulations require an air domain around the device, whereas contact simulations require accurate treatment of gap closure, non-penetration, and evolving interface geometry. A possible route is to explore third-medium contact formulations~\citep{wriggers2013third,faltus2024thirdmedium} coupled with electrostatics, where the contact interface and surrounding medium are treated in a more regularised computational setting. However, consideration of roughness for such methods opens up further branches in numerical development. Therefore, such approaches could support staggered or strongly coupled formulations in which deformation, contact pressure, contact area, surface charge, and electrostatic potential evolve consistently.

Agreement with measured voltage can remain ambiguous when both real contact area and surface charge density are adjustable inputs. A more discriminating validation strategy should test the modelling chain in stages: contact mechanics against force and contact-area measurements, electrostatics against capacitance and charge measurements, and circuit response against voltage and current under different electrical loads. Parameters calibrated under one condition should then be tested under other conditions within the intended operating range. Such comparisons would help identify which mechanisms and uncertainties control predictive accuracy. In this context, TENG-enabled digital twins have been proposed as a future route for linking sensing, modelling, and data-driven calibration~\citep{jin2020softrobotics,pang2024digitaltwins}. Such datasets and digital frameworks would allow numerical models to be calibrated and tested more systematically. Therefore, future TENG modelling should move towards scalable multiphysics frameworks that combine rough-surface contact mechanics, realistic material behaviour, contact-area-dependent charge generation and decay, electrostatic induction, and external circuit dynamics. This would reduce reliance on prescribed surface charge density and idealised geometry assumptions, enabling more reliable design and optimisation of TENGs under realistic operating conditions. Therefore, a practical next step is to establish application-specific benchmarks linking material characterisation, model predictions and device performance. For energy harvesters, these should include energy delivered to the intended load; for sensors, repeatability and calibration stability are central. In each case, loading history, environmental conditions and target operating lifetime should be specified. This would allow multiscale modelling to be judged by its ability to explain performance changes, guide design choices and predict behaviour beyond the conditions used for calibration.

\subsection{Scalable applications}
\label{sec:teng_applications}

The rapid development of triboelectric nanogenerators (TENGs) has significantly expanded their application spectrum, ranging from large-scale environmental energy harvesting to wearable healthcare systems and implantable biomedical devices. Despite these advances, most TENG technologies remain at the proof-of-concept stage, and only a limited number have demonstrated the reliability required for long-term practical deployment. Future research should therefore shift its focus from simply maximising electrical output toward developing robust, application-orientated systems with improved durability, manufacturing scalability, and seamless system integration~\cite{fan2012flexible,Shi2019MoreThanEnergy}.

At the macroscale, TENGs are expected to play an increasingly important role in smart infrastructure, transportation systems, industrial monitoring, and marine energy harvesting. Recent studies have demonstrated structurally engineered TENGs capable of harvesting low-frequency mechanical energy for applications such as intelligent flooring and distributed monitoring systems~\cite{so2023flexible,ghaffarinejad2025triboelectric}. However, practical deployment requires devices that can withstand prolonged exposure to moisture, dust, ultraviolet radiation, temperature fluctuations, and repeated mechanical loading. Moreover, the irregular nature of environmental mechanical excitation often results in unstable electrical output compared with laboratory conditions. Future research should therefore emphasise modular architectures, scalable manufacturing, and lifecycle assessment to ensure reliable long-term operation~\cite{fan2012flexible}.

At the millimetre scale, wearable TENGs have emerged as promising candidates for personalised healthcare, rehabilitation, and human--machine interaction. Smart gloves, textile-integrated sensors, respiratory monitoring devices, and rehabilitation systems have demonstrated the capability of converting subtle human motions into self-powered electrical signals~\cite{zhang20233d,ning2021flexible}. Nevertheless, practical wearable systems must satisfy multiple requirements simultaneously, including flexibility, breathability, user comfort, washability, and long-term mechanical stability. Perspiration, humidity, and repeated deformation can gradually reduce triboelectric performance, while delamination between electrodes and triboelectric layers remains a major reliability concern. Future efforts should therefore focus on breathable encapsulation, durable textile integration, adaptive calibration, and standardized evaluation methods. Furthermore, large-scale clinical validation will be necessary before TENG-based rehabilitation systems can be widely adopted~\cite{bhatia2021wearable}.

At the micro- and implantable scales, TENGs offer exciting opportunities for battery-free physiological monitoring, wireless power transfer, controlled drug delivery, neural stimulation, and bioresorbable therapeutic devices. Although implantable ligament sensors, magnetically actuated TENGs, and microscale triboelectric systems have demonstrated encouraging performance in animal models or proof-of-concept studies~\cite{kim2025p,sheng2022ultrastretchable,chen2020micro}, several critical challenges remain before clinical translation can be realized. Long-term biocompatibility, biofluid-induced degradation, immune responses, reliable encapsulation, sterilisation, and minimally invasive implantation procedures all require further investigation. Future studies should therefore prioritise chronic \textit{in vivo} evaluation together with clinically relevant safety and reliability assessments.

Across all length scales, another important challenge is the transition from individual TENG devices to complete self-powered systems. Efficient integration of TENGs with power-management circuits, energy-storage modules, wireless communication, and intelligent data-processing algorithms will be essential for practical applications~\cite{Shi2019MoreThanEnergy}. Equally important is the establishment of standardised testing protocols and reporting criteria, enabling meaningful comparison among different studies and accelerating commercialisation. By addressing these multidisciplinary challenges, TENGs are expected to evolve from laboratory demonstrations into reliable self-powered platforms for healthcare, intelligent infrastructure, environmental monitoring, and next-generation Internet of Things applications.

\section{Conclusion}
This review has examined TENG connecting material selection, computational modelling, and applications across length scales. The material survey shows that no single
triboelectric material can satisfy all performance requirements. Fluorinated polymers provide strong tribo-negative behaviour, while elastomers offer the mechanical compliance required for wearable and deformable devices. Thermoplastics and flexible films provide scalable processing routes, whereas inorganic fillers and hybrid composites can improve dielectric response, interfacial polarisation, and charge trapping. Natural and biodegradable
materials offer promising routes towards sustainable and transient devices,
although their electrical output, environmental stability, durability, and
manufacturing repeatability require further improvement. Consequently,
material polarity alone is insufficient as a design criterion; dielectric
properties, mechanical behaviour, surface morphology, charge retention,
environmental resistance, processability, and application-specific
requirements must be considered jointly.

Computational modelling has progressed from simplified analytical and
equivalent-circuit descriptions to finite-element electrostatics, mechanical contact simulations, rough-surface representations, molecular dynamics, and first-principles calculations. Each approach resolves a different part of the TENG response. Atomistic methods provide insight into interfacial electron or ion transfer but remain restricted in spatial and temporal scale. Continuum models can represent complete devices and external circuits, but commonly prescribe the surface charge density, material motion, or contact state. Furthermore, many models use nominal contact areas or idealised surface textures that do not capture the irregular and evolving contact patches of manufactured triboelectric surfaces. As a result, the direct relationship between material chemistry, real contact evolution, charge generation, electrostatic induction, and useful electrical power remains only partially resolved.

The length-scale classification further demonstrates that miniaturisation is not simply a geometric reduction of a macroscale TENG. Macroscale systems prioritise mechanical robustness, large-area energy harvesting, environmental protection, and integration with infrastructure. Milliscale devices require a balance between output performance, mechanical compliance, and compact integration for wearable and biomedical systems. Microscale TENGs introduce additional constraints associated with fabrication accuracy, biocompatibility, local mechanical interaction, and integration with microelectromechanical or therapeutic platforms. Material choice, modelling resolution, and performance measures must therefore be selected according to the characteristic dimensions and operating environment of the intended application.

At this stage, future progress requires scalable and experimentally validated multiscale frameworks in which surface charge density is no longer treated solely as a prescribed parameter. Such models should connect molecular-scale contact-electrification mechanisms with measured surface chemistry and topography, nonlinear and time-dependent material deformation, friction, adhesion, wear, humidity-dependent charging, charge decay, electrostatic induction, and external circuit dynamics. Validation should extend beyond voltage or current alone and include surface topography, real contact area, contact pressure, charge distribution, capacitance, transferred charge, and power output. Data-assisted constitutive descriptions and digital-twin approaches may complement physics-based models where direct multiscale coupling remains computationally prohibitive.

Overall, the continued investigation and experiments of TENGs will depend on treating materials, interfaces, device dimensions, numerical models, and application requirements as parts of a single design problem. Closer integration of these areas can reduce dependence on empirical trial-and-error development and enable more reliable optimisation of TENGs for sustainable energy harvesting, self-powered sensing, wearable electronics, biomedical systems, and large-area infrastructure applications.

\FloatBarrier



\bibliography{reference.bib}

@article{alghamdi2025triboelectric,
  author    = {Alghamdi, Mashael S and Morgan, Joseph James and Walsh, Kieran and Shin, Dong Wook and Nigmatullin, Rinat and Saadi, Zakaria and Routledge, Jack and Neves, Ana I S and Russo, Saverio and Eichhorn, Stephen James and Craciun, Monica F},
  title     = {Triboelectric nanogenerator based on cellulose nanocrystals and graphene for energy harvesting from piano playing motion},
  journal   = {Nano Energy},
  year      = {2025},
  volume    = {138},
  pages     = {110816},
  doi       = {10.1016/j.nanoen.2025.110816},
  url       = {https://www.sciencedirect.com/science/article/pii/S2211285525001752},
  issn      = {2211-2855}
}

@article{alzgool2022micro,
  author    = {Alzgool, Mohammad and Mousavi, Mohammad and Davaji, Benyamin and Towfighian, Shahrzad},
  title     = {Micro-triboelectric generator for zero-power shock detection},
  journal   = {Nano Energy},
  year      = {2022},
  volume    = {103},
  publisher = {Elsevier Ltd},
  doi       = {10.1016/j.nanoen.2022.107758},
  issn      = {22112855}
}

@article{an2021advances,
  author    = {An, Xuyao and Wang, Chunnan and Shao, Ruomei and Sun, Shuqing},
  title     = {Advances and prospects of triboelectric nanogenerator for self-powered system},
  journal   = {International Journal of Smart and Nano Materials},
  year      = {2021},
  volume    = {12},
  number    = {3},
  pages     = {233-255},
  publisher = {Taylor \& Francis},
  doi       = {10.1080/19475411.2021.1973143},
  issn      = {1947-5411}
}

@article{ankit2025engineering,
  author    = {Ankit, Raj and Kolte, Jayant},
  title     = {Engineering of Chitosan/poly (vinyl alcohol) nanofibrous membrane as triboelectric nanogenerator for energy harvesting and self-powered speed sensor},
  journal   = {Journal of Materials Science: Materials in Electronics},
  year      = {2025},
  volume    = {36},
  number    = {22},
  pages     = {1366},
  doi       = {10.1007/s10854-025-15399-2},
  issn      = {1573-482X}
}

@article{ankit2025scalable,
  author    = {Ankit, Raj and Chakraborty, Tirthankar and Kolte, Jayant},
  title     = {Scalable electrospun Chitosan/PVA nanofibers: A pathway for triboelectric nanogenerator-based energy harvesting},
  journal   = {Polymer},
  year      = {2025},
  volume    = {335},
  pages     = {128848},
  doi       = {10.1016/j.polymer.2025.128848},
  url       = {https://www.sciencedirect.com/science/article/pii/S0032386125008341},
  issn      = {0032-3861}
}

@article{ankit2026edible,
  author    = {Ankit, Raj and Kaur, Simranjeet and Athwal, Shinar and Kaur, Taranveer and Kolte, Jayant},
  title     = {An edible microbial cellulose-based triboelectric nanogenerator: a sustainable approach for energy harvesting},
  journal   = {Mater. Adv.},
  year      = {2026},
  volume    = {7},
  number    = {3},
  pages     = {1508-1518},
  publisher = {RSC},
  doi       = {10.1039/D5MA01186H}
}

@article{bairagi2026electro,
  author    = {Bairagi, Satyaranjan and Zada, Muhammad and Otesteanu, Corin and Menon, Carlo},
  title     = {Electro-active phase assisted all-fiber triboelectric nanogenerator (AF-TENG) for energy harvesting and human joint angle monitoring},
  journal   = {Nano Energy},
  year      = {2026},
  volume    = {150},
  pages     = {111784},
  doi       = {10.1016/j.nanoen.2026.111784},
  url       = {https://www.sciencedirect.com/science/article/pii/S2211285526000881},
  issn      = {2211-2855}
}

@article{bhatia2021wearable,
  author    = {Bhatia, Divij and Jo, Seong Hyeon and Ryu, Yeonhun and Kim, Yusung and Kim, Dong Hyun and Park, Hyung Soon},
  title     = {Wearable triboelectric nanogenerator based exercise system for upper limb rehabilitation post neurological injuries},
  journal   = {Nano Energy},
  year      = {2021},
  volume    = {80},
  publisher = {Elsevier Ltd},
  doi       = {10.1016/j.nanoen.2020.105508},
  issn      = {22112855}
}

@article{bhatia2022triboelectric,
  author    = {Bhatia, Divij and Lee, Kyoung Soub and Niazi, Muhammad Umer Khan and Park, Hyung Soon},
  title     = {Triboelectric nanogenerator integrated origami gravity support device for shoulder rehabilitation using exercise gaming},
  journal   = {Nano Energy},
  year      = {2022},
  volume    = {97},
  publisher = {Elsevier Ltd},
  doi       = {10.1016/j.nanoen.2022.107179},
  issn      = {22112855}
}

@article{bhattacharyya2025synthesis,
  author    = {Bhattacharyya, Mriganka and CP, Ajay Haridas and T, Muthamil Selvan and Mondal, Titash},
  title     = {Synthesis of organic-silicone hybrid elastomer-based pressure-sensitive adhesive for triboelectric nanogenerator},
  journal   = {Journal of Macromolecular Science, Part A},
  year      = {2025},
  volume    = {62},
  number    = {11},
  pages     = {1006-1015},
  publisher = {Taylor \& Francis},
  doi       = {10.1080/10601325.2025.2568986}
}

@article{bu2024submillimeter,
  author    = {Bu, Tianzhao and Deng, Wenli and Liu, Yaoyao and Wang, Zhong Lin and Chen, Xinchun and Zhang, Chi},
  title     = {Submillimeter-Scale Superlubric Triboelectric Nanogenerator},
  journal   = {Advanced Functional Materials},
  year      = {2024},
  volume    = {34},
  number    = {40},
  publisher = {John Wiley and Sons Inc},
  doi       = {10.1002/adfm.202404007},
  issn      = {16163028}
}

@article{chakraborty2026natural,
  author    = {Chakraborty, Sayan and Ray, Samit K},
  title     = {Natural silk fibroin -- MOF nanocomposite based wearable triboelectric nanogenerators for energy harvesting and real-time IoT-enabled healthcare monitoring},
  journal   = {Chemical Engineering Journal},
  year      = {2026},
  volume    = {532},
  pages     = {174424},
  doi       = {10.1016/j.cej.2026.174424},
  url       = {https://www.sciencedirect.com/science/article/pii/S1385894726018838},
  issn      = {1385-8947}
}

@article{chang2023triboelectric,
  author    = {Chang, Kuie and Parashar, Parag and Shen, Li Chien and Chen, An Rong and Huang, Yan Tsz and Pal, Arnab and Lim, Kee Chin and Wei, Po Han and Kao, Fu Cheng and Hu, Jin Jia and Lin, Zong Hong},
  title     = {A triboelectric nanogenerator-based tactile sensor array system for monitoring pressure distribution inside prosthetic limb},
  journal   = {Nano Energy},
  year      = {2023},
  volume    = {111},
  publisher = {Elsevier Ltd},
  doi       = {10.1016/j.nanoen.2023.108397},
  issn      = {22112855}
}

@article{chen2020micro,
  author    = {Chen, Chen and Wen, Zhen and Shi, Jihong and Jian, Xiaohua and Li, Peiyang and Yeow, John T.W. and Sun, Xuhui},
  title     = {Micro triboelectric ultrasonic device for acoustic energy transfer and signal communication},
  journal   = {Nature Communications},
  year      = {2020},
  volume    = {11},
  number    = {1},
  publisher = {Springer US},
  doi       = {10.1038/s41467-020-17842-w},
  issn      = {20411723}
}

@article{chen2020universal,
  author    = {Chen, Jinkai and Wang, Junchao and Xuan, Weipeng and Dong, Shurong and Luo, Jikui},
  title     = {Universal Triboelectric Nanogenerator Simulation Based on Dynamic Finite Element Method Model},
  journal   = {Sensors},
  year      = {2020},
  volume    = {20},
  number    = {17},
  doi       = {10.3390/s20174838},
  url       = {https://www.mdpi.com/1424-8220/20/17/4838},
  issn      = {1424-8220}
}

@article{bush1975elastic,
  title={The elastic contact of a rough surface},
  author={Bush, AW and Gibson, RD and Thomas, TR},
  journal={Wear},
  volume={35},
  number={1},
  pages={87--111},
  year={1975},
  publisher={Elsevier}
}

@article{chen2026all,
  author    = {Chen, Jiansong and Chen, Pengfei and Du, Haishun and Zhang, Xiaoxue and Wang, Yi-Cheng and Wang, Xudong and Pan, Xuejun},
  title     = {All-cellulose triboelectric nanogenerators for sustainable energy harvesting and self-powered sensing in smart home applications},
  journal   = {Matter},
  year      = {2026},
  volume    = {9},
  number    = {3},
  pages     = {102633},
  doi       = {10.1016/j.matt.2025.102633},
  url       = {https://www.sciencedirect.com/science/article/pii/S2590238525006769},
  issn      = {2590-2385}
}

@article{zhao2026solvent,
  title={Solvent-assisted running-in strategy enables improved triboelectric nanogenerator output},
  author={Zhao, Jun and Ge, Bin and Feng, Xiangyu and Nie, Yanjie and Zhu, Pengzhe and Zhang, Yajie and Wu, Wei and Li, Xueyuan and Wang, Yongfeng and Shi, Yijun},
  journal={Nature Communications},
  year={2026},
  publisher={Nature Publishing Group}
}

@article{nayak1971random,
  title={Random process model of rough surfaces},
  author={Nayak, P Ranganath},
  year={1971}
}

@article{fatti2026first,
  title={First-Principles Models of Triboelectrification},
  author={Fatti, Giulio and Cavin, John and Ko, Hyunseok and Mishra, Rohan and Dini, Daniele and Cho, Sung Beom},
  journal={Small Methods},
  volume={10},
  number={4},
  pages={e01782},
  year={2026},
  publisher={Wiley Online Library}
}

@article{carley2026multiscale,
  title={Multiscale Simulation of Asperity Flattening with Realistic Surface Topography and Microstructure},
  author={Carley-Clopton, Aiden and Winther, Grethe and Hazrati, Javad and Nielsen, Chris Valentin},
  journal={Materials Science Forum},
  volume={1184},
  pages={21--33},
  year={2026},
  publisher={Trans Tech Publications Ltd},
  doi={10.4028/p-9iISdK}
}

@article{yastrebov2015infinitesimal,
  title={From infinitesimal to full contact between rough surfaces: Evolution of the contact area},
  author={Yastrebov, Vladislav A and Anciaux, Guillaume and Molinari, Jean-Fran{\c{c}}ois},
  journal={International Journal of Solids and Structures},
  volume={52},
  pages={83--102},
  year={2015},
  publisher={Elsevier},
  doi={10.1016/j.ijsolstr.2014.09.019}
}

@article{chittibabu2026biomorphic,
  author    = {Chittibabu, Suresh Kumar and Vijayababu, Mullapudi and Natarajan, Karthick and Sharma, Nimmi and Chandrasekhar, Arunkumar and Chintagumpala, Krishnamoorthi and Lin, Zong-Hong},
  title     = {Biomorphic honeycomb-engineered Ti2C MXene/PDMS composite triboelectric nanogenerator for eco-conscious energy harvesting and autonomous analysis of perishable food freshness},
  journal   = {Chemical Engineering Journal},
  year      = {2026},
  volume    = {533},
  pages     = {174961},
  doi       = {10.1016/j.cej.2026.174961},
  url       = {https://www.sciencedirect.com/science/article/pii/S1385894726024204},
  issn      = {1385-8947}
}

@article{cho2022highly,
  author    = {Cho, Eunmi and Kim, Kyeong Nam and Yong, Hyungseok and Choi, Woo Jin and Park, Jin Seong and Lee, Sang Jin},
  title     = {Highly transparent and water-repellent hierarchical-wrinkled-architecture triboelectric nanogenerator with ultrathin plasma-polymer-fluorocarbon film for artificial triboelectric skin},
  journal   = {Nano Energy},
  year      = {2022},
  volume    = {103},
  publisher = {Elsevier Ltd},
  doi       = {10.1016/j.nanoen.2022.107785},
  issn      = {22112855}
}

@article{choi2023recent,
  author    = {Choi, Dongwhi and Lee, Younghoon and Lin, Zong-Hong and Cho, Sumin and Kim, Miso and Ao, Chi Kit and Soh, Siowling and Sohn, Changwan and Jeong, Chang Kyu and Lee, Jeongwan},
  title     = {Recent advances in triboelectric nanogenerators: from technological progress to commercial applications},
  journal   = {ACS nano},
  year      = {2023},
  volume    = {17},
  number    = {12},
  pages     = {11087-11219},
  publisher = {ACS Publications},
  issn      = {1936-0851}
}

@article{das2026microstructurally,
  author    = {Das, Tupan and Singh, Ajeet and Zhang, Jun-Jie and Li, Chieh-Jui and Lin, Meng-Fang and Kar, Manoranjan},
  title     = {Microstructurally engineered MXene/Metal organic framework based hybrid nanogenerator for water energy harvesting and self-powered gas sensor},
  journal   = {Journal of Power Sources},
  year      = {2026},
  volume    = {676},
  pages     = {239784},
  doi       = {10.1016/j.jpowsour.2026.239784},
  issn      = {0378-7753}
}

@article{davoudi2024review,
  author    = {Davoudi, Mohammadmahdi and An, Chi-Yoon and Kim, Dae-Eun},
  title     = {A Review on Triboelectric Nanogenerators, Recent Applications, and Challenges},
  journal   = {International Journal of Precision Engineering and Manufacturing-Green Technology},
  year      = {2024},
  volume    = {11},
  number    = {4},
  pages     = {1317-1340},
  doi       = {10.1007/s40684-023-00569-6},
  issn      = {2198-0810}
}

@article{deb2026zno,
  author    = {Deb, Anwesha and Dash, Anabadya and Das, Sudipta and Das, Binoy and Mondal, Suvra Prakash},
  title     = {ZnO Nanorods Modified Cotton Fabric-Based Triboelectric Nanogenerators for Self-Powered Energy Harvesting and Acetone Sensing Applications},
  journal   = {physica status solidi (RRL) -- Rapid Research Letters},
  year      = {2026},
  volume    = {20},
  number    = {2},
  pages     = {e202500375},
  publisher = {John Wiley \& Sons, Ltd},
  doi       = {10.1002/pssr.202500375},
  issn      = {1862-6254}
}

@article{dev2026advanced,
  author    = {Dev, Amar and Kumar, Amod and Das, Tupan and Kumar, Pawan and Pradhan, S K and Kour, P and Kar, Manoranjan},
  title     = {Advanced flexible composite films of Ti3C2TX MXene-CCTO-PVDF for high-performance TENG-based self-powered sensor applications},
  journal   = {Chemical Engineering Journal},
  year      = {2026},
  volume    = {536},
  pages     = {176098},
  doi       = {10.1016/j.cej.2026.176098},
  url       = {https://www.sciencedirect.com/science/article/pii/S138589472603559X},
  issn      = {1385-8947}
}

@article{dien2025energy,
  author    = {Dien, Xiong and Ramli, Nurulazlina and Thio, Tzer Hwai Gilbert and Yang, Zhuanqing and Hu, Siyu and He, Xiang},
  title     = {Energy Storage, Power Management, and Applications of Triboelectric Nanogenerators for Self-Powered Systems: A Review},
  journal   = {Micromachines},
  year      = {2025},
  volume    = {16},
  number    = {10},
  doi       = {10.3390/mi16101170},
  url       = {https://www.mdpi.com/2072-666X/16/10/1170},
  issn      = {2072-666X}
}

@article{dip2023triboelectric,
  title={Triboelectric nanogenerators for marine applications: Recent advances in energy harvesting, monitoring, and self-powered equipment},
  author={Dip, Tanvir Mahady and Arin, Md Reasat Aktar and Anik, Habibur Rahman and Uddin, Md Mazbah and Tushar, Shariful Islam and Sayam, Abdullah and Sharma, Suraj},
  journal={Advanced Materials Technologies},
  volume={8},
  number={21},
  pages={2300802},
  year={2023},
  publisher={Wiley Online Library}
}

@article{dong2025flexible,
  author    = {Dong, Wentao and Li, Mengyun and Chen, Chang and Xie, Kun and Hong, Jinhua and Yang, Lin},
  title     = {Flexible hybrid self-powered piezo-triboelectric nanogenerator based on BTO-PVDF/PDMS nanocomposites for human machine interaction},
  journal   = {Scientific Reports},
  year      = {2025},
  volume    = {15},
  number    = {1},
  pages     = {15991},
  doi       = {10.1038/s41598-025-00686-z},
  issn      = {2045-2322}
}

@article{duan2025fully,
  author    = {Duan, Zhengyin and Xu, Jiayue and Zhao, Han and Chen, Yinjie and Huang, Chunmei and Lu, Peng},
  title     = {A Fully Degradable and Enclosed Triboelectric Nanogenerator Based on Spent Coffee Grounds Triboelectric Material},
  journal   = {Advanced Functional Materials},
  year      = {2025},
  publisher = {John Wiley and Sons Inc},
  doi       = {10.1002/adfm.202523277},
  issn      = {16163028}
}

@article{esmeria2025triboelectric,
  author    = {Esmeria, Jose M and F. Ligutan, Dino Dominic and Abad, Alexander C},
  title     = {Triboelectric nanogenerator (TENG) energy harvester using polypropylene-based recycled food packaging materials},
  journal   = {Japanese Journal of Applied Physics},
  year      = {2025},
  volume    = {64},
  number    = {9},
  pages     = {09SP13},
  publisher = {IOP Publishing},
  doi       = {10.35848/1347-4065/adfd68}
}

@article{feng20220,
  author    = {Feng, Junrui and Zhou, Hanlin and Cao, Zhi and Zhang, Enyang and Xu, Shuxing and Li, Wangtao and Yao, Huilu and Wan, Linyu and Liu, Guanlin},
  title     = {0.5 m Triboelectric Nanogenerator for Efficient Blue Energy Harvesting of All‐Sea Areas},
  journal   = {Advanced Science},
  year      = {2022},
  volume    = {9},
  number    = {35},
  doi       = {10.1002/advs.202204407},
  issn      = {2198-3844}
}

@article{ghaffarinejad2025triboelectric,
  author    = {Ghaffarinejad, Ali and Garc{\'i}a-Casas, Xabier and N{\'u}{\~n}ez-G{\'a}lvez, Fernando and Budagosky, Jorge and Godinho, Vanda and L{\'o}pez-Santos, Carmen and S{\'a}nchez-Valencia, Juan Ram{\'o}n and Barranco, {\'A}ngel and Borr{\'a}s, Ana},
  title     = {Triboelectric pixels as building blocks for microscale and large-area integration of drop energy harvesters},
  journal   = {Device},
  year      = {2025},
  volume    = {3},
  number    = {2},
  publisher = {Cell Press},
  doi       = {10.1016/j.device.2024.100566},
  issn      = {26669986}
}

@article{gu2025biodegradable,
  author    = {Gu, Yuchen and Niu, Qianqian and Cao, Qunxiang and Zhou, Xinglu and Zhao, Weikun and Hu, Zhanao and Fan, Suna and Yao, Xiang and Zhang, Yaopeng},
  title     = {Biodegradable silk fibroin-based triboelectric nanogenerator with enhanced output regulated by interfacial and ionic polarization},
  journal   = {International Journal of Biological Macromolecules},
  year      = {2025},
  volume    = {304},
  pages     = {140832},
  doi       = {10.1016/j.ijbiomac.2025.140832},
  url       = {https://www.sciencedirect.com/science/article/pii/S0141813025013819},
  issn      = {0141-8130}
}

@article{hamid2020novel,
  author    = {Hamid, H. M.Ashfiqul and Çelik-Butler, Zeynep},
  title     = {A novel MEMS triboelectric energy harvester and sensor with a high vibrational operating frequency and wide bandwidth fabricated using UV-LIGA technique},
  journal   = {Sensors and Actuators, A: Physical},
  year      = {2020},
  volume    = {313},
  publisher = {Elsevier B.V.},
  doi       = {10.1016/j.sna.2020.112175},
  issn      = {09244247}
}

@article{han2021paper,
  author    = {Han, Jing and Xu, Nuo and Liang, Yuchen and Ding, Mei and Zhai, Junyi and Sun, Qijun and Wang, Zhong Lin},
  title     = {Paper-based triboelectric nanogenerators and their applications: a review},
  journal   = {Beilstein Journal of Nanotechnology},
  year      = {2021},
  volume    = {12},
  pages     = {151-171},
  publisher = {Beilstein-Institut},
  doi       = {10.3762/bjnano.12.12},
  issn      = {2190-4286}
}

@article{han2025thermal,
  author    = {Han, Gi Hyeon and Park, Sang Jeong and Lee, Jong Hyun and Lee, Seung Yong and Lee, Kyu Hyoung},
  title     = {Thermal stable cordierite-embedded polyimide-based triboelectric generator toward vibration sensors applications at high temperature},
  journal   = {Nano Energy},
  year      = {2025},
  volume    = {145},
  pages     = {111457},
  doi       = {10.1016/j.nanoen.2025.111457},
  url       = {https://www.sciencedirect.com/science/article/pii/S221128552500816X},
  issn      = {2211-2855}
}

@article{hao2025triboelectric,
  author    = {Hao, Yijia and Ji, Tuo and Zhang, Jing and Chen, Weilin},
  title     = {Triboelectric nanogenerator based on changing the nanomorphology of polyoxometalates for gait monitoring of teenagers},
  journal   = {Nano Research},
  year      = {2025},
  volume    = {18},
  number    = {2},
  pages     = {94907192},
  doi       = {10.26599/NR.2025.94907192},
  url       = {https://www.sciopen.com/article/10.26599/NR.2025.94907192}
}

@article{kim2017morphology,
  title   = {Morphology effect on the transferred charges in
             triboelectric nanogenerators: Numerical study using
             a finite element method},
  author  = {Kim, SeongMin and Ha, Jaewook and Kim, Jin-Baek},
  journal = {Integrated Ferroelectrics},
  volume  = {183},
  number  = {1},
  pages   = {19--25},
  year    = {2017},
  doi     = {10.1080/10584587.2017.1375820}
}

@article{he2021monitoring,
  author    = {He, Haijun and Guo, Jian and Illés, Balázs and Géczy, Attila and Istók, Balázs and Hliva, Viktor and Török, Dániel and Kovács, József Gábor and Harmati, István and Molnár, Kolos},
  title     = {Monitoring multi-respiratory indices via a smart nanofibrous mask filter based on a triboelectric nanogenerator},
  journal   = {Nano Energy},
  year      = {2021},
  volume    = {89},
  publisher = {Elsevier Ltd},
  doi       = {10.1016/j.nanoen.2021.106418},
  issn      = {22112855}
}

@article{hu2024triboelectric,
  author    = {Hu, Songtao and Li, Haoran and Lu, Wenhui and Han, Tianyu and Xu, Youzhong and Shi, Xi and Peng, Zhike and Cao, Xiaobao},
  title     = {Triboelectric Insoles with Normal-Shear Plantar Stress Perception},
  journal   = {Advanced Functional Materials},
  year      = {2024},
  volume    = {34},
  number    = {16},
  publisher = {John Wiley and Sons Inc},
  doi       = {10.1002/adfm.202313458},
  issn      = {16163028}
}

@article{hu2026development,
  author    = {Hu, Jiahao and Li, Chang-Heng and Zhang, Yingfan and Huang, Zhengyong and Sha, Zhou and Li, Jian},
  title     = {Development of composite fluorinated polyimide films with simultaneous improvement of wear resistance and triboelectric performance},
  journal   = {Applied Surface Science},
  year      = {2026},
  volume    = {718},
  pages     = {164893},
  doi       = {10.1016/j.apsusc.2025.164893},
  url       = {https://www.sciencedirect.com/science/article/pii/S0169433225026091},
  issn      = {0169-4332}
}

@article{huang2023real,
  author    = {Huang, Kangxu and Zhou, Yuhui and Zhang, Zhicheng and Zhang, He and Lü, Chaofeng and Luo, Jikui and Shen, Libin},
  title     = {A real-time quantitative acceleration monitoring method based on triboelectric nanogenerator for bridge cable vibration},
  journal   = {Nano Energy},
  year      = {2023},
  volume    = {118},
  pages     = {108960},
  doi       = {10.1016/j.nanoen.2023.108960},
  issn      = {22112855}
}

@article{jiang2022plant,
  author    = {Jiang, Chengmei and Zhang, Qi and He, Chengxin and Zhang, Chi and Feng, Xiaohui and Li, Xunjia and Zhao, Qiang and Ying, Yibin and Ping, Jianfeng},
  title     = {Plant-protein-enabled biodegradable triboelectric nanogenerator for sustainable agriculture},
  journal   = {Fundamental Research},
  year      = {2022},
  volume    = {2},
  number    = {6},
  pages     = {974-984},
  doi       = {10.1016/j.fmre.2021.09.010},
  url       = {https://www.sciencedirect.com/science/article/pii/S2667325821001783},
  issn      = {2667-3258}
}

@article{jin2025perovskite,
  author    = {Jin, Zhangsiyuan and Xing, Hening and Hu, Yian and Cho, Yuljae},
  title     = {Perovskite-Based Triboelectric Nanogenerator and Its Application Towards Self-Powered Devices},
  journal   = {Nanoenergy Advances},
  year      = {2025},
  volume    = {5},
  number    = {1},
  doi       = {10.3390/nanoenergyadv5010002},
  url       = {https://www.mdpi.com/2673-706X/5/1/2},
  issn      = {2673-706X}
}

@article{karimmirza2025high,
  author    = {Karimmirza, Erfan and Manavizadeh, Negin},
  title     = {High-performance flexible triboelectric nanogenerators enabled by metal nanoparticle--silicone composites for wearable biomedical applications},
  journal   = {Results in Engineering},
  year      = {2025},
  volume    = {28},
  pages     = {107593},
  doi       = {10.1016/j.rineng.2025.107593},
  url       = {https://www.sciencedirect.com/science/article/pii/S2590123025036461},
  issn      = {2590-1230}
}

@article{kim2025p,
  author    = {Kim, Junyeop and Yoo, Jeongmin and Seo, Hantae and Avila, Raudel and Chung, Gooyoon and Shin, Gyuri and Gwak, Sujeong and Han, Yongbin and Lee, Ju-Hyuck and Yoon, Hong-Joon and Park, Yoonseok},
  title     = {A P P L I E D S C I E N C E S A N D E N G I N E E R I N G Magnetically driven triboelectric nanogenerator for a wireless, versatile energy transfer system},
  howpublished= {Sci. Adv},
  year      = {2025},
  volume    = {11},
  url       = {https://www.science.org}
}

@article{kim2025sustainable,
  author    = {Kim, Seongwan and Akhtar, Farhan and Iqbal, Shahzad and Rehman, Muhammad Muqeet and Kim, Woo Young},
  title     = {Sustainable Cellulose- and Pectin-Rich Triboelectric Nanogenerator for Mechanical Energy Harvesting and Self-Powered Humidity Sensing},
  journal   = {Polymers},
  year      = {2025},
  volume    = {17},
  number    = {23},
  doi       = {10.3390/polym17233130},
  url       = {https://www.mdpi.com/2073-4360/17/23/3130},
  issn      = {2073-4360}
}

@article{arief2022elastomeric,
  title={Elastomeric microwell-based triboelectric nanogenerators by in situ simultaneous transfer-printing},
  author={Arief, Injamamul and Zimmermann, Philipp and Hait, Sakrit and Park, Hyeyoung and Ghosh, Anik Kumar and Janke, Andreas and Chattopadhyay, Santanu and Nagel, J{\"u}rgen and Heinrich, Gert and Wie{\ss}ner, Sven and others},
  journal={Materials Horizons},
  volume={9},
  number={5},
  pages={1468--1478},
  year={2022},
  publisher={The Royal Society of Chemistry}
}

@article{mandal2024self,
  title={Self-repairable hybrid piezoresistive-triboelectric sensor cum nanogenerator utilizing dual-dynamic reversible network in mechanically robust modified natural rubber},
  author={Mandal, Subhradeep and Arief, Injamamul and Chae, Soosang and Tahir, Muhammad and Hoang, Tung X and Heinrich, Gert and Wie{\ss}ner, Sven and Das, Amit},
  journal={Advanced Sensor Research},
  volume={3},
  number={10},
  pages={2400036},
  year={2024},
  publisher={Wiley Online Library}
}

@article{kim2025triboelectric,
  author    = {Kim, Inkyum and Yun, Jonghyeon and Kim, Geunchul and Kim, Daewon},
  title     = {Triboelectric-Enhanced Piezoelectric Nanogenerator with Pressure-Processed Multi-Electrospun Fiber-Based Polymeric Layer for Wearable and Flexible Electronics},
  journal   = {Polymers},
  year      = {2025},
  volume    = {17},
  number    = {17},
  doi       = {10.3390/polym17172295},
  url       = {https://www.mdpi.com/2073-4360/17/17/2295},
  issn      = {2073-4360}
}

@article{kim2026wireless,
  author    = {Kim, Young Jun and Kim, So Hee and Park, Byung Joon and Jeon, Jinyoung and Kang, Donghyeon and Chung, Youngwook and Hwang, Joon Ha and Yoon, Hong Joon and Lee, Kyu Hyoung and Choi, Byung Ok and Kim, Sang Woo},
  title     = {Wireless and bioresorbable triboelectric nerve block system for postoperative pain control},
  journal   = {Nature Biomedical Engineering},
  year      = {2026},
  publisher = {Nature Research},
  doi       = {10.1038/s41551-025-01579-2},
  issn      = {2157846X}
}

@article{kumar2026boosting,
  author    = {Kumar, Shailendra and Jha, Rajesh Kumar and Chitnis, Ujjwal and Goswami, Ankur},
  title     = {Boosting triboelectric charge generation using high-performance functional MAX phase nanofillers incorporated in silicone elastomers},
  journal   = {J. Mater. Chem. A},
  year      = {2026},
  pages     = {-},
  publisher = {The Royal Society of Chemistry},
  doi       = {10.1039/D5TA10576E}
}

@article{lee2025boosting,
  author    = {Lee, Kangpyo and Ryu, Jeong Ho and Han, HyukSu and Kim, Boyeol and Jung, Kyunghwan and Moon, In Yong and Kwon, Ohyung and Chung, Chan-Yeup and Mhin, Sungwook and Kim, Kang Min},
  title     = {Boosting triboelectric performance of PDMS-based nanogenerators through dual-filler embedded triboelectric layers},
  journal   = {Journal of Alloys and Compounds},
  year      = {2025},
  volume    = {1010},
  pages     = {177114},
  doi       = {10.1016/j.jallcom.2024.177114},
  url       = {https://www.sciencedirect.com/science/article/pii/S0925838824037010},
  issn      = {0925-8388}
}

@article{li2024review,
  author    = {Li, Xiaochuan and Yang, Qianxi and Ren, Dahu and Li, Qianying and Yang, Huake and Zhang, Xuemei and Xi, Yi},
  title     = {A review of material design for high performance triboelectric nanogenerators: performance improvement based on charge generation and charge loss},
  journal   = {Nanoscale Adv.},
  year      = {2024},
  volume    = {6},
  number    = {18},
  pages     = {4522-4544},
  publisher = {RSC},
  doi       = {10.1039/D4NA00340C}
}

@article{liu2019core,
  author    = {Liu, Jinmei and Gu, Long and Cui, Nuanyang and Bai, Suo and Liu, Shuhai and Xu, Qi and Qin, Yong and Yang, Rusen and Zhou, Feng},
  title     = {Core-Shell Fiber-Based 2D Woven Triboelectric Nanogenerator for Effective Motion Energy Harvesting},
  journal   = {Nanoscale Research Letters},
  year      = {2019},
  volume    = {14},
  number    = {1},
  publisher = {Springer New York LLC},
  doi       = {10.1186/s11671-019-3144-2},
  issn      = {1556276X}
}

@article{liu2023environmentally,
  author    = {Liu, Songling and Tong, Wangshu and Gao, Caixia and Liu, Yulun and Li, Xinnan and Zhang, Yihe},
  title     = {Environmentally friendly natural materials for triboelectric nanogenerators: a review},
  journal   = {J. Mater. Chem. A},
  year      = {2023},
  volume    = {11},
  number    = {17},
  pages     = {9270-9299},
  publisher = {The Royal Society of Chemistry},
  doi       = {10.1039/D2TA10024J}
}

@article{liu2025electronic,
  author    = {Liu, Zhirong and Tang, Chuyu and Han, Nannan and Jiang, Zhuoheng and Liang, Xi and Wang, Shaobo and Hu, Quanhong and Xiong, Cheng and Yao, Shuncheng and Wang, Zhuo and Wang, Zhong Lin and Zou, Duohong and Li, Linlin},
  title     = {Electronic vascular conduit for in situ identification of hemadostenosis and thrombosis in small animals and nonhuman primates},
  journal   = {Nature Communications},
  year      = {2025},
  volume    = {16},
  number    = {1},
  publisher = {Nature Research},
  doi       = {10.1038/s41467-025-58056-2},
  issn      = {20411723}
}

@article{liu2025self,
  author    = {Liu, Qiancheng and Chen, Xueye and Wei, XuDong and Sun, Junlong and Liu, Changxia},
  title     = {Self-powered wearable triboelectric nanogenerator based on MXene/ chitosan/carbon nanotube layered composite film},
  journal   = {Chemical Engineering Journal},
  year      = {2025},
  volume    = {517},
  pages     = {164570},
  doi       = {10.1016/j.cej.2025.164570},
  url       = {https://www.sciencedirect.com/science/article/pii/S1385894725054063},
  issn      = {1385-8947}
}

@article{liu2026research,
  author    = {Liu, Meng-Nan and Li, Xu and Zheng, Gang and Yin, Fang and Li, Ru and Zhang, Jun and Ramakrishna, Seeram and Long, Yun-Ze},
  title     = {Research progress in triboelectric nanogenerators based on new inorganic materials},
  journal   = {iScience},
  year      = {2026},
  publisher = {Elsevier},
  issn      = {2589-0042}
}

@article{liu2026unveiling,
  author    = {Liu, Jinmei and Zhang, Fanyu and Xu, Jiongyao and Jia, Shengyang and Cui, Nuanyang and Yang, Rusen and Wang, Weiwei and Gu, Long},
  title     = {Unveiling the synergistic mechanism of C-F and C-Cl bonds in enhancing the triboelectric performance of fluorinated polymers},
  journal   = {Nature Communications},
  year      = {2026},
  volume    = {17},
  number    = {1},
  pages     = {3698},
  doi       = {10.1038/s41467-026-71546-1},
  issn      = {2041-1723}
}

@article{mahapatra2026fluorinated,
  author    = {Mahapatra, Abhinav and Deepak, Deepak and Singh, Tanuja and Agarwal, Khushboo and Roy, Susanta Sinha},
  title     = {Fluorinated-PDMS enhanced single-electrode triboelectric nanogenerator for biomechanical energy harvesting and sensing applications},
  journal   = {APL Electronic Devices},
  year      = {2026},
  volume    = {2},
  number    = {1},
  pages     = {016121},
  doi       = {10.1063/5.0320510},
  issn      = {2995-8423}
}

@article{moon2026enhanced,
  author    = {Moon, Chae-Ju and Choi, Geon-Ju and Sohn, Sang-Hyun and Kim, Se-Jin and Park, Il-Kyu},
  title     = {Enhanced Performance of Triboelectric Nanogenerator Based on TiO2 Nanoparticle/Nylon 66 Composite Nanofibers},
  journal   = {International Journal of Precision Engineering and Manufacturing-Green Technology},
  year      = {2026},
  volume    = {13},
  number    = {1},
  pages     = {181-193},
  doi       = {10.1007/s40684-025-00736-x},
  issn      = {2198-0810}
}

@article{mousavi2023event,
  author    = {Mousavi, Mohammad and Alzgool, Mohammad and Davaji, Benyamin and Towfighian, Shahrzad},
  title     = {Event-driven MEMS vibration sensor: Integration of triboelectric nanogenerator and low-frequency switch},
  journal   = {Mechanical Systems and Signal Processing},
  year      = {2023},
  volume    = {187},
  publisher = {Academic Press},
  doi       = {10.1016/j.ymssp.2022.109921},
  issn      = {10961216}
}

@article{mousavi2024high,
  author    = {Mousavi, Mohammad and Alzgool, Mohammad and Davaji, Benyamin and Towfighian, Shahrzad},
  title     = {High Signal-to-Noise Ratio Event-Driven MEMS Motion Sensing},
  journal   = {Small},
  year      = {2024},
  volume    = {20},
  number    = {10},
  publisher = {John Wiley and Sons Inc},
  doi       = {10.1002/smll.202304591},
  issn      = {16136829}
}

@article{ning2021flexible,
  author    = {Ning, Chuan and Dong, Kai and Cheng, Renwei and Yi, Jia and Ye, Cuiying and Peng, Xiao and Sheng, Feifan and Jiang, Yang and Wang, Zhong Lin},
  title     = {Flexible and Stretchable Fiber-Shaped Triboelectric Nanogenerators for Biomechanical Monitoring and Human-Interactive Sensing},
  journal   = {Advanced Functional Materials},
  year      = {2021},
  volume    = {31},
  number    = {4},
  pages     = {1-9},
  doi       = {10.1002/adfm.202006679},
  issn      = {16163028}
}

@article{ning2022helical,
  author    = {Ning, Chuan and Cheng, Renwei and Jiang, Yang and Sheng, Feifan and Yi, Jia and Shen, Shen and Zhang, Yihan and Peng, Xiao and Dong, Kai and Wang, Zhong Lin},
  title     = {Helical Fiber Strain Sensors Based on Triboelectric Nanogenerators for Self-Powered Human Respiratory Monitoring},
  journal   = {ACS Nano},
  year      = {2022},
  volume    = {16},
  number    = {2},
  pages     = {2811-2821},
  publisher = {American Chemical Society},
  doi       = {10.1021/acsnano.1c09792},
  issn      = {1936086X}
}

@article{panda2026triboelectric,
  author    = {Panda, Swati and Hajra, Sugato and Bhosale, Premkumar Sharad and Belal, Mohammed Ahmed and Kaja, Kushal Ruthvik and Kim, Hoe Joon and Lee, Kyoungtae},
  title     = {Triboelectric energy harvesting from rice paper/PDMS under extreme humidity environments},
  journal   = {Materials Letters},
  year      = {2026},
  volume    = {406},
  pages     = {139971},
  doi       = {10.1016/j.matlet.2025.139971},
  url       = {https://www.sciencedirect.com/science/article/pii/S0167577X25020014},
  issn      = {0167-577X}
}

@article{pandey2024recent,
  author    = {Pandey, Puran and Maharjan, Pukar and Seo, Min-Kyu and Thapa, Keshav and Sohn, Jung Inn},
  title     = {Recent Progress in Wearable Triboelectric Nanogenerator for Advanced Health Monitoring and Rehabilitation},
  journal   = {International Journal of Energy Research},
  year      = {2024},
  volume    = {2024},
  number    = {1},
  doi       = {10.1155/2024/5572736},
  issn      = {0363-907X}
}

@article{potu2023enhancing,
  author    = {Potu, Supraja and Navaneeth, M and Bhadoriya, Anshika and Bora, Arbacheena and Sivalingam, Yuvaraj and Babu, Anjaly and Velpula, Mahesh and Gollapelli, Buchaiah and Rajaboina, Rakesh Kumar and Khanapuram, Uday Kumar},
  title     = {Enhancing Triboelectric Nanogenerator Performance with Metal--Organic-Framework-Modified ZnO Nanosheets for Self-Powered Electronic Devices and Energy Harvesting},
  journal   = {ACS Applied Nano Materials},
  year      = {2023},
  volume    = {6},
  number    = {24},
  pages     = {22701-22710},
  publisher = {ACS Publications},
  issn      = {2574-0970}
}

@article{pu2017p,
  author    = {Pu, Xianjie and Guo, Hengyu and Chen, Jie and Wang, Xue and Xi, Yi and Hu, Chenguo and Zhong, T and Wang, Lin},
  title     = {Eye motion triggered self-powered mechnosensational communication system using triboelectric nanogenerator},
  year      = {2017},
  url       = {https://www.science.org}
}

@article{pu2023triboelectric,
  author    = {Pu, Xiong and Zhang, Chi and Wang, Zhong Lin},
  title     = {Triboelectric nanogenerators as wearable power sources and self-powered sensors},
  journal   = {National Science Review},
  year      = {2023},
  volume    = {10},
  number    = {1},
  doi       = {10.1093/nsr/nwac170},
  issn      = {2095-5138}
}

@article{rajasekar2025recent,
  author    = {Rajasekar, Mani and Sivakumar, Meenambigai},
  title     = {Recent trends and future perspectives of triboelectric nanogenerators (TENGs) and applications},
  journal   = {RSC Adv.},
  year      = {2025},
  volume    = {15},
  number    = {57},
  pages     = {49600-49642},
  publisher = {The Royal Society of Chemistry},
  doi       = {10.1039/D5RA06645J}
}

@article{ramaraj2024biocompatible,
  author    = {Ramaraj, Sankar Ganesh and Elamaran, Durgadevi and Tabata, Hitoshi and Zhang, Fuchun and Liu, Xinghui},
  title     = {Biocompatible triboelectric energy generators (BT-TENGs) for energy harvesting and healthcare applications},
  journal   = {Nanoscale},
  year      = {2024},
  volume    = {16},
  number    = {39},
  pages     = {18251-18273},
  doi       = {10.1039/D4NR01987C},
  issn      = {2040-3364}
}

@proceedings{ref20192019,
  title     = {2019 IEEE Sensors Applications Symposium},
  year      = {2019},
  publisher = {IEEE}
}

@article{shvarts2018trapped,
  author  = {Shvarts, A. G. and Yastrebov, V. A.},
  title   = {Trapped fluid in contact interface},
  journal = {Journal of the Mechanics and Physics of Solids},
  volume  = {119},
  pages   = {140--162},
  year    = {2018},
  doi     = {10.1016/j.jmps.2018.06.016}
}

@article{shvarts2021computational,
  author  = {Shvarts, Andrei G. and Vignollet, Julien and
             Yastrebov, Vladislav A.},
  title   = {Computational framework for monolithic coupling
             for thin fluid flow in contact interfaces},
  journal = {Computer Methods in Applied Mechanics and Engineering},
  volume  = {379},
  pages   = {113738},
  year    = {2021},
  doi     = {10.1016/j.cma.2021.113738}
}

@article{alavi2026numerical,
  author  = {Alavi, P. and Anciaux, G. and Rocchi, L. and
             Richard, J. and Molinari, J. F.},
  title   = {Numerical modeling of rough contact interfaces
             with trapped compressive liquid pockets},
  journal = {Tribology International},
  volume  = {214},
  pages   = {111142},
  year    = {2026},
  doi     = {10.1016/j.triboint.2025.111142}
}

@article{ren2026edge,
  author    = {Ren, Shan and Wang, Wenpeng and Cai, Tingting and Zhang, Zimu and Li, Xiankai and Wang, Nannan and Wang, Daoai},
  title     = {Edge-effect based liquid--metal solid--liquid triboelectric nanogenerator with enhanced electrostatic induction},
  journal   = {Applied Surface Science},
  year      = {2026},
  volume    = {738},
  pages     = {166952},
  doi       = {10.1016/j.apsusc.2026.166952},
  url       = {https://www.sciencedirect.com/science/article/pii/S0169433226011566},
  issn      = {0169-4332}
}

@article{shao2020theoretical,
  author    = {Shao, Jiajia and Willatzen, Morten and Wang, Zhong Lin},
  title     = {Theoretical modeling of triboelectric nanogenerators (TENGs)},
  journal   = {Journal of Applied Physics},
  year      = {2020},
  volume    = {128},
  number    = {11},
  pages     = {111101},
  doi       = {10.1063/5.0020961},
  issn      = {0021-8979}
}

@article{sheng2022ultrastretchable,
  author    = {Sheng, Feifan and Zhang, Bo and Zhang, Yihan and Li, Yanyan and Cheng, Renwei and Wei, Chuanhui and Ning, Chuan and Dong, Kai and Wang, Zhong Lin},
  title     = {Ultrastretchable Organogel/Silicone Fiber-Helical Sensors for Self-Powered Implantable Ligament Strain Monitoring},
  journal   = {ACS Nano},
  year      = {2022},
  volume    = {16},
  number    = {7},
  pages     = {10958-10967},
  publisher = {American Chemical Society},
  doi       = {10.1021/acsnano.2c03365},
  issn      = {1936086X}
}

@article{shivanna2026pbat,
  author    = {Shivanna, Kariyappa Gowda Guddenahalli and Thammannagowda, Vishnu Kadabahalli and Shankaregowda, Smitha Ankanahalli and Kalappa, Prashantha},
  title     = {PBAT-based biodegradable triboelectric nanogenerators for sustainability in energy, environment, and health},
  journal   = {Hybrid Advances},
  year      = {2026},
  volume    = {12},
  pages     = {100602},
  doi       = {10.1016/j.hybadv.2026.100602},
  url       = {https://www.sciencedirect.com/science/article/pii/S2773207X26000047},
  issn      = {2773-207X}
}

@article{so2023flexible,
  author    = {So, Mei Yi and Xu, Bingang and Li, Zihua and Lai, Cheuk Lam and Jiang, Chenghanzhi},
  title     = {Flexible corrugated triboelectric nanogenerators for efficient biomechanical energy harvesting and human motion monitoring},
  journal   = {Nano Energy},
  year      = {2023},
  volume    = {106},
  publisher = {Elsevier Ltd},
  doi       = {10.1016/j.nanoen.2022.108033},
  issn      = {22112855}
}

@article{tran2025high,
  author    = {Tran, Gia Huy and Nguyen, Quang Binh and Pham, Anh Quoc and Bui, Van-Tan and Do, Xuan Huy and Bui, Van-Tien},
  title     = {High-performance no-gap triboelectric nanogenerator derived from hierarchical polyimide-copper mesh hybrid for advanced force sensing and energy harvesting},
  journal   = {Chemical Engineering Journal},
  year      = {2025},
  volume    = {519},
  pages     = {164806},
  doi       = {10.1016/j.cej.2025.164806},
  url       = {https://www.sciencedirect.com/science/article/pii/S1385894725056426},
  issn      = {1385-8947}
}

@article{troncoso2023lignocellulosic,
  author    = {Troncoso, Omar P and Corman-Hijar, Jim I and Torres, Fernando G},
  title     = {Lignocellulosic Biomass for the Fabrication of Triboelectric Nano-Generators (TENGs)---A Review},
  journal   = {International Journal of Molecular Sciences},
  year      = {2023},
  volume    = {24},
  number    = {21},
  doi       = {10.3390/ijms242115784},
  url       = {https://www.mdpi.com/1422-0067/24/21/15784},
  issn      = {1422-0067}
}

@article{tushar2023triboelectric,
  title={Triboelectric nanogenerator assisted synthesis and detection of chemical compounds},
  author={Tushar, Shariful Islam and Sayam, Abdullah and Uddin, Md Mazbah and Dip, Tanvir Mahady and Anik, Habibur Rahman and Aktar Arin, Md Reasat and Sharma, Suraj},
  journal={Journal of Materials Chemistry A},
  volume={11},
  number={36},
  pages={19244--19280},
  year={2023},
  publisher={The Royal Society of Chemistry}
}

@misc{uddin2025triboelectric,
  author    = {Uddin, Md Mazbah and Dip, Tanvir Mahady and Tushar, Shariful Islam and Sayam, Abdullah and Anik, Habibur Rahman and Aktar Arin, Md Reasat and Talukder, Amit and Sharma, Suraj},
  title     = {Triboelectric Nanogenerators for Self-Powered Degradation of Chemical Pollutants},
  howpublished= {ACS Omega},
  year      = {2025},
  volume    = {10},
  number    = {1},
  pages     = {26-54},
  publisher = {American Chemical Society},
  doi       = {10.1021/acsomega.4c07889},
  issn      = {24701343}
}

@article{udofa2025recent,
  author    = {Udofa, Emaediong Sylvanus and Mohamed Ali, Mohamed Sultan and Leow, Pei Ling},
  title     = {Recent development in triboelectric nanogenerators: a review},
  journal   = {Journal of Materials Science: Materials in Electronics},
  year      = {2025},
  volume    = {37},
  number    = {1},
  pages     = {6},
  doi       = {10.1007/s10854-025-16223-7},
  issn      = {1573-482X}
}

@article{viswanathan2025stretch,
  author    = {Viswanathan, Pramila and Chandrasekhar, Arunkumar},
  title     = {Stretch-responsive triboelectric nanogenerator based on conductive PVA/AC hydrogel for human activity monitoring},
  journal   = {Results in Engineering},
  year      = {2025},
  volume    = {28},
  pages     = {107734},
  doi       = {10.1016/j.rineng.2025.107734},
  url       = {https://www.sciencedirect.com/science/article/pii/S2590123025037879},
  issn      = {2590-1230}
}

@article{wang2023multifunctional,
  author    = {Wang, Yu and Zhang, Xiaoxuan and Chen, Guopu and Lu, Minhui and Zhao, Yuanjin},
  title     = {Multifunctional structural color triboelectric microneedle patches for psoriasis treatment},
  journal   = {Matter},
  year      = {2023},
  volume    = {6},
  number    = {5},
  pages     = {1555-1568},
  publisher = {Cell Press},
  doi       = {10.1016/j.matt.2023.03.005},
  issn      = {25902385}
}

@article{wang2025enhancing,
  author    = {Wang, Feijie and Wang, Suyang and Wang, Hao and Liu, Yichi and Ouyang, Shiqiang and Zhang, Shenzhuo and Hu, Yueming and Zhao, Junhua and Ma, Shufeng and Wu, Zhen and Wang, Liqiang},
  title     = {Enhancing the output of cellulose-based triboelectric nanogenerators via dual interfacial polarization effects},
  journal   = {Carbohydrate Polymers},
  year      = {2025},
  volume    = {368},
  pages     = {124093},
  doi       = {10.1016/j.carbpol.2025.124093},
  url       = {https://www.sciencedirect.com/science/article/pii/S0144861725008781},
  issn      = {0144-8617}
}

@article{wang2025flexible,
  author    = {Wang, Keping and Cui, Juan and Li, Xin and Liu, Huizi and Li, Yake and Hao, Jiachao and Zheng, Yongqiu and Xue, Chenyang},
  title     = {A flexible MWNT/PVDF electrospun membrane-based triboelectric nanogenerator with a spatially ordered hollow array for musculoskeletal disorder monitoring},
  journal   = {Chemical Engineering Journal},
  year      = {2025},
  volume    = {523},
  pages     = {168847},
  doi       = {10.1016/j.cej.2025.168847},
  url       = {https://www.sciencedirect.com/science/article/pii/S1385894725096895},
  issn      = {1385-8947}
}

@article{wang2026electric,
  author    = {Wang, Yunfeng and Mu, Haibao and Yao, Huanmin and He, Shasha and Wang, Shuai and Yang, Yiyun and Chen, Yaxin and Chen, Xinran and Zhang, Guanjun},
  title     = {Electric field-assisted construction of bionic directional channels enhances charge storage in MXene/Ecoflex triboelectric nanogenerators for self-powered sensing},
  journal   = {Nano Energy},
  year      = {2026},
  volume    = {150},
  pages     = {111750},
  doi       = {10.1016/j.nanoen.2026.111750},
  url       = {https://www.sciencedirect.com/science/article/pii/S2211285526000546},
  issn      = {2211-2855}
}

@article{wang2026flexible,
  author    = {Wang, Hongyu and Li, Yuanzheng and Liu, Jianhua and Wan, Xingfu and Li, Yuchen and Xu, Gengchen and Yuan, Haichao and Wang, Siyuan and Xu, Peng and Xu, Minyi},
  title     = {A flexible wearable single-electrode triboelectric sensor based on self-healing ionic elastomer for reliable underwater motion sensing},
  journal   = {Chemical Engineering Journal},
  year      = {2026},
  volume    = {537},
  pages     = {176192},
  doi       = {10.1016/j.cej.2026.176192},
  url       = {https://www.sciencedirect.com/science/article/pii/S1385894726036533},
  issn      = {1385-8947}
}

@article{wei2026ambient,
  author    = {Wei, Yi and Tian, Yushu and Chen, Xiaokang and Chu, Junfeng and Wang, Junjie and Chen, Xiangyu and Wu, Wenjie and Zhang, Liqun},
  title     = {Ambient Stable Triboelectric Nanogenerator Based on Conductive Filler Modified Silicone Rubber with Gas Barrier Encapsulation for Footstep Energy Conversion},
  journal   = {Advanced Science},
  year      = {2026},
  volume    = {13},
  number    = {10},
  pages     = {e19523},
  publisher = {John Wiley \& Sons, Ltd},
  doi       = {10.1002/advs.202519523},
  issn      = {2198-3844}
}

@article{xu2022dynamics,
  author    = {Xu, Guoqiang and Li, Chuanyang and Chen, Chaojie and Fu, Jingjing and Hou, Tingting and Zi, Yunlong},
  title     = {Dynamics of triboelectric nanogenerators: A review},
  journal   = {International Journal of Mechanical System Dynamics},
  year      = {2022},
  volume    = {2},
  number    = {4},
  pages     = {311-324},
  publisher = {John Wiley \& Sons, Ltd},
  doi       = {10.1002/msd2.12058},
  issn      = {2767-1399}
}

@article{xu2023multi,
  author    = {Xu, Shuai and Nie, Wenqi and Sun, Jiangdong and Sun, Pengfei and Jia, Hao and Zheng, Xianhong and Sun, Yanyan and Xu, Zhenzhen and Liu, Lianmei},
  title     = {Multi-mode and durable fiber triboelectric nanogenerator for power and sensor enabled by Hookean vascular stent structure},
  journal   = {Chemical Engineering Journal},
  year      = {2023},
  volume    = {472},
  publisher = {Elsevier B.V.},
  doi       = {10.1016/j.cej.2023.145088},
  issn      = {13858947}
}

@article{xue2025triboelectric,
  author    = {Xue, Na and Wu, Junpeng and Liu, Yujia and Zheng, Yang and Ni, Yingjuan and Liu, Lu and Liu, Tengfei and Qu, Dingrong and Xu, Xiaofeng and Cui, Hongzhi and Li, Xiaoyi},
  title     = {Triboelectric nanogenerator based on degradable materials for efficient self-powered seed growth promotion system},
  journal   = {Nano Research},
  year      = {2025},
  volume    = {18},
  number    = {8},
  pages     = {94907583},
  doi       = {10.26599/NR.2025.94907583},
  url       = {https://www.sciopen.com/article/10.26599/NR.2025.94907583}
}

@article{yamklang2026agricultural,
  author    = {Yamklang, Wimonsiri and Luechar, Pawonpart and Kaeochana, Walailak and Bunriw, Weeraya and Chimsida, Praphadsorn and Sintusiri, Jirapan and Jaisutti, Rawat and Harnchana, Viyada},
  title     = {Agricultural-waste-derived cellulose paper modified with PVDF/Ag nanowire composite for triboelectric nanogenerator and multifunctional sensing applications},
  journal   = {Materials Science and Engineering: B},
  year      = {2026},
  volume    = {325},
  pages     = {119082},
  doi       = {10.1016/j.mseb.2025.119082},
  url       = {https://www.sciencedirect.com/science/article/pii/S0921510725011067},
  issn      = {0921-5107}
}

@article{yang2026mechanisms,
  author    = {Yang, Zhe and Wu, Ning and Chen, Muqi and Yu, Zeyang and Liu, Jianming and Zhao, Juanli and Jiang, Tao and Pang, Yaokun and Xiong, Zhihua and Willatzen, Morten and Luo, Jianjun and Wang, Zhong Lin},
  title     = {Mechanisms of Polarity-Driven and Material-Dependent Charge Transfer at PVDF/Polymer Interfaces for High-Performance Triboelectric Nanogenerators},
  journal   = {Carbon Neutralization},
  year      = {2026},
  volume    = {5},
  number    = {3},
  pages     = {e70152},
  publisher = {John Wiley \& Sons, Ltd},
  doi       = {10.1002/cnl2.70152},
  issn      = {2769-3333}
}

@article{ye2025noncontact,
  author    = {Ye, Jiahao and Cinque, Gianfelice and Donà, Lorenzo and Tan, Jin-Chong},
  title     = {Noncontact triboelectric nanogenerators based on fluorinated metal--organic frameworks for rotational energy harvesting and sensing},
  journal   = {APL Electronic Devices},
  year      = {2025},
  volume    = {1},
  number    = {3},
  pages     = {036109},
  doi       = {10.1063/5.0273138},
  issn      = {2995-8423}
}

@article{jorgensen1996opls,
  author  = {Jorgensen, William L. and Maxwell, David S. and Tirado-Rives, Julian},
  title   = {Development and Testing of the OPLS All-Atom Force Field
             on Conformational Energetics and Properties of Organic Liquids},
  journal = {Journal of the American Chemical Society},
  year    = {1996},
  volume  = {118},
  number  = {45},
  pages   = {11225--11236},
  doi     = {10.1021/ja9621760}
}

@article{cornell1995forcefield,
  author  = {Cornell, Wendy D. and Cieplak, Piotr and Bayly, Christopher I.
             and Gould, Ian R. and Merz, Kenneth M. and Ferguson, David M.
             and Spellmeyer, David C. and Fox, Thomas and Caldwell, James W.
             and Kollman, Peter A.},
  title   = {A Second Generation Force Field for the Simulation of Proteins,
             Nucleic Acids, and Organic Molecules},
  journal = {Journal of the American Chemical Society},
  year    = {1995},
  volume  = {117},
  number  = {19},
  pages   = {5179--5197},
  doi     = {10.1021/ja00124a002}
}

@article{verners2022smooth,
  author  = {Verners, Osvalds and Lap{\v{c}}inskis, Linards and {\c{G}}ermane, L{\={i}}va and Kasikov, Aarne and Timusk, Martin and Pudzs, Kaspars and Ellis, Amanda V. and Sherrell, Peter C. and {\v{S}}utka, Andris},
  title   = {Smooth polymers charge negatively: Controlling contact electrification polarity in polymers},
  journal = {Nano Energy},
  year    = {2022},
  volume  = {104},
  pages   = {107914},
  doi     = {10.1016/j.nanoen.2022.107914}
}

@article{xu2019slips,
  author  = {Xu, Wanghuai and Zhou, Xiaofeng and Hao, Chonglei and Zheng, Huanxi and Liu, Yuan and Yan, Xiantong and Yang, Zhengbao and Leung, Michael and Zeng, Xiao Cheng and Xu, Ronald X. and Wang, Zuankai},
  title   = {{SLIPS-TENG}: Robust triboelectric nanogenerator with optical and charge transparency using a slippery interface},
  journal = {National Science Review},
  year    = {2019},
  volume  = {6},
  number  = {3},
  pages   = {540--550},
  doi     = {10.1093/nsr/nwz025}
}

@article{meng2024droplet,
  author  = {Meng, Huan and Zhang, Jingjing and Zhu, Rui and Wang, Jingjing and Ge, Ying and Liu, Huimin and Feng, Can and Zhou, Zunkang and Meng, Yao and Huang, Zanying and Yang, Ke and Jia, Yu and Du, Zuliang and Cui, Peng and Cheng, Gang},
  title   = {Elevating Outputs of Droplet Triboelectric Nanogenerator through Strategic Surface Molecular Engineering},
  journal = {ACS Energy Letters},
  year    = {2024},
  volume  = {9},
  number  = {6},
  pages   = {2670--2676},
  doi     = {10.1021/acsenergylett.4c00532}
}

@article{wu2023liquidsolid,
  author  = {Wu, Jun and Cao, Jiapeng and Bi, Hailin and Zhang, Jun and Cao, Qing},
  title   = {Liquid-solid contact electrification and its effect on the formation of electric double layer: An atomic-level investigation},
  journal = {Nano Energy},
  year    = {2023},
  volume  = {111},
  pages   = {108442},
  doi     = {10.1016/j.nanoen.2023.108442}
}

@article{verners2023mechanisms,
  author  = {Verners, Osvalds and Das, Amit},
  title   = {Comparison of Contact Electrification Mechanisms of Selected Polymers and Surface-Functionalized Molecules},
  journal = {The Journal of Physical Chemistry B},
  year    = {2023},
  volume  = {127},
  number  = {46},
  pages   = {10035--10042},
  doi     = {10.1021/acs.jpcb.3c04817}
}

@article{yoon2025creating,
  author    = {Yoon, Jae Uk and Woo, Insun and Gajula, Prasad and Bae, Jin Woo},
  title     = {Creating Smart Washable Flooring: Sandwich-Style Single-Electrode Triboelectric Nanogenerator with Barium Titanate and Graphite-Fluorinated Polymer-Infused Ecoflex Hybrid Composites for Enhanced Safety and Security},
  journal   = {Advanced Functional Materials},
  year      = {2025},
  volume    = {35},
  number    = {24},
  pages     = {2421977},
  publisher = {John Wiley \& Sons, Ltd},
  doi       = {10.1002/adfm.202421977},
  issn      = {1616-301X}
}

@article{zainalabidin2025current,
  author    = {Zainal Abidin, Ahmad Razin and Zambry, Nor Syafirah and Ibrahim, Fatimah and Nor Rashid, Nurshamimi and Jamaluddin, Nurul Fauzani and Wan Kamarul Zaman, Wan Safwani},
  title     = {Current advances in biocompatibility assessment for MEMS in biomedical applications},
  journal   = {Microfluidics and Nanofluidics},
  year      = {2025},
  volume    = {29},
  number    = {6},
  pages     = {32},
  doi       = {10.1007/s10404-025-02806-7},
  issn      = {1613-4982}
}

@article{zhang2020material,
  author    = {Zhang, Renyun and Olin, Håkan},
  title     = {Material choices for triboelectric nanogenerators: A critical review},
  journal   = {EcoMat},
  year      = {2020},
  volume    = {2},
  number    = {4},
  pages     = {e12062},
  publisher = {John Wiley \& Sons, Ltd},
  doi       = {10.1002/eom2.12062}
}

@article{zhang2020theories,
  author    = {Zhang, He and Yao, Linjie and Quan, Liwei and Zheng, Xianglong},
  title     = {Theories for triboelectric nanogenerators: A comprehensive review},
  journal   = {Nanotechnology Reviews},
  year      = {2020},
  volume    = {9},
  number    = {1},
  pages     = {610-625},
  doi       = {10.1515/ntrev-2020-0049}
}

@article{zhang2021sponge,
  author    = {Zhang, He and Su, Xinjie and Quan, Liwei and Jiang, Jiqing and Dong, Beibei and Wei, Gang},
  title     = {Sponge-Supported Triboelectric Nanogenerator for Energy Harvesting from Rail Vibration},
  journal   = {Journal of Energy Engineering},
  year      = {2021},
  volume    = {147},
  number    = {3},
  doi       = {10.1061/(ASCE)EY.1943-7897.0000751},
  issn      = {0733-9402}
}

@article{zhang20233d,
  author    = {Zhang, Shipeng and Rana, SM Sohel and Bhatta, Trilochan and Pradhan, Gagan Bahadur and Sharma, Sudeep and Song, Hyesu and Jeong, Seonghoon and Park, Jae Yeong},
  title     = {3D printed smart glove with pyramidal MXene/Ecoflex composite-based toroidal triboelectric nanogenerators for wearable human-machine interaction applications},
  journal   = {Nano Energy},
  year      = {2023},
  volume    = {106},
  publisher = {Elsevier Ltd},
  doi       = {10.1016/j.nanoen.2022.108110},
  issn      = {22112855}
}

@article{zhang2025flexible,
  author    = {Zhang, Hao and Zhang, Dongzhi and Mao, Ruiyuan and Zhu, Liang and Zhang, Chengyu and Zhou, Lina and Yang, Chunqing and Ji, Yuncheng and Mao, Yuling},
  title     = {A flexible triboelectric nanogenerator based on Ecoflex/BaTiO3/carbon spheres for active and passive tactile intelligent sensing systems},
  journal   = {Chemical Engineering Journal},
  year      = {2025},
  volume    = {506},
  pages     = {159840},
  doi       = {10.1016/j.cej.2025.159840},
  url       = {https://www.sciencedirect.com/science/article/pii/S1385894725006394},
  issn      = {1385-8947}
}

@article{zhao2025fluorinated,
  author    = {Zhao, Dandan and Li, Long and Cai, Zhuanghong and Gao, Kexin and Liu, Weibing and Su, Lei and Nie, Shixuan and Yu, Xiutong and Sun, Xiaoyu and Shao, Zhichao},
  title     = {A fluorinated-coordination polymer eutectic crystal as high output triboelectric nanogenerator material for self-powered system},
  journal   = {Journal of Molecular Structure},
  year      = {2025},
  volume    = {1330},
  pages     = {141571},
  doi       = {10.1016/j.molstruc.2025.141571},
  url       = {https://www.sciencedirect.com/science/article/pii/S0022286025002601},
  issn      = {0022-2860}
}

@article{zhao2026fluorinated,
  author    = {Zhao, He and Xu, Chu and Zhou, Caobin and Zhang, Ling and Li, Chunzhong},
  title     = {Fluorinated bi-layer E-textile enabled by electrospinning rotary collection for self-powered dual-functional electrochemical sweat sensing and motion detection},
  journal   = {Chemical Engineering Journal},
  year      = {2026},
  volume    = {530},
  pages     = {173580},
  doi       = {10.1016/j.cej.2026.173580},
  url       = {https://www.sciencedirect.com/science/article/pii/S1385894726010399},
  issn      = {1385-8947}
}

@article{zhu2024biopolymer,
  author    = {Zhu, Qiliang and Sun, Enqi and Zhao, Zequan and Wu, Tong and Meng, Shuchang and Ma, Zimeng and Shoaib, Muhammad and Ur Rehman, Hafeez and Cao, Xia and Wang, Ning},
  title     = {Biopolymer Materials in Triboelectric Nanogenerators: A Review},
  journal   = {Polymers},
  year      = {2024},
  volume    = {16},
  number    = {10},
  doi       = {10.3390/polym16101304},
  url       = {https://www.mdpi.com/2073-4360/16/10/1304},
  issn      = {2073-4360}
}

@article{zou2019bionic,
  author    = {Zou, Yang and Tan, Puchuan and Shi, Bojing and Ouyang, Han and Jiang, Dongjie and Liu, Zhuo and Li, Hu and Yu, Min and Wang, Chan and Qu, Xuecheng and Zhao, Luming and Fan, Yubo and Wang, Zhong Lin and Li, Zhou},
  title     = {A bionic stretchable nanogenerator for underwater sensing and energy harvesting},
  journal   = {Nature Communications},
  year      = {2019},
  volume    = {10},
  number    = {1},
  publisher = {Nature Publishing Group},
  doi       = {10.1038/s41467-019-10433-4},
  issn      = {20411723}
}

@article{niu2013theoretical,
  title={Theoretical study of contact-mode triboelectric nanogenerators as an effective power source},
  author={Niu, Simiao and Wang, Sihong and Lin, Long and Liu, Ying and Zhou, Yu Sheng and Hu, Youfan and Wang, Zhong Lin},
  journal={Energy \& Environmental Science},
  volume={6},
  number={12},
  pages={3576--3583},
  year={2013},
  publisher={Royal Society of Chemistry}
}

@article{dharmasena2017triboelectric,
  title={Triboelectric nanogenerators: providing a fundamental framework},
  author={Dharmasena, Randunu Devage Ishara Gihan and Jayawardena, KDGI and Mills, CA and Deane, JHB and Anguita, JV and Dorey, RA and Silva, SRP},
  journal={Energy \& Environmental Science},
  volume={10},
  number={8},
  pages={1801--1811},
  year={2017},
  publisher={Royal Society of Chemistry}
}

@article{dharmasena2018unified,
  title={A unified theoretical model for Triboelectric Nanogenerators},
  author={Dharmasena, RDIG and Jayawardena, KDGI and Mills, CA and Dorey, RA and Silva, SRP},
  journal={Nano Energy},
  volume={48},
  pages={391--400},
  year={2018},
  publisher={Elsevier}
}

@article{dharmasena2018nature,
  title={Nature of power generation and output optimization criteria for triboelectric nanogenerators},
  author={Dharmasena, RD Ishara G and Deane, Jonathan HB and Silva, S Ravi P},
  journal={Advanced Energy Materials},
  volume={8},
  number={31},
  pages={1802190},
  year={2018},
  publisher={Wiley Online Library}
}

@article{XUVoc,
title = {A unified contact force-dependent model for triboelectric nanogenerators accounting for surface roughness},
journal = {Nano Energy},
volume = {76},
pages = {105067},
year = {2020},
author = {Yang Xu and Guanbo Min and Nikolaj Gadegaard and Ravinder Dahiya and Daniel M. Mulvihill}
}

@article{wang2013sliding,
  title={Sliding-triboelectric nanogenerators based on in-plane charge-separation mechanism},
  author={Wang, Sihong and Lin, Long and Xie, Yannan and Jing, Qingshen and Niu, Simiao and Wang, Zhong Lin},
  journal={Nano letters},
  volume={13},
  number={5},
  pages={2226--2233},
  year={2013},
  publisher={ACS Publications}
}

@article{bai2013cylindrical,
  title={Cylindrical rotating triboelectric nanogenerator},
  author={Bai, Peng and Zhu, Guang and Liu, Ying and Chen, Jun and Jing, Qingshen and Yang, Weiqing and Ma, Jusheng and Zhang, Gong and Wang, Zhong Lin},
  journal={ACS nano},
  volume={7},
  number={7},
  pages={6361--6366},
  year={2013},
  publisher={ACS Publications}
}

@article{zhang2020non,
  title={Non-contact cylindrical rotating triboelectric nanogenerator for harvesting kinetic energy from hydraulics},
  author={Zhang, Nan and Qin, Cheng and Feng, Tianxing and Li, Jun and Yang, Zhirui and Sun, Xiupeng and Liang, Erjun and Mao, Yanchao and Wang, Xudong},
  journal={Nano Research},
  volume={13},
  number={7},
  pages={1903--1907},
  year={2020},
  publisher={Springer}
}

@article{yan2022design,
  title={Design of a soft-contact triboelectric nanogenerator for vibrational energy collection and its output performance},
  author={Yan, Jin and Mei, Naerduo and Zhang, Dapeng and Zhong, Yinghao},
  journal={Frontiers in Energy Research},
  volume={10},
  pages={1014983},
  year={2022},
  publisher={Frontiers Media SA}
}

@article{shang2023high,
  title={High storable power density of triboelectric nanogenerator within centimeter size},
  author={Shang, Yurui and Li, Chengyu and Yu, Gao and Yang, Yuhan and Zhao, Wenting and Tang, Wei},
  journal={Materials},
  volume={16},
  number={13},
  pages={4669},
  year={2023},
  publisher={MDPI}
}

@article{qin2023ball,
  title={Ball-mill-inspired durable triboelectric nanogenerator for wind energy collecting and speed monitoring},
  author={Qin, Qinghao and Cao, Xia and Wang, Ning},
  journal={Nanomaterials},
  volume={13},
  number={5},
  pages={939},
  year={2023},
  publisher={Multidisciplinary Digital Publishing Institute}
}

@article{zhang2024simulation,
  title={A simulation study on the contact-separation triboelectric nano-generator for magnetic energy harvester},
  author={Zhang, Yibing and Li, Fenghui and Wang, Yagang and Tong, Shanlin and Li, Yuhao and Li, Xuhui and Lu, Wu and Liu, Yongsheng},
  journal={IET Nanodielectrics},
  volume={7},
  number={4},
  pages={189--202},
  year={2024},
  publisher={Wiley Online Library}
}

@article{venugopal2022effective,
  title={Effective modeling and numerical simulation of triboelectric nanogenerator for blood pressure measurement based on wrist pulse signal using comsol multiphysics software},
  author={Venugopal, Karthikeyan and Shanmugasundaram, Vivekanandan},
  journal={ACS omega},
  volume={7},
  number={30},
  pages={26863--26870},
  year={2022},
  publisher={ACS Publications}
}

@article{shuvo2024finite,
  title={Finite element modeling of MXene/PVDF-based high-performance triboelectric nanogenerators for self-powered wearable electronics},
  author={Shuvo, Arefin Ahamed and Bhuiyan, Ashraful Ghani and Islam, Md Sherajul},
  journal={ACS Applied Electronic Materials},
  volume={6},
  number={6},
  pages={4478--4488},
  year={2024},
  publisher={ACS Publications}
}

@article{liu2024biodegradable,
  title={Biodegradable flexible triboelectric nanogenerator for winter sports monitoring},
  author={Liu, Bocong and Xie, Zhenning and Feng, Qingyang and Wang, Yunlu and Zhang, Mengqi and Lu, Zhuo and Mao, Yupeng and Zhang, Shouwei},
  journal={Frontiers in Materials},
  volume={11},
  pages={1400258},
  year={2024},
  publisher={Frontiers Media SA}
}

@article{niu2014theoretical,
  title={Theoretical investigation and structural optimization of single-electrode triboelectric nanogenerators},
  author={Niu, Simiao and Liu, Ying and Wang, Sihong and Lin, Long and Zhou, Yu Sheng and Hu, Youfan and Wang, Zhong Lin},
  journal={Advanced Functional Materials},
  volume={24},
  number={22},
  pages={3332--3340},
  year={2014},
  publisher={Wiley Online Library}
}

@article{niu2013theory,
  title={Theory of sliding-mode triboelectric nanogenerators},
  author={Niu, Simiao and Liu, Ying and Wang, Sihong and Lin, Long and Zhou, Yu Sheng and Hu, Youfan and Wang, Zhong Lin},
  journal={Advanced materials},
  volume={25},
  number={43},
  pages={6184--6193},
  year={2013},
  publisher={Wiley Online Library}
}

@article{xie2014grating,
  title={Grating-structured freestanding triboelectric-layer nanogenerator for harvesting mechanical energy at 85\% total conversion efficiency},
  author={Xie, Yannan and Wang, Sihong and Niu, Simiao and Lin, Long and Jing, Qingshen and Yang, Jin and Wu, Zhengyun and Wang, Zhong Lin},
  journal={Advanced materials},
  volume={26},
  number={38},
  pages={6599--6607},
  year={2014},
  publisher={Wiley Online Library}
}

@incollection{gravesen2023quantitative,
  title     = {Quantitative Calculation of Output Power and Spherical Triboelectric Nanogenerators},
  author    = {Gravesen, Jens and Willatzen, Morten and Shao, Jiajia and Wang, Zhong Lin},
  booktitle = {Handbook of Triboelectric Nanogenerators},
  publisher = {Springer},
  address   = {Cham},
  pages     = {203--241},
  year      = {2023},
  doi       = {10.1007/978-3-031-28111-2_6}
}

@article{heo2024ultrahigh,
  title   = {Ultra-High Peak Power Generation for Rotational Triboelectric Nanogenerator via Simple Charge Control and Boosted Discharge Design},
  author  = {Heo, Deokjae and Son, Jin-Ho and Hur, Jiwoong and Yong, Hyungseok and Cha, Kyunghwan and Hwang, Patrick T. J. and Koo, Bonwook and Gwak, Yunki and Jin, Youngho and Kim, Dongseob and Hong, Jinkee and Lee, Sangmin},
  journal = {Advanced Functional Materials},
  volume  = {34},
  number  = {41},
  pages   = {2406032},
  year    = {2024},
  doi     = {10.1002/adfm.202406032}
}

@article{seol2015impact,
  title={Impact of contact pressure on output voltage of triboelectric nanogenerator based on deformation of interfacial structures},
  author={Seol, Myeong-Lok and Lee, Sang-Han and Han, Jin-Woo and Kim, Daewon and Cho, Gyu-Hyeong and Choi, Yang-Kyu},
  journal={Nano Energy},
  volume={17},
  pages={63--71},
  year={2015},
  publisher={Elsevier}
}

@article{jin2016contact,
  title={On the contact behavior of micro-/nano-structured interface used in vertical-contact-mode triboelectric nanogenerators},
  author={Jin, Congrui and Kia, Danial Sharifi and Jones, Matthew and Towfighian, Shahrzad},
  journal={Nano Energy},
  volume={27},
  pages={68--77},
  year={2016},
  publisher={Elsevier}
}

@article{yang2018comprehensive,
  title={Comprehensive contact analysis for vertical-contact-mode triboelectric nanogenerators with micro-/nano-textured surfaces},
  author={Yang, Weixu and Wang, Xiaoli and Li, Hanqing and Wu, Jun and Hu, Yanqiang},
  journal={Nano Energy},
  volume={51},
  pages={241--249},
  year={2018},
  publisher={Elsevier}
}

@article{yang2019fundamental,
  title={Fundamental research on the effective contact area of micro-/nano-textured surface in triboelectric nanogenerator},
  author={Yang, Weixu and Wang, Xiaoli and Li, Hanqing and Wu, Jun and Hu, Yanqiang and Li, Zhihao and Liu, Hui},
  journal={Nano Energy},
  volume={57},
  pages={41--47},
  year={2019},
  publisher={Elsevier}
}

@article{wang2021investigation,
  title={Investigation on the adhesive contact and electrical performance for triboelectric nanogenerator considering polymer viscoelasticity},
  author={Wang, Chenfei and Wang, Xiaoli and Hu, Yanqiang and Li, Lizhou and Li, Zhihao and Wu, Heng and Zhao, Zirui},
  journal={Nano Research},
  volume={14},
  number={12},
  pages={4625--4633},
  year={2021},
  publisher={Springer}
}

@article{dong2022stretchable,
  title   = {3D stretchable and self-encapsulated multimaterial fibers for multifunctional sensors},
  author  = {Dong, Cong and Leber, Alexander and Das Gupta, Tura and Chandran, Ravi and Volpi, Marco and Qu, Yi and Nguyen-Dang, Tung and Bartolomei, N{\'e}stor and Yan, Wen and Sorin, Fabien},
  journal = {Science Advances},
  volume  = {8},
  number  = {45},
  pages   = {eabo0869},
  year    = {2022},
  doi     = {10.1126/sciadv.abo0869}
}

@article{cao2018lantern,
  title   = {A Self-Powered Lantern Based on a Triboelectric--Photovoltaic Hybrid Nanogenerator},
  author  = {Cao, Ran and Wang, Jiaona and Xing, Yi and Song, Weixing and Li, Nianwu and Zhao, Shuai and Zhang, Chi and Li, Cheng},
  journal = {Advanced Materials Technologies},
  volume  = {3},
  number  = {4},
  pages   = {1700371},
  year    = {2018},
  doi     = {10.1002/admt.201700371}
}

@article{song2018textile,
  title   = {Ultra-Flexible and Large-Area Textile-Based Triboelectric Nanogenerators with a Sandpaper-Induced Surface Microstructure},
  author  = {Song, Jinlong and Yang, Biao and Zeng, Wenzheng and Peng, Zhi-Hui and Lin, Shuai and Li, Jian and Tao, Xiaoming},
  journal = {Materials},
  volume  = {11},
  number  = {11},
  pages   = {2120},
  year    = {2018},
  doi     = {10.3390/ma11112120}
}

@article{zhou2020ultrastretchable,
  title   = {Ultra-Stretchable Triboelectric Nanogenerator as High-Sensitive and Self-Powered Electronic Skins for Energy Harvesting and Tactile Sensing},
  author  = {Zhou, Kangkang and Zhao, Yi and Sun, Xiupeng and Yuan, Zeng and Zheng, Guoqiang and Dai, Kai and Mi, Liwei and Pan, Chao and Liu, Chuntai and Shen, Cangjie},
  journal = {Nano Energy},
  volume  = {70},
  pages   = {104546},
  year    = {2020},
  doi     = {10.1016/j.nanoen.2020.104546}
}

@article{karthikeyan2025optimising,
  title   = {Optimising the materials for triboelectric nanogenerator to harvest the wrist pulse signal: A numerical study},
  author  = {Karthikeyan, V. and Vivekanandan, Shanmugasundaram},
  journal = {RSC Advances},
  year    = {2025},
  doi     = {10.1039/D5RA06066D}
}

@article{yang2021controlled,
  title   = {On the controlled adhesive contact and electrical performance of vertical contact-separation mode triboelectric nanogenerators with micro-grooved surfaces},
  author  = {Yang, Weixu and Wang, Xiaoli and Chen, Ping and Hu, Yanqiang and Li, Lizhou and Sun, Zhen},
  journal = {Nano Energy},
  volume  = {85},
  pages   = {106037},
  year    = {2021},
  doi     = {10.1016/j.nanoen.2021.106037}
}

@article{li2019viscoelasticink,
  title   = {3D printed flexible triboelectric nanogenerator with viscoelastic inks for mechanical energy harvesting},
  author  = {Li, Hui and Li, Ruihuan and Fang, Xiaoting and Jiang, Hongwei and Ding, Xinrui and Tang, Biao and Zhou, Guofu and Zhou, Rui and Tang, Yong},
  journal = {Nano Energy},
  volume  = {58},
  pages   = {447--454},
  year    = {2019},
  doi     = {10.1016/j.nanoen.2019.01.066}
}

@article{xiong2019selfrestoring,
  title   = {Self-restoring, waterproof, tunable microstructural shape memory triboelectric nanogenerator for self-powered water temperature sensor},
  author  = {Xiong, Jiaqing and Luo, Hongsheng and Gao, Dace and Zhou, Xinran and Cui, Peng and Thangavel, Gurunathan and Parida, Kaushik and Lee, Pooi See},
  journal = {Nano Energy},
  volume  = {61},
  pages   = {584--593},
  year    = {2019},
  doi     = {10.1016/j.nanoen.2019.04.089}
}

@article{cai2022selfhealable,
  title   = {Self-healable, super-stretchable and shape-adaptive triboelectric nanogenerator based on double cross-linked PDMS for electronic skins},
  author  = {Cai, Ya-Wei and Wang, Gui-Gen and Mei, Yi-Cheng and Zhao, Da-Qiang and Peng, Jian-Jing and Sun, Ning and Zhang, Hai-Yan and Han, Jian-Chun and Yang, Yu},
  journal = {Nano Energy},
  volume  = {102},
  pages   = {107683},
  year    = {2022},
  doi     = {10.1016/j.nanoen.2022.107683}
}

@article{niu2022thermadapt,
  title   = {Self-Healing, Thermadapt Triple-Shape Memory Ionomer Vitrimer for Shape Memory Triboelectric Nanogenerator},
  author  = {Niu, Xiling and Wang, Min and Li, Xiaolong and Zhang, Yijun and Zhang, Zhe and Wang, Xin},
  journal = {ACS Applied Materials \& Interfaces},
  volume  = {14},
  number  = {44},
  pages   = {50269--50278},
  year    = {2022},
  doi     = {10.1021/acsami.2c13294}
}

@article{ye2023vitrimer,
  title   = {Flexible Vitrimers for Self-Healable Triboelectric Nanogenerators},
  author  = {Ye, Chongnan and Li, Wenjian and Voet, Vincent S. D. and Folkersma, Rudy and Pei, Yutao and Loos, Katja},
  journal = {Advanced Materials Technologies},
  volume  = {8},
  number  = {8},
  pages   = {2201670},
  year    = {2023},
  doi     = {10.1002/admt.202201670}
}

@article{kumar2023multiscale,
  title   = {Multiscale in-situ quantification of the role of surface roughness and contact area using a novel Mica-PVS triboelectric nanogenerator},
  author  = {Kumar, Charchit and Perris, Jack and Bairagi, Satyaranjan and Min, Guanbo and Xu, Yang and Gadegaard, Nikolaj and Mulvihill, Daniel M.},
  journal = {Nano Energy},
  volume  = {107},
  pages   = {108122},
  year    = {2023},
  doi     = {10.1016/j.nanoen.2022.108122}
}

@article{hossain2013more,
  title={More hyperelastic models for rubber-like materials: consistent tangent operators and comparative study},
  author={Hossain, Mokarram and Steinmann, Paul},
  journal={Journal of the Mechanical Behavior of Materials},
  volume={22},
  number={1-2},
  pages={27--50},
  year={2013},
  publisher={De Gruyter}
}

@article{wu2018insights,
  title={Insights into the mechanism of metal-polymer contact electrification for triboelectric nanogenerator via first-principles investigations},
  author={Wu, Jun and Wang, Xiaoli and Li, Hanqing and Wang, Feng and Yang, Weixu and Hu, Yanqiang},
  journal={Nano Energy},
  volume={48},
  pages={607--616},
  year={2018},
  publisher={Elsevier},
  doi={10.1016/j.nanoen.2018.04.025}
}

@article{wu2019amorphous,
  title={First-principles investigations on the contact electrification mechanism between metal and amorphous polymers for triboelectric nanogenerators},
  author={Wu, Jun and Wang, Xiaoli and Li, Hanqing and Wang, Feng and Hu, Yanqiang},
  journal={Nano Energy},
  volume={63},
  pages={103864},
  year={2019},
  publisher={Elsevier},
  doi={10.1016/j.nanoen.2019.103864}
}

@article{li2020humidity,
  title={The electron transfer mechanism between metal and amorphous polymers in humidity environment for triboelectric nanogenerator},
  author={Li, Lizhou and Wang, Xiaoli and Zhu, Pengzhe and Li, Hanqing and Wang, Feng and Wu, Jun},
  journal={Nano Energy},
  volume={70},
  pages={104476},
  year={2020},
  publisher={Elsevier},
  doi={10.1016/j.nanoen.2020.104476}
}

@article{tan2021corrugated,
  title={Electron transfer in the contact-electrification between corrugated 2D materials: A first-principles study},
  author={Tan, Dan and Willatzen, Morten and Wang, Zhong Lin},
  journal={Nano Energy},
  volume={79},
  pages={105386},
  year={2021},
  publisher={Elsevier},
  doi={10.1016/j.nanoen.2020.105386}
}

@article{zhang2024thermodynamic,
  title={Thermodynamic driving forces in contact electrification between polymeric materials},
  author={Zhang, Hang and Sundaresan, Sankaran and Webb, Michael A},
  journal={Nature Communications},
  volume={15},
  number={1},
  pages={2616},
  year={2024},
  publisher={Nature Publishing Group UK London},
  doi={10.1038/s41467-024-46932-2}
}

@article{ratanaporn2024reaxff,
  title={Electrostatic energy-driven contact electrification mechanism from the ReaxFF molecular dynamics perspective},
  author={Ratanaporn, Sart and Bunriw, Weeraya and others},
  journal={The Journal of Chemical Physics},
  volume={161},
  number={9},
  pages={094702},
  year={2024},
  publisher={AIP Publishing},
  doi={10.1063/5.0223350}
}

@article{persson2002elastic,
  title={Elastic contact between randomly rough surfaces: comparison of theory with numerical results},
  author={Persson, BNJ and Bucher, F and Chiaia, Bernardino},
  journal={Physical review B},
  volume={65},
  number={18},
  pages={184106},
  year={2002},
  publisher={APS}
}

@inproceedings{shvarts2021finite,
  title={Finite-element modelling of triboelectric nanogenerators accounting for surface roughness},
  author={Shvarts, Andrei G. and Xu, Yang and Min, Guanbo and Gadegaard, Nikolaj and Pearce, Chris J. and Mulvihill, Daniel M.},
  booktitle={Engineering Mechanics Institute Conference and Probabilistic Mechanics \& Reliability Conference},
  year={2021}
}

@article{sanglap2026multiphysics,
  title={A multiphysics model for triboelectric nanogenerator design with explicit surface roughness representation},
  author={Sanglap, MD Tanzib Ehsan and Perris, Jack and Mukherjee, Rudra and Kumar, Charchit and Mulvihill, Daniel M and Pearce, Chris J and Kaczmarczyk, Lukasz and Shvarts, Andrei G},
  journal={Nano Energy},
  pages={112314},
  year={2026},
  publisher={Elsevier}
}

@article{wen2021load,
  title={A load-dependent model of triboelectric nanogenerators for surface roughness sensing},
  author={Wen, Jian and Niu, Chunping and He, Hailong and Han, Weilong and Zhong, Yilin and Wu, Yi},
  journal={IEEE Sensors Journal},
  volume={21},
  number={18},
  pages={20220--20228},
  year={2021},
  publisher={IEEE}
}

@article{hu2024review,
  title={A review of contact electrification at diversified interfaces and related applications on triboelectric nanogenerator},
  author={Hu, Jun and Iwamoto, Mitsumasa and Chen, Xiangyu},
  journal={Nano-Micro Letters},
  volume={16},
  number={1},
  pages={7},
  year={2024},
  publisher={Springer}
}

@article{liu2022standardized,
  title={Standardized measurement of dielectric materials' intrinsic triboelectric charge density through the suppression of air breakdown},
  author={Liu, Di and Zhou, Linglin and Cui, Shengnan and Gao, Yikui and Li, Shaoxin and Zhao, Zhihao and Yi, Zhiying and Zou, Haiyang and Fan, Youjun and Wang, Jie and Wang, Zhong Lin},
  journal={Nature Communications},
  volume={13},
  number={1},
  pages={6019},
  year={2022},
  publisher={Nature Publishing Group UK London},
  doi={10.1038/s41467-022-33766-z}
}

@article{li2024visualization,
  title={Visualization and standardized quantification of surface charge density for triboelectric materials},
  author={Li, Yi and Luo, Yi and Xiao, Song and Zhang, Cheng and Pan, Cheng and Zeng, Fuping and Cui, Zhaolun and Huang, Bangdou and Tang, Ju and Shao, Tao and Zhang, Xiaoxing and Xiong, Jiaqing and Wang, Zhong Lin},
  journal={Nature Communications},
  volume={15},
  number={1},
  pages={6004},
  year={2024},
  publisher={Nature Publishing Group UK London},
  doi={10.1038/s41467-024-49660-9}
}

@article{wriggers2013third,
  title={A finite element method for contact using a third medium},
  author={Wriggers, Peter and Schr{\"o}der, J{\"o}rg and Schwarz, Alexander},
  journal={Computational Mechanics},
  volume={52},
  number={4},
  pages={837--847},
  year={2013},
  publisher={Springer},
  doi={10.1007/s00466-013-0848-5}
}

@article{faltus2024thirdmedium,
  title={Third medium finite element contact formulation for pneumatically actuated systems},
  author={Faltus, Ond{\v r}ej and Hor{\'a}k, Martin and Do{\v s}k{\'a}{\v r}, Martin and Roko{\v s}, Ond{\v r}ej},
  journal={Computer Methods in Applied Mechanics and Engineering},
  volume={431},
  pages={117262},
  year={2024},
  publisher={Elsevier},
  doi={10.1016/j.cma.2024.117262}
}

@article{jin2020softrobotics,
  title={Triboelectric nanogenerator sensors for soft robotics aiming at digital twin applications},
  author={Jin, Tao and Sun, Zhong and Li, Lilei and Zhang, Qing and Zhu, Ming and Zhang, Zengxing and Yuan, Gaowen and Chen, Tianyiyi and Tian, Yunliang and Hou, Xiaoyuan and Lee, Chwee Teck},
  journal={Nature Communications},
  volume={11},
  number={1},
  pages={5381},
  year={2020},
  publisher={Nature Publishing Group UK London},
  doi={10.1038/s41467-020-19059-3}
}

@article{pang2024digitaltwins,
  title={Triboelectric Nanogenerator-Enabled Digital Twins in Civil Engineering Infrastructure 4.0: A Comprehensive Review},
  author={Pang, Yuhang and He, Tao and Liu, Shiquan and others},
  journal={Advanced Science},
  volume={11},
  number={20},
  pages={2306574},
  year={2024},
  publisher={Wiley},
  doi={10.1002/advs.202306574}
}

@article{mizzi2022flexoelectricity,
  title={When flexoelectricity drives triboelectricity},
  author={Mizzi, Christopher A and Marks, Laurence D},
  journal={Nano letters},
  volume={22},
  number={10},
  pages={3939--3945},
  year={2022},
  publisher={ACS Publications}
}

@article{silavnieks2025flexoelectricity,
  title={Flexoelectricity driven elastic contact-separation model for triboelectrification},
  author={Silavnieks, Ulvis and Jing, Qingshen and Gadegaard, Nikolaj and Mulvihill, Daniel M and Xu, Yang},
  journal={Friction},
  year={2025},
  publisher={Tsinghua University Press}
}

@article{qiao2021mixed,
  title={Mixed triboelectric and flexoelectric charge transfer at the nanoscale},
  author={Qiao, Huimin and Zhao, Pin and Kwon, Owoong and Sohn, Ahrum and Zhuo, Fangping and Lee, Dong-Min and Sun, Changhyo and Seol, Daehee and Lee, Daesu and Kim, Sang-Woo and others},
  journal={Advanced Science},
  volume={8},
  number={20},
  pages={2101793},
  year={2021},
  publisher={Wiley Online Library}
}

@article{mcbride2020modelling,
  title={Modelling the flexoelectric effect in solids: A micromorphic approach},
  author={McBride, AT and Davydov, Denis and Steinmann, Paul},
  journal={Computer Methods in Applied Mechanics and Engineering},
  volume={371},
  pages={113320},
  year={2020},
  publisher={Elsevier}
}

@article{deng2023mixed,
  title={A mixed finite element method for large deformation of flexoelectric materials},
  author={Deng, Feng and Yu, Wenshan and Liang, Xu and Shen, Shengping},
  journal={Applied Mathematical Modelling},
  volume={118},
  pages={303--321},
  year={2023},
  publisher={Elsevier}
}

@article{liu2026mixed,
  title={Mixed finite element modeling and analysis of piezoelectric plate coupled with flexoelectric effect},
  author={Liu, Shen and Fu, Shi-Bo and Lin, Cheng-Hui and Zhu, Qing-Hua and Zhang, Shun-Qi},
  journal={International Journal of Heat and Mass Transfer},
  volume={268},
  pages={129051},
  year={2026},
  publisher={Elsevier}
}

@book{bonet2008nonlinear,
  title={Nonlinear Continuum Mechanics for Finite Element Analysis},
  author={Bonet, Javier and Wood, Richard D.},
  edition={2},
  year={2008},
  publisher={Cambridge University Press},
  address={Cambridge},
  doi={10.1017/CBO9780511755446}
}

@book{popov2019handbook,
  title={Handbook of Contact Mechanics: Exact Solutions of Axisymmetric Contact Problems},
  author={Popov, Valentin L. and He{\ss}, Markus and Willert, Emanuel},
  year={2019},
  publisher={Springer},
  address={Berlin, Heidelberg},
  doi={10.1007/978-3-662-58709-6}
}

@book{wriggers2006computational,
  title={Computational Contact Mechanics},
  author={Wriggers, Peter},
  year={2006},
  publisher={Springer},
  address={Berlin, Heidelberg},
  doi={10.1007/978-3-540-32609-0}
}

@book{kim2014introduction,
  title={Introduction to nonlinear finite element analysis},
  author={Kim, Nam-Ho},
  year={2014},
  publisher={Springer Science \& Business Media}
}

@article{hencky1928uber,
  title={Uber die Form des Elastizitatsgesetzes bei ideal elastischen Stoffen},
  author={Hencky, Heinrich},
  journal={Zeit. Tech. Phys.},
  volume={9},
  pages={215--220},
  year={1928}
}

@book{ogden1984nonlinear,
  title={Non-Linear Elastic Deformations},
  author={Ogden, Raymond W.},
  year={1984},
  publisher={Ellis Horwood},
  address={Chichester}
}

@article{mooney1940,
  author  = {Mooney, Melvin},
  title   = {A theory of large elastic deformation},
  journal = {Journal of Applied Physics},
  year    = {1940},
  volume  = {11},
  number  = {9},
  pages   = {582--592}
}

@article{rivlin1951,
  author  = {Rivlin, R. S. and Saunders, D. W.},
  title   = {Large elastic deformations of isotropic materials. VII. Experiments on the deformation of rubber},
  journal = {Philosophical Transactions of the Royal Society of London. Series A, Mathematical and Physical Sciences},
  year    = {1951},
  volume  = {243},
  number  = {865},
  pages   = {251--288}
}

@article{yeoh1990characterization,
  title={Characterization of elastic properties of carbon-black-filled rubber vulcanizates},
  author={Yeoh, O. H.},
  journal={Rubber Chemistry and Technology},
  volume={63},
  number={5},
  pages={792--805},
  year={1990},
  publisher={Rubber Division, ACS},
  doi={10.5254/1.3538289}
}

@article{arruda1993three,
  title={A three-dimensional constitutive model for the large stretch behavior of rubber elastic materials},
  author={Arruda, Ellen M. and Boyce, Mary C.},
  journal={Journal of the Mechanics and Physics of Solids},
  volume={41},
  number={2},
  pages={389--412},
  year={1993},
  publisher={Elsevier},
  doi={10.1016/0022-5096(93)90013-6}
}

@article{gent1996new,
  title={A new constitutive relation for rubber},
  author={Gent, Alan N.},
  journal={Rubber Chemistry and Technology},
  volume={69},
  number={1},
  pages={59--61},
  year={1996},
  publisher={Rubber Division, ACS},
  doi={10.5254/1.3538357}
}

@article{fan2012flexible,
  title={Flexible triboelectric generator},
  author={Fan, Feng-Ru and Tian, Zhong-Qun and Wang, Zhong Lin},
  journal={Nano energy},
  volume={1},
  number={2},
  pages={328--334},
  year={2012},
  publisher={Elsevier}
}

@inproceedings{mandriota2024modelling,
  title={Modelling viscoelastic adhesion and friction in sliding contact mechanics},
  author={Mandriota, Cosimo and Carbone, Giuseppe and Menga, Nicola},
  booktitle={International Tribology Symposium of IFToMM},
  pages={406--415},
  year={2024},
  organization={Springer}
}

@article{ogden1972large,
  title={Large deformation isotropic elasticity: On the correlation of theory and experiment for incompressible rubberlike solids},
  author={Ogden, Raymond W.},
  journal={Proceedings of the Royal Society of London. A. Mathematical and Physical Sciences},
  volume={326},
  number={1567},
  pages={565--584},
  year={1972},
  publisher={The Royal Society},
  doi={10.1098/rspa.1972.0026}
}

@misc{comsol_bem_fem_electrostatics_learning_center,
  author       = {{COMSOL, Inc.}},
  title        = {Fundamentals of Modeling Electrostatics with a BEM--FEM Approach},
  year         = {n.d.},
  howpublished = {\url{https://www.comsol.com/support/learning-center/article/fundamentals-of-modeling-electrostatics-with-a-bemfem-approach-26991}},
  note         = {COMSOL Learning Center},
}

@article{mastropasqua2025adhesion,
  title={Adhesion and friction in viscoelastic rough contacts},
  author={Mastropasqua, Michele and Ronchi, Andrea and Serra, Antonio and Menga, Nicola and Carbone, Giuseppe},
  journal={Tribology International},
  pages={111000},
  year={2025},
  publisher={Elsevier}
}

@article{cavalieri2013augmented,
  title={An augmented Lagrangian technique combined with a mortar algorithm for modelling mechanical contact problems},
  author={Cavalieri, Federico Jos{\'e} and Cardona, Alberto},
  journal={International Journal for Numerical Methods in Engineering},
  volume={93},
  number={4},
  pages={420--442},
  year={2013},
  publisher={Wiley Online Library}
}

@article{laursen1993continuum,
  title={A continuum-based finite element formulation for the implicit solution of multibody, large deformation-frictional contact problems},
  author={Laursen, TA and Simo, JC1242927},
  journal={International Journal for numerical methods in engineering},
  volume={36},
  number={20},
  pages={3451--3485},
  year={1993},
  publisher={Wiley Online Library}
}

@book{popov2010contact,
  title={Contact Mechanics and Friction: Physical Principles and Applications},
  author={Popov, Valentin L and others},
  year={2010},
  publisher={Springer},
  doi={10.1007/978-3-642-10803-7}
}

@article{he2020boosting,
  title   = {Boosting output performance of sliding mode triboelectric nanogenerator by charge space-accumulation effect},
  author  = {He, Wencong and Liu, Wenlin and Chen, Jie and Wang, Zhao and Liu, Yike and Pu, Xianjie and Yang, Hongmei and Tang, Qian and Yang, Huake and Guo, Hengyu and Hu, Chenguo},
  journal = {Nature Communications},
  volume  = {11},
  year    = {2020},
  doi     = {10.1038/s41467-020-18086-4}
}

@article{yang2024anisotropic,
  title   = {Anisotropic tribology and electrification properties of sliding-mode triboelectric nanogenerator with groove textures},
  author  = {Yang, Weixu and Wang, Jieyang and Wang, Xiaoli and Chen, Ping},
  journal = {Friction},
  volume  = {12},
  number  = {8},
  pages   = {1828--1837},
  year    = {2024},
  doi     = {10.1007/s40544-024-0861-z}
}

@inproceedings{kumar2023mechanics,
  title={Mechanics of Sliding Triboelectric Nanogenerators for Sustainable Energy Harvesting},
  author={Kumar, Charchit and Laughtland, Kaspar and Bairagi, Satyaranjan and Gadegaard, Nikolaj and Mulvihill, Daniel},
  booktitle = {17th International Conference on Advances in Experimental Mechanics}, 
  address = {Glasgow, UK},
  note = {Conference abstract},
  year={2023}
}

@article{mulvihill2025test,
  title={How to Test Triboelectric Nanogenerators: Key Factors for Standardized Performance Evaluation},
  author={Mulvihill, Daniel M and Mukherjee, Rudra and Xu, Yang and Kumar, Charchit and Khandelwal, Gaurav and Perris, Jack and Bairagi, Satyaranjan and Aliyana, Akshaya Kumar and Stylios, George K and Pillai, Suresh C and others},
  journal={Advanced Energy Materials},
  volume={15},
  number={44},
  pages={e02920},
  year={2025},
  publisher={Wiley Online Library}
}

@article{vsutka2026electric,
  title={Electric fields control static charge during contact-friction-separation},
  author={{\v{S}}utka, Andris and Lap{\v{c}}inskis, Linards and Linarts, Artis and Sun, Qiang and Lungevics, Janis and Egl{\=\i}tis, Raivis and Knite, M{\=a}ris and Dassanayaka, Dumindu G and Ellis, Amanda V and Sherrell, Peter C},
  journal={Nature Communications},
  volume={17},
  number={1},
  pages={9971},
  year={2026},
  publisher={Nature Publishing Group UK, London}
}

@article{kim2025coupling,
  title={Coupling of finite and boundary element methods for contact analysis of dielectric solids immersed in electrostatic medium},
  author={Kim, Moonhong and Sohn, Dongwoo},
  journal={Computers \& Structures},
  volume={307},
  pages={107591},
  year={2025},
  publisher={Elsevier}
}

@book{landau1984electrodynamics,
  title     = {Electrodynamics of Continuous Media},
  author    = {Landau, L. D. and Lifshitz, E. M.},
  publisher = {Pergamon Press},
  edition   = {2},
  year      = {1984}
}

@book{jackson1999classical,
  title     = {Classical Electrodynamics},
  author    = {Jackson, John David},
  publisher = {John Wiley \& Sons},
  edition   = {3},
  year      = {1999}
}

@article{suo2008nonlinear,
  title   = {A nonlinear field theory of deformable dielectrics},
  author  = {Suo, Zhigang and Zhao, Xuanhe and Greene, William H.},
  journal = {Journal of the Mechanics and Physics of Solids},
  volume  = {56},
  number  = {2},
  pages   = {467--486},
  year    = {2008},
  doi     = {10.1016/j.jmps.2007.05.021}
}

@article{pelrine2000high,
  title   = {High-field deformation of elastomeric dielectrics for actuators},
  author  = {Pelrine, Ronald and Kornbluh, Roy and Pei, Qibing and Joseph, Jose},
  journal = {Materials Science and Engineering: C},
  volume  = {11},
  number  = {2},
  pages   = {89--100},
  year    = {2000},
  doi     = {10.1016/S0928-4931(00)00128-4}
}

@article{gao2011modeling,
  title   = {Modeling and simulation of the coupled mechanical--electrical response of dielectric elastomer actuators},
  author  = {Gao, Zhigang and Wang, Lizhi and Hu, Hongyuan},
  journal = {Journal of Intelligent Material Systems and Structures},
  volume  = {22},
  number  = {10},
  pages   = {1109--1118},
  year    = {2011}
}

@article{tuncer2005modeling,
  title   = {Modeling electromechanical properties of layered electrets: Application of the finite-element method},
  author  = {Tuncer, Enis and Wegener, Michael and Gerhard-Multhaupt, Reimund},
  journal = {Journal of Applied Physics},
  volume  = {98},
  number  = {2},
  pages   = {024106},
  year    = {2005},
  doi     = {10.1063/1.1989449}
}

@article{jia2016numerical,
  title   = {Numerical study on the electromechanical behavior of dielectric elastomer with finite element method},
  author  = {Jia, Kun and Li, Tao and Wang, Fei and Chen, Xiaodong},
  journal = {International Journal of Smart and Nano Materials},
  volume  = {7},
  number  = {1},
  pages   = {1--14},
  year    = {2016},
  doi     = {10.1080/19475411.2016.1170074}
}

@article{dorfmann2017nonlinear,
  title   = {Nonlinear electroelasticity: material properties, continuum theory and applications},
  author  = {Dorfmann, Luis and Ogden, Ray W.},
  journal = {Proceedings of the Royal Society A},
  volume  = {473},
  number  = {2204},
  pages   = {20170311},
  year    = {2017},
  doi     = {10.1098/rspa.2017.0311}
}

@article{wissler2007electromechanical,
  title   = {Electromechanical coupling in dielectric elastomer actuators},
  author  = {Wissler, Michael and Mazza, Edoardo},
  journal = {Sensors and Actuators A: Physical},
  volume  = {138},
  number  = {2},
  pages   = {384--393},
  year    = {2007},
  doi     = {10.1016/j.sna.2007.05.029}
}

@article{yastrebov2016sliding,
  title={Sliding without slipping under Coulomb friction: opening waves and inversion of frictional force},
  author={Yastrebov, Vladislav A},
  journal={Tribology Letters},
  volume={62},
  number={1},
  pages={1},
  year={2016},
  publisher={Springer}
}

@article{yu2004adhesive,
  title={Adhesive contact based on the Lennard--Jones potential: a correction to the value of the equilibrium distance as used in the potential},
  author={Yu, Ning and Polycarpou, Andreas A},
  journal={Journal of Colloid and Interface Science},
  volume={278},
  number={2},
  pages={428--435},
  year={2004},
  publisher={Elsevier}
}

@article{Shi2019MoreThanEnergy,
  author  = {Shi, Q. and He, T. and Lee, C.},
  title   = {More than energy harvesting -- Combining triboelectric nanogenerator and flexible electronics technology for enabling novel micro-/nano-systems},
  journal = {Nano Energy},
  volume  = {57},
  pages   = {851--871},
  year    = {2019},
  doi     = {10.1016/j.nanoen.2019.01.002}
}

@article{hossain2015eight,
  title={Eight-chain and full-network models and their modified versions for rubber hyperelasticity: a comparative study},
  author={Hossain, Mokarram and Amin, AFMS and Kabir, Muhammad Nomani},
  journal={Journal of the mechanical behavior of materials},
  volume={24},
  number={1-2},
  pages={11--24},
  year={2015},
  publisher={De Gruyter}
}

@article{hossain20203d,
  title={3D printed elastomeric polyurethane: Viscoelastic experimental characterizations and constitutive modelling with nonlinear viscosity functions},
  author={Hossain, Mokarram and Navaratne, Rukshan and Peri{\'c}, Djordje},
  journal={International Journal of Non-Linear Mechanics},
  volume={126},
  pages={103546},
  year={2020},
  publisher={Elsevier}
}

@article{zhou2025modified,
  title={A modified semi-soft model of liquid crystal elastomers: Application to elastic and viscoelastic responses},
  author={Zhou, Yu and Wei, Chen and Jin, Lihua},
  journal={Journal of the Mechanics and Physics of Solids},
  volume={196},
  pages={106027},
  year={2025},
  publisher={Elsevier}
}

@article{wineman2009nonlinear,
  title={Nonlinear viscoelastic solids—a review},
  author={Wineman, A25126821197},
  journal={Mathematics and mechanics of solids},
  volume={14},
  number={3},
  pages={300--366},
  year={2009},
  publisher={Sage Publications Sage UK: London, England}
}

@article{pipkin1968nonlinear,
  title   = {A non-linear integral representation for viscoelastic behaviour},
  author  = {Pipkin, A. C. and Rogers, T. G.},
  journal = {Journal of the Mechanics and Physics of Solids},
  volume  = {16},
  number  = {1},
  pages   = {59--72},
  year    = {1968},
  doi     = {10.1016/0022-5096(68)90016-1}
}

@article{schapery1969characterization,
  title   = {On the characterization of nonlinear viscoelastic materials},
  author  = {Schapery, R. A.},
  journal = {Polymer Engineering \& Science},
  volume  = {9},
  number  = {4},
  pages   = {295--310},
  year    = {1969},
  doi     = {10.1002/pen.760090410}
}

@article{simo1987finite,
  title   = {On a fully three-dimensional finite-strain viscoelastic damage model: Formulation and computational aspects},
  author  = {Simo, J. C.},
  journal = {Computer Methods in Applied Mechanics and Engineering},
  volume  = {60},
  number  = {2},
  pages   = {153--173},
  year    = {1987},
  doi     = {10.1016/0045-7825(87)90107-1}
}

@article{reese1998theory,
  title   = {A theory of finite viscoelasticity and numerical aspects},
  author  = {Reese, Stefanie and Govindjee, Sanjay},
  journal = {International Journal of Solids and Structures},
  volume  = {35},
  number  = {26--27},
  pages   = {3455--3482},
  year    = {1998},
  doi     = {10.1016/S0020-7683(97)00217-5}
}

@article{bergstrom1998constitutive,
  title   = {Constitutive modeling of the large strain time-dependent behavior of elastomers},
  author  = {Bergstr{\"o}m, J. S. and Boyce, M. C.},
  journal = {Journal of the Mechanics and Physics of Solids},
  volume  = {46},
  number  = {5},
  pages   = {931--954},
  year    = {1998},
  doi     = {10.1016/S0022-5096(97)00075-6}
}

@article{arief2026metasurface,
  title={Large-Area 2D Metasurface-Based Triboelectric E-Skin Arrays: Contact \& Proximity Tactile Mapping with Broadband Acoustic Readouts},
  author={Arief, Injamamul and Sarkar, Swagato and Ghosh, Anik Kumar and Verners, Osvalds and Meena, Kamal Kumar and Lee, Su-Hyeong and Chae, Soosang and K{\"o}nig, Tobias AF and Krause, Beate and Fery, Andreas and others},
  journal={Advanced Materials},
  volume={38},
  number={36},
  pages={e21525},
  year={2026},
  publisher={Wiley Online Library}
}

\end{document}